\documentclass[11pt]{article}
\usepackage{jheppubmod}
\pdfoutput=1
\usepackage[showonlyrefs]{mathtools}
\usepackage{psfrag}
\usepackage{array}
\usepackage{amssymb}
\usepackage{graphicx}
\usepackage{slashed}
\usepackage[labelsep=quad]{subcaption}
\mathtoolsset{showonlyrefs}
\usepackage{epstopdf}
	
\usepackage{epsfig}
\usepackage[punctsep]{collref}
\collectsep[]{;}

\def\al{A}

\def\tr{{\rm Tr}}

\def\lvac{\langle 0 |}
\def\rvac{|0 \rangle}

\def\Or[#1]{{\text{O}}\left({#1}\right)}
\def\dotl[#1,#2]{\left\langle #1,\, #2 \right\rangle}
\def\dotlb[#1,#2]{\left\langle #1,\, #2 \right\rangle}
\def\dotlm[#1,#2]{\left[ #1,\, #2 \right]}
\def\dotp[#1,#2]{(\vect{#1} \cdot\vect{#2})}
\def\aff[#1,#2]{\hat{#1}(#2)}
\def\n4sym{{\cal N}=4 SYM}
\def\>{\rangle}
\def\<{\langle}
\def\weight[#1,#2,#3]{\{(#1),#2,#3\}}
\def\ads[#1]{$\text{AdS}_{#1}$}
\def\cft[#1]{$\text{CFT}_{#1}$}

\newcommand{\op}{{\cal O}} 
\newcommand{\lads}{\ell_{\text{ads}}}
\newcommand{\mtr}{M^{\text{tr}}}
\newcommand{\alset}{{\cal A}}
\newcommand{\lpl}{\ell_{\text{pl}}}
\newcommand{\mpl}{M_{\text{pl}}}
\newcommand{\dmwp}{d \mu_{\text{WP}}}
\newcommand{\lplp}{\ell_{\text{pl},p}}
\newcommand{\be}{\begin{equation}}
\newcommand{\ee}{\end{equation}}
\newcommand{\ba}{\begin{align}}
\newcommand{\ea}{\end{align}}
\newcommand{\bs}{\begin{split}}
\newcommand{\detp}{\text{det}'}
\def\sess\end{split}

\newcommand{\vect}[1]{{\vec{#1}}}

\newcommand{\non}{\nonumber}
\notoc
\title{Loss of locality in gravitational correlators with a large number of insertions}
\author{Sudip Ghosh and Suvrat Raju}
\affiliation{International Centre for Theoretical Sciences, Tata Institute of Fundamental Research, Shivakote, Bengaluru 560089, India.}
\emailAdd{sudip.ghosh@icts.res.in}
\emailAdd{suvrat@icts.res.in}

\date{}

\abstract{We review lessons from the AdS/CFT correspondence that indicate that the emergence of locality in quantum gravity is contingent on considering observables with a small number of insertions. Correlation functions where the number of insertions scales with a power of the central charge of the CFT  are sensitive to nonlocal effects in the bulk theory, which arise from a combination of the effects of the bulk Gauss law and a breakdown of perturbation theory. To examine whether a similar effect occurs in flat space,  we consider the scattering of massless particles in the bosonic string and the superstring in the limit where the number of external particles, n,  becomes very large. We use estimates of the volume of the Weil-Petersson moduli space of punctured Riemann surfaces to argue that string amplitudes grow factorially in this limit.  We  verify this factorial behaviour through an extensive numerical analysis of string amplitudes at large n. Our numerical calculations rely on the observation that, in the large n limit, the string scattering amplitude localizes on the Gross-Mende saddle points, even though individual particle energies are small. This factorial growth implies the breakdown of string perturbation theory for $n \sim \left({M_{\text{pl}} \over E} \right)^{d-2}$ in $d$ dimensions where $E$ is the typical individual particle energy.  We explore the implications of this breakdown for the black hole information paradox. We show that the loss of locality suggested by this breakdown is precisely sufficient to resolve the cloning and strong subadditivity paradoxes.}

\begin{document}
\maketitle

\tableofcontents
\section{Introduction}
It is generally recognized that quantum gravity cannot be an exactly local theory due to the difficulty of localizing operators in spacetime to an accuracy better than the Planck length, $\lpl$. In this paper, we would like to present evidence that, for at least some observables, these nonlocal effects can spread over {\em macroscopic distance scales}. We will argue that the observables that are sensitive
to these effects are correlation functions with a very ``large'' number of insertions. A significant part of our analysis will be devoted to quantifying what we mean by ``large'' here.

The initial motivation to consider such effects came from the results of  \cite{Papadodimas:2013jku,Papadodimas:2013wnh,Papadodimas:2015xma,Papadodimas:2015jra}. These papers proposed a representation of the interior of  large black holes in the AdS/CFT correspondence \cite{Maldacena:1997re,Gubser:1998bc,Witten:1998qj}. But this representation had the remarkable property that the operators that described degrees of freedom in the interior of the black-hole were complicated combinations
of operators that described the exterior of the black hole. This implied that a suitably complicated combination of exterior operators would fail to commute with an interior operator even if the exterior and interior operators were separated by a distance that was large in Planck units.

However, it is natural to expect that if these nonlocal effects are real, then they should be visible even in empty space, in the absence of black holes. This is indeed the case. A  simple example of these nonlocal effects was examined in a controlled setting in \cite{Banerjee:2016mhh}.  The paper \cite{Banerjee:2016mhh} considered an operator localized in the center of  empty AdS.\footnote{We describe what we mean by an operator ``localized'' at a point in greater detail in section \ref{localopclarify}.}
In AdS/CFT, operators in the bulk of AdS can be mapped to the CFT using the standard HKLL mapping \cite{Hamilton:2006az,Hamilton:2005ju,Hamilton:2006fh}. The authors of \cite{Banerjee:2016mhh} then showed that this operator could be explicitly rewritten as a complicated combination of other operators that were localized near the boundary of AdS on the same time slice. We review this construction in section \ref{sectoycomp}, where we show that the operator in the center of AdS, $\phi(0)$, can be written as
\be
\label{bulkrewriting}
\phi(0) = \sum_{n, m} c_{n m} X_n {\cal P}_0 X_m^{\dagger}.
\ee
Here $X_n$ and $X_m$ are simple polynomials in operators localized near the boundary, $c_{n m}$ are c-number coefficients, and ${\cal P}_0$, which we calculate explicitly in \ref{sectoycomp}, is a complicated polynomial that involves the bulk graviton fluctuations near the boundary and has a degree as large as $N$, where $N^2$ is the central charge of the theory. 

The central feature of \eqref{bulkrewriting} is that the operators  that appear in the polynomials on the right hand side are all localized at points that are {\em spacelike} to the operator that appears on the left hand side. This equation obviously implies non-zero commutators at spacelike separation. For example an operator that contains the momentum conjugate to $\phi(0)$ will fail to commute with the operators on the right hand side of \eqref{bulkrewriting}.
However, the relation \eqref{bulkrewriting} provides us with a stronger statement: it tells us that the information in the center of AdS can be entirely recovered by making a suitably complicated measurement, on the same time slice,  near the boundary of AdS. Thus \eqref{bulkrewriting} provides a  toy-model of the phenomenon of black-hole complementarity  \cite{'tHooft:1984re,Susskind:1993if}. 

There are two physical effects that allow the relation \eqref{bulkrewriting} to hold. One of them is the bulk Gauss law, which tells us that the energy of a local operator can be measured through a Hamiltonian that is entirely defined through a surface integral at infinity. The Gauss law itself leads to small nonvanishing commutators between operators at spacelike separations. In AdS, these commutators are suppressed by factors of ${1 \over N}$. However, the key to \eqref{bulkrewriting} is that by taking complicated enough combinations of localized operators, we can enhance these ${1 \over N}$ effects to an $\Or[1]$ effect. This requires the {\em breakdown} of ${1 \over N}$ perturbation theory. Thus the nonlocal effects that we are looking for arise from a combination of a kinematical effect --- the Gauss law --- and a dynamical effect --- the breakdown of gravitational perturbation theory. 

In this paper, our main focus is on exploring whether similar effects exist in flat space and, if so, on the implications that such effects might have for black-hole evaporation. Of course, the kinematic ingredient that was present in AdS --- the Gauss law --- operates in flat space as well.  In flat space, the Gauss law leads to commutators between local operators that are suppressed by a power of $\left({E \over M_{\text{pl}}} \right)$ where $E$ is a measure of the energy of the configuration of operators and $M_{\text{pl}} = {1 \over \lpl}$ is the Planck scale.

So, in this paper, we will focus on the dynamical effect that was required above: the breakdown of perturbation theory.  In flat space, the analogue of the  breakdown of the ${1 \over N}$ expansion is the 
breakdown of gravitational perturbation theory for correlation functions where the number of insertions becomes very large, even if these insertions are well separated. 

The fact that the breakdown of perturbation theory corresponds to a loss of bulk locality in theories of gravity is also natural from the path-integral viewpoint.  The reason that gravity is approximately local for simple observables, even though the path-integral sums over all metrics, is because the path-integral is dominated by a saddle-point in many circumstances. This saddle-point provides a dominant metric and when we refer to approximate locality, we are referring to locality as defined by this metric.

The breakdown of perturbation theory is a sign that the saddle-point approximation is no longer valid for a particular observable. So it is natural that perturbative breakdown in gravity corresponds to either a loss or a change in the notion of locality.\footnote{We caution the reader that, in doing the quantum-gravity path integral, we hold the asymptotic geometry fixed.  Therefore, in this entire discussion, loss of locality only refers to a loss of ``bulk locality'' and not any change in the causal relationship of points on the asymptotic boundary to each other. This is true in AdS/CFT as well where the boundary theory remains exactly local even though the bulk theory is not exactly local.}

To study this perturbative breakdown in a precise setting we focus on S-matrix elements rather than correlation functions since scattering amplitudes are naturally gauge invariant.  We also study scattering in string theory--- both the bosonic theory and superstring theory --- rather than pure gravity. This helps us ensure that the breakdown of perturbation theory that we describe here is not cured by stringy effects. However, it is also true --- perhaps somewhat surprisingly --- that the technical analysis of the breakdown of perturbation theory turns out to be easier in string theory than in pure gravity for reasons that we explain  in section \ref{puregravcomments}.

Our results are as follows. We consider the scattering of massless particles in the bosonic string and type II superstring theory with external polarization tensors chosen so that these particles correspond
to linear combinations of the graviton and the dilaton. We take the limit where the number of external particles, $n \rightarrow \infty$, but where the energy per particle is taken to zero in string units $E \sqrt{\alpha'} = {1 \over n^{\gamma}}$ with $0 < \gamma < {1 \over d - 2}$,   and the dimensionless string coupling constant, $g_s$ is also taken to be very small.  Then we argue that string perturbation theory breaks down for $n \propto \left( {1 \over g_s^{2}} \right)^{1 \over 1 - (d-2) \gamma}$. This threshold for the breakdown of perturbation theory  can also be rewritten as $n \propto \left({M_{\text{pl}} \over E}\right)^{d-2}$.

To derive this bound, we first derive some simple bounds on the growth of tree amplitudes in any perturbative theory in section \ref{secbounds}. These bounds state that if tree amplitudes grow factorially in the number of external particles then, provided the energy per particle does not fall too rapidly, perturbation theory eventually breaks down for a large enough number of external particles. In section \ref{analyticgrowth} and section \ref{numericalgrowth} we then argue that tree-level scattering amplitudes, both in the bosonic string and the superstring, do grow factorially. 

Our arguments in section \ref{analyticgrowth} rely on the growth of the volume of moduli space of punctured Riemann surfaces. This is a subject that has attracted some recent attention in the mathematical literature. We are able to utilize these results by arguing that at large $n$, the size of the string amplitudes is bounded below by the volume of moduli space.

Our arguments in section \ref{numericalgrowth} are independent and rely on a numerical study of string scattering at large $n$. Here, we exploit the fact that at large $n$, the integral over moduli space localizes to a set of saddle points that are solutions of the so-called ``scattering equations.'' By solving these equations numerically, we are able to numerically estimate the growth of amplitudes. This numerical estimate precisely matches the estimate from the volume of moduli space that we derive in section \ref{analyticgrowth}.

The arguments above then suggest that scattering experiments with a large enough number of external particles may be sensitive to nonlocal effects in the bulk. In section \ref{secinfo}, we explore the implications of such effects for the information paradox. In particular, we consider two versions of the information paradox --- the ``cloning paradox'' and the ``strong subadditivity paradox.''

The cloning paradox is based on the observation that the Hawking radiation that emerges from a black hole at late times carries information about the infalling matter. However, the geometry of the evaporating black hole suggests that we can draw a spacelike nice slice that intersects both the infalling matter and the outgoing Hawking radiation. This seems to suggest that the same information is present in two places, which would violate the linearity of quantum mechanics. 

However, when we carefully examine the observables that are required to extract information from the outgoing radiation, it turns out that we need to measure $S$-point correlators of the outgoing Hawking quanta, at intervals of size ${1 \over E}$, where $S$ is the entropy of the black hole and $E$ is energy of a typical Hawking quanta. Our analysis of the breakdown of perturbation theory tells us that gravitational perturbation theory breaks down {\em precisely} for such correlators.

This suggests an elegant resolution to the cloning paradox. It is {\em not} the case that two different operators can extract the same information from the state. Rather the simple operator that acts on the infalling matter should be {\em equated} to a complicated operator that acts on the outgoing Hawking radiation. Even though it seems that these operators are distinct, because they seem to act at different points in space, they may still be related through a relation of the form \eqref{bulkrewriting}. Therefore our resolution to the cloning paradox is that the linearity of quantum mechanics is preserved and it is our notion of locality that must be modified.

The strong subadditivity paradox is closely related to the cloning paradox. As we review in section \ref{secinfo}, this paradox relies on splitting up the black-hole spacetime into three regions on a spacelike slice. Plausible arguments about the von Neumann entropies of each of these three regions then suggest that these entropies violate the strong subadditivity of entropy. However, the von Neumann entropy is a fine-grained probe of a state and we argue in \ref{secprotocol} that measuring this quantity is equivalent to measuring a $S$-point correlator in the Hawking radiation. Therefore, nonlocal effects may be important for such quantities. In particular, we should not expect that the Hilbert space of the theory factorizes into a tensor product of the Hilbert spaces corresponding to different local regions on a spacelike slice. Thus the strong subadditivity paradox may also be resolved by recognizing limits on locality in gravity. 

In section \ref{sectoyinfo} we show how the simple setup of empty AdS considered in section \ref{sectoycomp} can also be used to produce toy models of both the cloning and the strong subadditivity paradoxes. In this setting, the resolution to both of these paradoxes is absolutely clear and involves, as we expect, a loss of bulk locality rather than any modification of quantum mechanics.

A summary of the main results in this paper was provided in \cite{Ghosh:2016fvm}. The scattering of a large number of particles was also studied in \cite{Dvali:2014ila,Addazi:2016ksu} although the motivation and perspective of these papers was different from ours. The significance of $\Or[S]$ point correlators for the information paradox, and the fact that they might deviate strongly from naive expectations, was also  discussed in \cite{ArkaniHamed:2007ky}.

Our conventions in the rest of the paper are as follows. We set  $\alpha' = 2$. We will use the string coupling constant $4 \pi^2 g_s^2  = \kappa^2 \left({\alpha' \over 2} \right)^{2 - d \over 2}$. With our choice of units, this simply becomes $g_s = {\kappa \over 2 \pi}$ where $\kappa$ is the $d$-dimensional gravitational coupling. We also use $\kappa^2 = 8 \pi G =  \lpl^{d-2}$. 
 
\section{Locality in gravity and perturbation theory \label{secpertloc}}
In this section, we describe the relation between approximate notions of locality in gravity and the validity of perturbation theory. First, we clarify what we mean by an approximately local operator in a theory of gravity. Then we review the results of \cite{Banerjee:2016mhh}. This analysis provides an explicit example of nonlocality in quantum gravity in a controlled setting.  Then we abstract some lessons from this  example, and show how appreciable nonlocal effects in correlators result from a {\em combination} of the effect of the Gauss law and the breakdown of perturbation theory. We then provide some additional arguments for nonlocality based on the path-integral. Finally, we emphasize the distinction between asymptotic notions of locality, which we expect to be exact, and bulk locality, which we expect is only approximate. 

\subsection{Approximately local operators in gravity \label{localopclarify}}
Since this paper is devoted to  the study of nonlocal effects in gravity, it is important to clarify what we mean by an approximately local operator. For simplicity we will consider scalar operators. Under a diffeomorphism, $x^{\mu}  \rightarrow  x^{\mu} + \eta^{\mu}$, a  scalar operator transforms as $\phi(x) \rightarrow \phi(x) - \eta^{\mu} \partial_{\mu} \phi(x)$. Therefore, unless we provide additional information, this operator does not provide us with gauge invariant information.

The simplest way to resolve this issue is just to fix gauge. Alternately, it is possible to use a relational prescription that fixes the location of the operator with reference to an asymptotic boundary. An example of such a relational prescription is given in section 4 of \cite{Papadodimas:2015xma} or section 3.1.1 of \cite{Papadodimas:2015jra}. 

Both alternatives then lead to operators that have the following important property. If $x_1, x_2$ are spacelike separated then in the limit where $M_{\text{pl}} \rightarrow \infty$, and within low point correlators these operators satisfy
\be
\label{spacelikecomm}
\lvac [\phi({x_1}), \phi({x_2})] \phi(x_3) \ldots \phi(x_{2 n}) \rvac = 0,~\text{as}~\mpl \rightarrow \infty,
\ee
where $n$ does not scale with any power of $M_{\text{pl}}$ and $|x_{i j}| \gg {1 \over \mpl}, \forall i,j$.

The property \eqref{spacelikecomm} {\em defines} what we mean by a local operator in this paper. Note that the commutator of such operators does {\em not} vanish in general when $\mpl$ is finite.  Second,  as we describe in great detail in the rest of this paper various subtle effects arise when we keep $M_{\text{pl}}$ finite and scale the number of insertions in a correlator with a power of $M_{\text{pl}}$. Consequently, the notion of locality described above is only {\em approximate}. We will sometimes refer to  will refer to such operators as  ``quasilocal'' operators.

\subsection{Small algebras and locality in AdS/CFT \label{sectoycomp}}
If nonlocal effects are present in nature, we would expect that they should be observable even in empty space without black-hole horizons which are, after all, teleological objects. The AdS/CFT correspondence provides us with a simple setting to study such effects, and it is indeed true that these effects are present even in empty AdS. This was shown in \cite{Banerjee:2016mhh}. We review this example below, and it will serve as a prototype for the nonlocal effects that we will later invoke while considering the information paradox in flat space.

\begin{figure}
\begin{center}
\includegraphics[height=0.3\textheight]{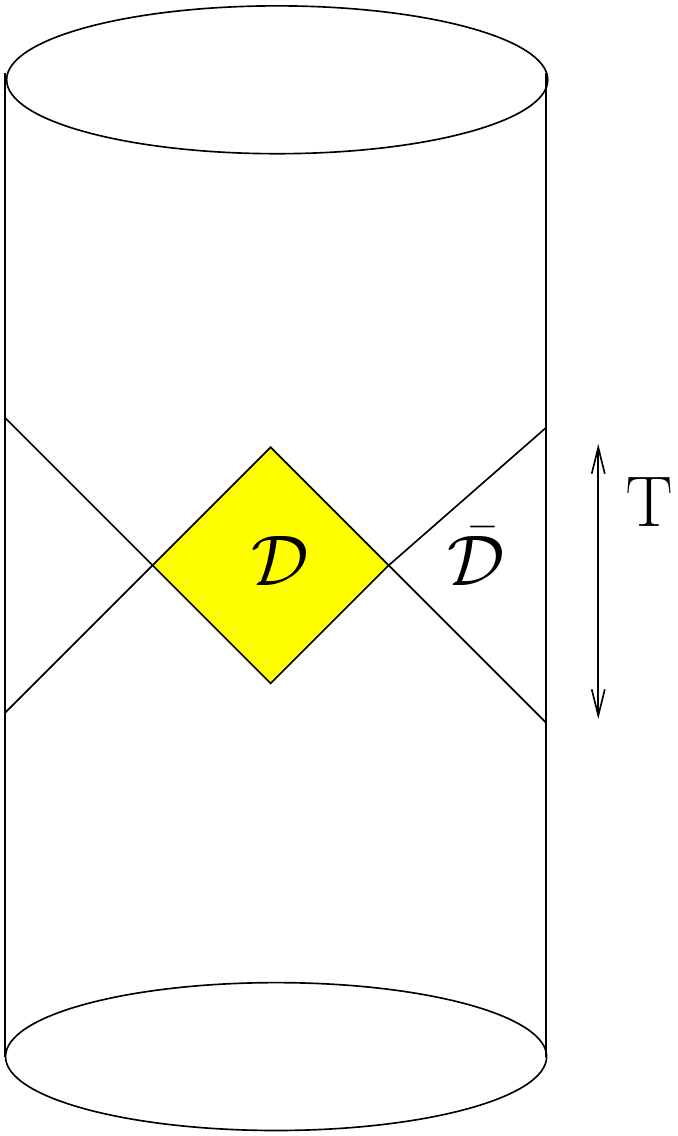}
\caption{\em A band of length $T < \pi$ on the boundary. A causal diamond $D$ in the middle of AdS is out of causal contact with the band. \label{figadsband}}
\end{center}
\end{figure}

Consider empty global AdS$_{d+1}$, where we set the AdS radius $\lads = 1$ so that the metric is
\be
ds^2 =  -\left(1+r^2\right) dt^2 +{ dr^2 \over 1+r^2} + r^2 d\Omega_{d-1}^2.
\ee
 We also consider  a band of length $T$ on its boundary, where $T < {\pi}$ so that the band is smaller than the light-crossing time of AdS. This gives rise to the setup shown in Figure \ref{figadsband}. We denote the band itself by ${\cal B}$, by which we mean the set of all points on the boundary with time coordinate between ${-T \over 2}$ and ${T \over 2}$. A causal diamond ${\cal D}$, in the bulk, is out of causal contact with ${\cal B}$. We also have the ``causal wedge'' ${\bar{\cal D}}$ in the bulk. Here, by causal wedge we mean the region in the bulk such that from each point in this region, it is possible to send {\em both} a future directed light ray and a past-directed light-ray to the band ${\cal B}$. In the literature, the term ``causal wedge'' is often only applied to bulk regions that are dual to an entire causal diamond on the boundary, but this terminology also makes sense in the setting considered here. The significance of the causal wedge is that perturbative fields propagating in the causal wedge can be related to simple operators in ${\cal B}$ using the equations of motion.

 The bulk AdS may have various propagating, weakly coupled fields. We will take $\phi$ to be one such field, dual to an operator $\op$ of dimension $\Delta$. The standard AdS/CFT dictionary then relates the boundary value of $\phi$ to the value of $\op$ through
\be
\lim_{r \rightarrow \infty} r^{\Delta} \phi(t, r, \Omega) = \op(t, \Omega).
\ee
If the field $\phi$ is weakly coupled in the bulk, then the operator $\op(t,\Omega)$ is a  generalized free-field in the conformal field theory that is dual to this bulk theory. 

We would now like to consider the ``algebra'' of ``low-order polynomials'' in these generalized free-fields. This means that we consider the set of polynomials in generalized free-fields
\be
\label{alsetdef}
\alset = \{\op(t_1, \Omega_1), \op(t_2, \Omega_2) \op(t_3, \Omega_3), \ldots \op(t_m, \Omega_m) \op(t_{m+1} \Omega_{m+1})  \ldots \op(t_{m+n}, \Omega_{m+n}) \},
\ee
where $(t_i, \Omega_i) \in {\cal B}$. We put a cutoff on the number of operators that such a polynomial can contain by imposing $n \ll N$, where $N^2 \equiv \left({\ell_{\text{ads}} \over \ell_{\text{pl}}} \right)^{d-1}$. Here when we write $n \ll N$ we mean that in the limit where $N \rightarrow \infty$, $n$ does not scale as any finite power of $N$. Note that, as a result of this cutoff,  $\alset$ is, strictly speaking, just a set and not an algebra since it is not closed under multiplication. Nevertheless we will continue to use the phrase ``algebra'' below although this caveat should be kept in mind.

Now, by solving the bulk equations of motion, the set of operators $\alset$ can be related to the algebra of truncated polynomials in the bulk fields $\phi$ in the region $\bar{\cal D}$ in figure \ref{figadsband}. The explicit mapping is given in \cite{Banerjee:2016mhh}. While respecting the constraints of bulk locality, we clearly cannot relate operators in ${\cal D}$ to operators in ${\cal B}$ since all points in ${\cal D}$ are spacelike to all points in ${\cal B}$. 

Nevertheless, it was shown in \cite{Banerjee:2016mhh} that it is possible to construct operators in ${\cal D}$ using operators in the band ${\cal B}$, provided that we consider {\em complicated} polynomials of generalized free-fields that are {\em not} elements of $\alset$. This can be done in two steps as follows. To be specific, consider an operator $\phi(r=0, t = 0, \Omega)$ localized in the diamond ${\cal D}$ at some point on the sphere. We will denote this operator by $\phi(0)$ to lighten the notation below. Now, this operator can be approximated as
\be
\label{completeset}
\phi(0) \approx \sum_{n,m \ll N} c_{nm} |n \rangle \langle m|,
\ee
where the states $|n \rangle$ correspond to energy eigenstates in AdS and $c_{n m}$ are c-number coefficients. At energies below the Planck scale, $N$, these energy eigenvalues are approximately quantized in units of the AdS scale. So far, in  \eqref{completeset}, we
have not done anything except expand the operator in a complete set of states and place a cutoff since we are not interested in the matrix elements of $\phi(0)$ for ultra-Planckian energies.

Now, we note that the states $|n \rangle$ can be written as
\be
\label{reehschlieder}
|n \rangle = X_n |0 \rangle,
\ee
where $X_n \in \alset$. This means that all low-energy eigenstates in anti-de Sitter space can be created by acting on the vacuum with the algebra of {\em simple} polynomials of generalized free-field operators in the band, ${\cal B}$, or by simple polynomials of bulk fields in ${\bar{\cal D}}$. The reader may think of this as an application of the Reeh-Schlieder theorem \cite{Haag:1992hx} as applied to the set of simple bulk operators in the region $\bar{{\cal D}}$. However, in \cite{Banerjee:2016mhh}, this theorem was directly proved by using the properties of generalized free fields in a CFT and by considering the algebra of such fields in the band, ${\cal B}$. Explicit expressions for the operators, $X_n$, are also available in \cite{Banerjee:2016mhh}.

Although \eqref{reehschlieder} is a surprising statement, so far there has been no violation of locality since \eqref{reehschlieder} would hold even in a theory without gravity.  However, we now note that we can also write
\be
P_0 = |0 \rangle \langle 0| = \lim_{\alpha \rightarrow \infty} e^{-\alpha H}.
\ee
Here, $H$ is the Hamiltonian and given by
\be
\label{hstress}
H = \int T^{00}(\Omega) d^{d-1} \Omega - E_0,
\ee
where,
\be
E_0 = \langle \Omega| \int T^{00}(\Omega) d^{d-1} \Omega | \Omega \rangle.
\ee
The key point is that, in a theory of gravity in the bulk, $H$ is an operator in ${\cal B}$ and the boundary value of a bulk operator in $\bar{\cal D}$. This property is evidently true in AdS/CFT since the stress-tensor is a generalized free-field in ${\cal B}$ and is dual to the fluctuations of the graviton in the bulk.  However, this fact is independent of AdS/CFT and relies on a fundamental property that emerges from the canonical quantization of gravity: the Hamiltonian, in any theory of quantum gravity, is a boundary term. 

In fact, the projector above can be approximated using a {\em very complicated polynomial}. In particular, we can  write
\be
\label{polynomialproject}
P_0 \approx {\cal P}_0  = \sum_{p=0}^{p_c} {(\alpha_c H)^p  \over p!},
\ee
by choosing a particular large value of $\alpha = \alpha_c$ and expanding the exponential in a power series and cutting that off at $p_c$. We can see that a choice of $\alpha_c = \log(N)$ is sufficient to ensure that for the lowest possible non-zero energy eigenvalue, $E_1$, we have ${\cal P}_0 |E_1 \rangle = \Or[{1 \over N}]$. Second, an exponential $e^{x}$ can be expanded in a power series as long as we keep $\Or[x]$ terms. Therefore, if we choose $p_c = N \log(N)$, we ensure that this polynomial approximates the true projector on the vacuum, ${\cal P}_0 \approx P_0$, as long as it is inserted only in states with energy eigenvalues much smaller than $N$. 

Putting these observations together we obtain the following {\em formula}
\be
\label{nonlocalformula}
\phi(0) = \sum_{n,m \ll N} c_{nm} X_n {\cal P}_0 X_m^{\dagger}.
\ee

We thus see that we have succeeded in representing an operator in the center of AdS in terms of a complicated polynomial of operators that are uniformly spatially separated from the original operator. This is an important example, since it serves as a concrete prototype for the nonlocal effects that we expect are important for black-hole evaporation.

We return to this toy model in \ref{secinfo} showing how it also provides a toy model of various examples of the information paradox, which can then
be resolved in this setting. 

\subsection{Breakdown of perturbation theory and locality}
The example above shows how operators at one point in anti-de Sitter space can be represented entirely in terms of operators at other points. We can abstract two important elements from this analysis.
The root of the nonlocality visible in the formula above lies in the Gauss law. The fact that the energy of a local insertion in the bulk can be measured at infinity leads to the fact that the Hamiltonian is purely a boundary term in gravity. This is what allowed us to construct the projector on the vacuum as a complicated polynomial  of boundary operators in \eqref{polynomialproject}. 

It is important to realize that, naively, the Gauss law seems to lead to very small nonlocal commutators that are suppressed by factors of ${1 \over N}$. The stress-tensor, as it appears in \eqref{hstress} has a two-point function\footnote{The precise form of this function is not relevant for our discussion but is given by \cite{Osborn:1993cr}
\be
\langle T_{i j}(x) T_{l m}(0) \rangle = {N^{2} \over x^{2 d}} I_{i j, l m} (x),
\ee
where the tensor $I_{i j, l m}(x)$ is given by
\be
I_{i j, l m} = {1 \over 2} \left(I_{i l}(x) I_{j m}(x)  + I_{i m}(x) I_{j l}(x) \right) - {1 \over d} \delta_{i j} \delta_{l m},
\ee
and $I_{i j}(x) = \delta_{i j} - 2 {x_{i} x_{j}  \over x^2}$.}  that is proportional to $N^2$. In particular, the canonically normalized bulk graviton field is dual to $h_{i j} \leftrightarrow {T_{i j} \over N}$. Therefore the nonlocal commutator $i [H, \phi(t, r, \Omega)] = \dot{\phi}(t, r, \Omega)$ leads to non-local commutators between the canonically normalized graviton field and other bulk fields that are suppressed by ${1 \over N}$. 
It is important to note that the effect above, where we were able to completely rewrite a bulk operator in terms of other operators near the boundary was obtained by suitably ``enhancing'' this ${1 \over N}$ effect to an order $1$ effect.
This can only happen when ${1 \over N}$ perturbation theory breaks down. This is why it is important that the polynomials that appear in \eqref{nonlocalformula} contain  $\Or[N]$ insertions.

We thus find two underlying features in our toy-model that lead to the nonlocality that is visible there.  These are 
\begin{enumerate}
\item
Nonlocal perturbative commutators due to the Gauss law.
\item
The enhancement of these small commutators due to the breakdown of perturbation theory.
\end{enumerate}

Precisely the same analysis applies in flat space. As we review below,  the Gauss law leads to commutators between two quasilocal operators that are suppressed by a power of ${E \over M_{\text{pl}}}$, where $E$ is a measure of the energy of the operators involved. The breakdown of perturbation theory may enhance these commutators to give rise to physically significant effects.

Recall that the Hamiltonian in gravity, in asymptotic flat space, can also be written as a boundary term. If we choose $n^k$ to be the unit normal to the sphere at infinity, then we have 
\be
H = {1 \over 16 \pi G}  \int_{\infty} d^{d-2} \Omega n^k (g_{i k,i} - g_{i i,k}),
\ee
where the repeated indices are summed over the spatial directions \cite{Arnowitt:1962hi,Regge:1974zd}. The Hamiltonian generates translations of asymptotic time. 

Now consider studying a quasilocal operator in the interior of spacetime, $\phi(t, \vec{x})$ where we have separated the time $t$ from spatial coordinates, $\vec{x}$.  To define what we mean by the coordinates $(t,\vec{x})$,  we need to fix gauge or use a relational prescription. But provided that our prescription for localizing the operator satisfies the property that large diffeomorphisms that translate the asymptotic time coordinate by $\delta t$ also lead to translations of the local time coordinate $t \rightarrow t  + \delta t$, then it is guaranteed that that $i[H, \phi(t,\vec{x})] = {d \phi(t,\vec{x}) \over d t}$. This commutator is nonlocal since the Hamiltonian can be defined by integration on a surface that is entirely spacelike to the point $(t,\vec{x})$. 

This is simply the Gauss law in action again.  In fact, it is well known that, in gravity, as a result of the Gauss law, there are {\em no} exactly gauge-invariant local operators. Nevertheless, it is possible to define quasilocal observables, since the nonlocality induced by the commutators above is small, as we now explain.  

Note that, in terms of the canonically normalized fluctuations of the metric, $g_{\mu \nu} = \eta_{\mu \nu} + \sqrt{8 \pi G} \tilde{h}_{\mu \nu}$, the expression for the ADM Hamiltonian can be written as
\be
H = {1 \over 2 \sqrt{8 \pi G}}  \int_{\infty} d^{d-2} \Omega n^k (\tilde{h}_{i k,i} - \tilde{h}_{i i,k}),
 \ee
where we recognize $\sqrt{8 \pi G} = \lpl^{d - 2 \over 2}$. Therefore, purely on dimensional grounds we expect that if smear the metric fluctuations to obtain a unit-normalized operator, and consider its commutator with another unit-normalized operator then this commutator will be suppressed by $\left({E \over M_{\text{pl}}} \right)^{d - 2 \over 2}$, where $E$ is a measure of the energy of the configuration of the two operators. Commutators between other field operators (not the metric) may be suppressed by further factors.

The precise commutators depend on how we define our quasilocal operators. Equivalently, the precise nonlocal commutators induced by the Gauss law depend on a choice of gauge. But, a concrete example was analyzed in  \cite{Donnelly:2015hta}, and we can use their results to verify our expectations. The authors of \cite{Donnelly:2015hta} worked in $d = 4$ so we will use this value below and then indicate the generalization to arbitrary $d$. With the choice made in \cite{Donnelly:2015hta}, the authors found that the commutator between two quasilocal scalar operators outside the light cone was given by
\be
[\phi(t, \vec{x}), \phi(t', \vect{y})] = {-i G \over 8} \left( \dot{\phi}(t, \vec{x}) \partial_i \phi(t', \vect{y}) + \partial_{i} \phi(t, \vec{x}) \dot{\phi}(t', \vect{y})  \right){x^i - y^i \over  \Big((t - t')^2 + (\vec{x} - \vect{y})^2 \Big)^{1 \over 2}}.
\ee
We emphasize that the precise form of the commutator above depends on the {\em choice} made in \cite{Donnelly:2015hta} of how to localize the operator. We remind the reader that this is similar to quantum electrodynamics, where our choice of how we place the Wilson lines on a local charged field controls the commutators of that field with local currents.

To estimate the size of this commutator evaluated in \cite{Donnelly:2015hta}, we smear both fields to generate operators with a unit-normalized two-point function. We choose
\be
\begin{split}
\phi(f) = \int \phi(t, \vec{x}) f(t, \vec{x}) d t \, d^{3} \vec{x}; \quad \phi(g) = \int \phi(t, \vec{x}) g(t, \vec{x}) d t d^{3} \vec{x},
\end{split}
\ee
with the constraint that $\langle \phi(f) \phi(f) \rangle = \langle \phi(g) \phi(g) \rangle = 1$. This leads to simple constraints on the functions $f$ and $g$:
\be
\int {f(t, \vec{x}) f(t', \vect{y})  d^{3} \vec{x} d^{3} \vect{y} d t d t' \over  (t - t')^2 + (\vec{x} - \vect{y})^2 }= 1, \quad \int {g(t, \vec{x}) g(t', \vect{y})   d^{3} \vec{x} d^{3} \vect{y} d t d t' \over (t - t')^2 + (\vec{x} - \vect{y})^2 }= 1.
\ee
These conditions are calculated at leading order since we have put in the leading order two-point function for $\phi$ but they can be corrected order by order in perturbation theory if required. 
We also demand that the two functions have no points of common support: $f(t, \vec{x}) \neq 0 \Rightarrow g(t, \vec{x}) = 0$ and $g(t, \vec{x}) \neq 0 \Rightarrow f(t, \vec{x}) = 0$.
The expectation value of the commutator then becomes
\be
\langle [\phi(f), \phi(g)] \rangle = i \left({E \over M_{\text{pl}}}\right)^{2}, \quad d = 4
\ee
where the ``energy'' of this configuration of operators is defined through
\be
\label{edeffourd}
E^{2} \equiv \int   {(t - t') (\vec{x} - \vec{y})^2 \over \Big((t - t')^2 + (\vec{x} - \vec{y})^2 \Big)^{7 \over 2}} f(t, \vec{x}) g(t', \vec{y}) d^{3} \vec{x} d^{3} \vec{y} d t d t',
\ee
in $d = 4$. The expression is not covariant due to various choices made in defining the operators; these choices can be thought of as a choice of gauge. We have used the term ``energy'' for this quantity because it is a measure of the inverse distance scale in this configuration of operators. 

In arbitrary $d$, we expect an entirely analogous result to hold 
\be
\langle [\phi(f), \phi(g)] \rangle = i \left({E \over M_{\text{pl}}}\right)^{d-2},
\ee
with $E$ defined in analogy to  \eqref{edeffourd}, up to numerical prefactors, and with the exponent of ${7 \over 2}$ in the denominator replaced by ${2 d - 1 \over 2}$.

The fact that, for separations much larger than the Planck length, we have $E \ll M_{\text{pl}}$ tells us that the nonlocality induced by the Gauss law is small, and explains why we observe approximate locality in nature.

However, the perturbative parameter for gravitational perturbation theory in flat space is also $\left({E \over M_{\text{pl}}}\right)^{d - 2}$.\footnote{In any specific calculation, it may be more convenient to choose a definition of the coupling constant that differs from this by an $\Or[1]$ numerical prefactor. However, what is important here is that if all the distances are scaled by $\lambda$, the gravitational interactions fall by $\lambda^{2-d}$.} This suggests that the limits in which perturbation theory breaks down in flat space may also be interesting from the point of view of the loss of locality. We caution the reader that unlike the case of empty AdS above, we will not be able to demonstrate this effect explicitly in flat space. However, we believe that it is extremely likely that, at least in some settings, a combination of the fundamental nonlocality induced by the Gauss law, and the breakdown of gravitational perturbation theory leads to large-scale nonlocal effects. We now describe why this is also natural from a path-integral viewpoint. 

\subsection{Locality in path integrals and perturbation theory}
A consideration of the quantum-gravity path integral provides another heuristic argument for the claim that the breakdown of locality is concomitant with the breakdown of perturbation theory.  In the Euclidean theory, we can consider some quasilocal observables $\phi(t_i^E, \vec{x}_i)$ at some value of Euclidean time, $t_i^E$ and position $\vec{x}_i$. As mentioned above, to define these observables precisely we need to choose a gauge or a relational prescription. We can then imagine inserting these operators into a path-integral to compute a Euclidean correlation function. For example,
\be
\label{multilocalcorr}
\langle \phi(t_1^E, \vec{x}_1) \ldots \phi(t_n^E, \vec{x}_n) \rangle = \int e^{-S} \phi(t_1^E, \vec{x}_1) \ldots \phi(t_n^E, \vec{x}_n)  {\cal D} g_{\mu \nu} {\cal D} \phi,
\ee
where we integrate over all bulk metrics with some specified asymptotic boundary conditions. 

Now, if we want this multi-local correlator to conform to some notions of locality, then we need a notion of when two points $(t_1^E, \vec{x}_1)$ and $(t_2^E, \vec{x}_2)$ are close to each other. Such a notion is predicated on a metric. However, in the path-integral we only specify asymptotic boundary conditions for the metric.

Nevertheless, it is possible to define an approximate notion of locality when the path-integral is dominated by a {\em saddle point}. In the saddle point approximation, some particular metric $g_{\mu \nu}^0$ dominates the path-integral and this metric allows us to define the distance between two points. We may continue to use this metric to specify our notion of locality, provided that the correlator in \eqref{multilocalcorr} can be computed in an asymptotic series expansion about this saddle point. 

However, if perturbation theory breaks down in the computation, this is a sign that the saddle-point approximation to the path-integral has broken down. In this case, either the quantity \eqref{multilocalcorr} is dominated by another metric saddle point, $\tilde{g}^0_{\mu \nu} \neq g^{0}_{\mu \nu}$, or else perhaps it cannot be computed in a saddle-point approximation at all. In either case, our original notion of locality, which was predicated on the metric $g^{0}_{\mu \nu}$ is invalidated. 

Therefore, from a path-integral point of view, it is natural for the breakdown of perturbation theory in gravity to be a signal of the loss of locality. This analysis also helps us emphasize that, in a non-gravitational theory, where the metric does not fluctuate there is no such link between perturbative breakdown and the loss of locality. It is {\em only in a theory of gravity} that locality is tied to the the dominance of a particular background metric as a saddle-point in the path-integral, which in turn is tied to the validity of perturbation theory. 

\subsection{Boundary vs bulk locality \label{bulkvsboundary}}
The path-integral analysis also helps us clarify another issue. It is very important that that the effects we describe here do {\em not} lead to a violation of boundary causality. For example, in AdS/CFT the boundary theory satisfies microcausality and locality in the boundary theory is {\em not} lost even if we consider arbitrarily high-point correlators. Rather, our claim is that very high-point correlators may not have a simple {\em bulk} local interpretation.

The situation is similar in flat space. In doing the path-integral in \eqref{multilocalcorr}, we keep the asymptotic geometry fixed and we do not expect to violate asymptotic notions of locality, as defined in \cite{Gao:2000ga}. For example we may consider the situation where we take the points $(t_i, x_i)$ to either future or past null infinity: ${\cal I}^{\pm}$. On both ${\cal I}^{\pm}$ we can specify these points through a null coordinate --- which we denote by $u_i$ and $v_i$ respectively ---  and a point on the sphere at infinity $\Omega_i$. Then  the points $(u_1, \Omega_1)$ and $(u_2, \Omega_2)$ are spacelike to each other if $\Omega_1 \neq \Omega_2$. In this situation, for instance, we expect that 
\be
\langle [\phi(u_1, \Omega_1), \phi(u_2, \Omega_2)] \phi(u_3, \Omega_3) \ldots \phi(u_n, \Omega_n) \rangle = 0, \quad \text{even~as~}n \rightarrow \infty.
\ee
So, we expect that this commutator vanishes even if it is inserted in a correlator with an arbitrarily large number of insertions. 

Our point, in this paper, is simply that if we try and define quasilocal operators that are {\em not} asymptotic operators, then it is possible that even approximate notions of locality for such operators may break down with the breakdown of perturbation theory. With this motivation we now turn to a detailed study of the breakdown of string perturbation theory.

\section{Bounds on perturbation theory \label{secbounds}}
We will start our analysis of the breakdown of perturbation theory by reviewing some simple bounds on the rate of growth of tree-amplitudes in perturbation theory. Consider the scattering of ${n \over 2}$-particles.  (We assume $n$ is even.) Unitarity of the S-matrix tells us that 
\be
\label{unitarityrelation}
\sum_f \int d \Pi_f {|M(\{k_1 \ldots k_{n \over 2} \}  \rightarrow \{f\})|^2}  =  2 \text{Im}\left[M(\{k_1 \ldots k_{n \over 2} \}  \rightarrow \{k_1 \ldots k_{n \over 2}\}) \right].
\ee
Here  $d \Pi_f$ is the measure on phase space and the sum over $f$ schematically indicates the sum over all possible final states.

Since the left hand side is a sum over positive terms, we can restrict the sum to the case where the final state also consists of only ${n \over 2}$ particles to obtain an inequality. This leads to
\be
\label{unitarityinequality}
 \int d \Pi_{n \over 2} {|M(\{k_1 \ldots k_{n \over 2}\}  \rightarrow \{k_{{n \over 2} + 1} \ldots k_n\})|^2}  \leq  2 \text{Im}\left[M(\{k_1 \ldots k_{n \over 2} \}  \rightarrow \{k_1 \ldots k_{n \over 2}\}) \right].
\ee
In a theory with coupling constant $g$, we can  expand the amplitude as
\be
M(\{k_1 \ldots k_{n \over 2}\}  \rightarrow \{k_{{n \over 2} + 1} \ldots k_n\}) = \sum_{\ell = 0}^{\infty} g^{n-2 + 2 \ell} M^{\ell}(\{k_1 \ldots k_{n \over 2}\}  \rightarrow \{k_{{n \over 2} + 1} \ldots k_n\}),
\ee
where $\ell$ is the loop-order. Later, the relevant coupling constant will turn out to be the string coupling $g_s$ but for now we do not need to specify any particular value for $g$. While perturbation theory is valid, the inequality \eqref{unitarityinequality} must then hold {\em order by order} in perturbation theory. Within this perturbative expansion, the first few terms simply lead to some positivity constraints but the first non-trivial term in the inequality \eqref{unitarityinequality} is
\be
\resizebox{\textwidth}{!}{$\int d \Pi_{n \over 2} g^{2 n-4} {\big| M^0(\{k_1 \ldots k_{n \over 2}\}  \rightarrow \{k_{{n \over 2} + 1} \ldots k_n\}) \big|^2} \leq 2 g^{2 n - 4}  \text{Im}\left[M^{n - 2 \over 2}(\{k_1 \ldots k_{n \over 2} \}  \rightarrow \{k_1 \ldots k_{n \over 2} \}) \right],$}
\ee
where on the right hand side we have a high-order loop amplitude with ${n-2  \over 2}$ loops. Now, the validity of perturbation theory requires that loop-amplitudes be smaller than tree-amplitudes. 
From this criterion, we obtain the relation
\be
g^{2 n - 4}  \text{Im}\left[M^{n - 2 \over 2}(\{k_1 \ldots k_{n \over 2} \}  \rightarrow \{k_1 \ldots k_{n \over 2} \}) \right] \leq g^{n - 2} {|M^0(\{k_1 \ldots k_{n \over 2} \}  \rightarrow \{k_1 \ldots k_{n \over 2}\})|}.
\ee
We emphasize that this is a very {\em weak} condition because we have included factors of the coupling constant in the inequality. A much higher power of the coupling constant appears on the left-hand side. This relation can fail to hold only if a huge factor in the amplitude overcomes this large power of the coupling constant. When this happens, perturbation theory breaks down. Combining the relations above, we find that
\be
\int d \Pi_{n \over 2} g^{2 n-4} {|M^0(\{k_1 \ldots k_{n \over 2} \}  \rightarrow \{k_{{n \over 2} + 1} \ldots k_{n} \})|^2} \leq 2 g^{n-2}  \text{Im}\left[M^0(\{k_1 \ldots k_{n \over 2} \}  \rightarrow \{k_1 \ldots k_{n \over 2}\}) \right].
\ee
Defining
\be
\mtr(\{k_1 \ldots k_{n \over 2} \} \rightarrow \{k_{{n \over 2} + 1} \ldots k_{n}\}) = g^{n - 2} M^0(\{k_1 \ldots k_{n \over 2} \} \rightarrow \{k_{{n \over 2} + 1} \ldots k_{n}\}),
\ee
which is just the tree level amplitude including all powers of the coupling, we see that this relation becomes 
\be
\label{perturbinequality}
\int d \Pi_{n \over 2}  {|\mtr(\{k_1 \ldots k_{n \over 2}   \rightarrow  \{k_{{n \over 2} + 1} \ldots k_{n}\})|^2} \leq  2 \left|\mtr(\{k_1 \ldots k_{n \over 2} \}  \rightarrow \{k_1 \ldots k_{n \over 2} \}) \right|.
\ee
At the point where the bound is violated, we expect that perturbation theory breaks down and all orders in the perturbative answer become as important as the tree-level answer to the S-matrix.

To give concrete form to this inequality, we also need the phase-space factor. We will consider massless particles so that the phase space factor is simply given by
\be
\label{phasespacemeasure}
d \Pi_{n \over 2} = {(2 \pi)^d \delta({n E \over 2} - \sum |k_l|) \delta^{d-1}(\sum k_l) \over (n/2)!} \prod_{t} {d^{d-1} k_t \over (2 \pi)^{d-1} 2 |k_t|}.
\ee
Note the factor of $\left({n \over 2}\right)!$ which appears in the denominator. This simply arises from the conventional normalization of  scattering amplitudes. Here, $E$ is the center-of-mass energy per particle and $l,t \in \{{n \over 2} + 1, \ldots n\}$. 

It is not too difficult to work out the total volume of phase space, which we do in Appendix \ref{phasespacevol}. The result is
\be
\label{volphasespace}
\int d \Pi_{n \over 2} = v \frac{ E^{{(d-2) n \over 2} - d}}{(n/2)!}.
\ee
Here,
\be
\label{phasespacefactor}
v  = \frac{2 \pi \Gamma({d \over 2} - 1)^{n \over 2} ({n \over 2}) ^{n({d \over 2} -1) - d}}{(4 \pi)^{(n-2) {d \over 4}} \Gamma\big({(d-2)(n-2)\over4}\big)\Gamma\big({(d-2)n\over4}\big)}.
\ee

In this paper we are concerned with the situation where, at large $n$, tree amplitudes grow as
\be
\label{factgrowth}
\mtr(k_1 \ldots k_{n\over 2} \rightarrow k_{{n\over2}+1} \ldots k_{n})= {n! \over \Lambda^{{(d-2) n \over 2} - d}}.
\ee
Here, $\Lambda$ is a physically important energy scale, and its appearance on the right hand side can be understood through dimensional considerations. Momentum eigenstates are normalized as $\langle k | k' \rangle = (2 \pi)^{d-1} (2 |k|) \delta^{d-1}(k - k')$.  The amplitude is given by the overlap of an in-state with $n/2$ particles with an out state with $n/2$ particles. In our analysis we do {\em not} include the overall momentum conserving delta function in the amplitude.  Hence the mass-dimension of the amplitude is $d - {n (d-2) \over 2}$ and the power of $\Lambda$ ensures that the right hand side also has the correct dimension. We will return to the significance of $\Lambda$ below.

When \eqref{factgrowth} holds,  we see that inequality \eqref{perturbinequality} is violated at a value of $n$ that satisfies
\be
v {(n!)^2  E^{{(d-2) n \over 2} - d} \over \Lambda^{(d-2) n  - 2 d}} {1 \over (n/2)!} = {2 n! \over \Lambda^{{(d-2) n \over 2} - d}},
\ee
or
\be
{v \over 2} {(n!) \over (n/2)!} \left({E \over \Lambda} \right)^{{(d-2) n \over 2} - d}  = 1.
\ee
Keeping only the leading terms, and using Stirling's approximation for the factorial: $\log(n!) \approx n \log(n) - n + \Or[\log(n)]$, we find that perturbation theory breaks down at a value of $n$ that satisfies
\be
{(2 - d) \log{E \over \Lambda} \over \log(n)} = 1 + \Or[{1 \over \log(n)}].
\ee
We see that although the prefactor $v$  grows exponentially in $n$, it  turns out to be irrelevant in the final answer because the dominant terms grow factorially with $n$.
More precisely, we have $\log(v)/(n \log n) \rightarrow 0$ in the large $n$ limit.

The breakdown of perturbation theory at large $n$ is not specific to string theory or gravity. In fact, it is well known in ordinary quantum field theories. For example, amplitudes grow factorially even in the $\lambda \phi^4$ theory in four dimensions.  Moreover, we note that two particles with momentum components of $\Or[E]$ can be added to the amplitude at the cost of an additional propagator that contributes a term of  $\Or[1 \over E^2]$ and a single coupling constant factor $\lambda$. Therefore the energy scale that enters \eqref{factgrowth} is $\Lambda = {E/\sqrt{\lambda}}$. Note that $\Lambda$ depends on the energy per particle. These arguments suggest that perturbation theory breaks down for $n=\Or[1/\lambda]$ in $d=4$ and this is indeed the result that was found in \cite{Libanov:1997nt}.

In sections \ref{analyticgrowth} and \ref{numericalgrowth} we will now argue that tree amplitudes in string theory also display at least a factorial growth where the energy scale in \eqref{factgrowth} is $\Lambda =  \mpl = \lpl^{-1}$. The fact that this energy scale appears can be seen easily since the coupling constant in string theory is $g_s$. In the units that we have adopted here, we have $2 \pi g_s = {\lpl^{d-2 \over 2}}.$ A factor of $g_s^{n - 2}$ appears with each $n$ point tree-level string amplitude. When dimensions are restored this is equivalent to a factor of $\lpl^{(d - 2) (n - 2) \over 2}$.  We then find that string perturbation theory breaks down for values of $n$ and $E$ that satisfy
\be
\label{thresholdlpl}
\frac{\log( E \lpl)}{\log(n)} = \frac{1}{2-d} +\Or[{1 \over \log(n)}].
\ee

If we take the energy per particle to scale with $n$ as $E \propto {1 \over n^{\gamma}}$ then this threshold can also be written as 
\be
\frac{\log(g_s^2)}{\log(n)} = (d-2) \gamma - 1 + \Or[{1 \over \log(n)}].
\ee
It is important that we take $\gamma > 0$ (for reasons that we explain in section \ref{boundgreen}) and from the relations above we see that we must also take $\gamma < {1 \over d - 2}$. 

\paragraph{Calculation with compact extra dimensions \\}

Before we proceed to string amplitudes, we would like to describe some simple extensions of the bound above. One situation that is of physical interest is when the string theory lives in $d$ dimensions but some of these dimensions are compactified. For simplicity, we take $m$ of the extra dimensions to be compactified on a torus, where each side has radius $\rho$; it is easy to generalize our calculation to more general compact manifolds. We consider a regime where ${1 \over E} \gg \rho \gg \sqrt{\alpha'}$ but we do not scale $\rho$ with $n$.  We also define $p = d - m$, the number of non-compact dimensions. 

In general, as we will see below, the estimate for the amplitude is not altered by the compactification. This is because that estimate for the growth of the amplitude relies on the volume of moduli space that depends on the structure of the worldsheet and not on whether the spacetime is compactified. In section \ref{numericalgrowth}, we will also estimate the amplitude through a sum of the solutions of the scattering equations. But the fact that the number of solutions to the  scattering equations grows factorially is independent of whether some of the spacetime dimensions are compact or not. 

However, in the compact extra dimensions case, the volume of phase space given above in \eqref{volphasespace} is modified.  The momentum in the compact extra dimension is quantized, and string theory also generates new winding-sector states. However, in the regime under consideration the lowest mass of a state in the winding sector is $m^2 = {\rho^2 \over \alpha'^2} \gg {1 \over \alpha'}$; so these states are heavier than the lightest string excitations and can consistently be ignored. Including the Kaluza-Klein excitations, the new volume of phase-space can be calculated as follows.
\begin{eqnarray*}
\int d \Pi^{m}_{n \over 2} &=& \sum_{n_{q t}} \int  \left[( 2 \pi)^p \delta({n E \over 2} - \sum_{l} k_l^0) \delta^{p-1}(\sum_{l} k_l^i) {(2 \pi \rho)^m \delta^m(\sum_t {n_{q t}}) \over (n/2)!} \right. \\ &\times& \left. \prod_{t} ( 2 \pi) \delta(k_t^2 - \sum_{q=1}^m {n_{q t}^2 \over \rho^2} ) {d^{p} k_t \over (2 \pi)^{p} (2 \pi \rho)^m} \right],
\end{eqnarray*}
where the components of the momentum, $k_t$ in the compact directions are specified by $n_{q t}$, where $q = 1 \ldots m$; the measure $d^{p} k_t$ runs over the non-compact dimensions including time and the $(p-1)$-dimensional delta function runs over the spatial non-compact dimensions. The delta function imposing momentum conservation in the compact directions is, of course, a discrete delta function. We have also placed a superscript $m$ on the measure on the volume of phase space to indicate that $m$ dimensions are compact. 

We are not aware of any method of evaluating this integral and sum exactly. However, fortunately, in the limit under consideration where ${1 \over \rho} \gg E$, the sum is dominated by the term with $n_{q t} = 0$. In this limit, the volume of phase space is given by
\[
\int d \Pi^{m}_{n \over 2} = {v_m \over (2 \pi \rho)^{m {n \over 2} - m}} {  E^{{(p-2) n \over 2} - p} \left(1 + \Or[{E \rho}] \right)  \over (n/2)! } 
\]
where 
\[
v_m =  {2 \pi \Gamma({p \over 2} - 1)^{n \over 2} (n/2) ^{n({p \over 2} -1) - p} \over (4 \pi)^{(n-2) {p \over 4}} \Gamma\left({(p-2)(n-2)\over4}\right)\Gamma\left({(p-2)n\over4}\right)}
\]

Repeating the analysis above, we find that perturbation theory breaks down for 
\begin{equation}
\label{breakdown}
{\log( E \lplp ) \over \log(n)} = {1 \over 2-p} + {\rm O}\left({1 \over \log(n)}\right).
\end{equation}
where $\lplp$ is the $p$-dimensional Planck length that is related to the $d$-dimensional Planck length through $\lplp^{p-2} (2 \pi \rho)^{m}= \lpl^{d-2}$.

Thus we see that our $d$-dimensional bound on the validity of string-perturbation theory naturally generalizes to a lower-dimensional bound, when some of the dimensions are compactified.

\paragraph{\bf Other combinations of incoming and outgoing particles \\}
In the analysis above, we considered a process where ${n \over 2}$ particles were incoming and ${n \over 2}$ were outgoing. Of course, it is also possible to consider other processes such as $2 \longrightarrow n$ scattering.  We do not consider these combinations in this paper to avoid any possible complications that may arise if one of the ingoing or outgoing particles has trans-Planckian energy. However, we
emphasize that, assuming that the factorial growth outlined above continues to hold, these kinematical configurations would not yield a bound that is any stronger than the bound above. This is easy to see as follows. Consider scattering from $\alpha n \rightarrow  {\beta n}$ particles, where $\alpha, \beta$ are some fractions. Then we see that the unitarity bound above is saturated at
\be
{v \over 2} \left[((\alpha + \beta)n)! \right]^2  \left({E \over \Lambda} \right)^{(d-2)\beta n - d} = (2 \alpha n)! (\beta n)!.
\ee
Simplifying this expression and disregarding all subleading terms, we see once again that this leads to a breakdown of perturbation theory at 
\be
{({2 - d}) \log{E \over \Lambda}  \over \log(n)} = 1 + \Or[\log(n)]
\ee
which is precisely the same expression as the one above.

\section{Analytic estimates of the growth of string amplitudes \label{analyticgrowth}}
In this section, we will argue that $n$ point tree-level scattering amplitudes of massless states in closed bosonic string theory as well as in type II superstring theories grow at least as fast as $n! g_s^{n-2}$ for large n. Our argument is based on the formulation of string scattering amplitudes as integrals over the moduli space of punctured Riemann surfaces. We then provide some evidence that these moduli-space integrals are dominated by the volume of moduli space, which allows us to utilize results from the mathematics literature on these volumes. Since the analysis for the bosonic string and the superstring is similar, we provide several details for the bosonic string and then briefly describe the generalization to the superstring.

\subsection{Closed bosonic string amplitudes }
Scattering amplitudes in closed bosonic string theory can be formulated as integrals over the moduli space of punctured Riemann surfaces.  This representation may be somewhat unfamiliar to the reader, since the textbook formulation of string scattering leads to a formula for the amplitude where the positions of the vertex operators are integrated over a non-singular 
worldsheet \cite{Polchinski:1998rq}. So we first review the equivalence of the two prescriptions. 

We claim that the string scattering amplitude  may be written as
\be
\label{polyakovpunctures}
M(k_1 \ldots k_n) = {\cal N} g_s^{n-2 + 2 g} \int [d m] {\det \langle \mu_{\alpha}, \phi_{\beta} \rangle  \over \left( \det \langle \phi_{\alpha}, \phi_{\beta} \rangle \right)^{1 \over 2}} (\detp P_1^{\dagger} P_1)^{1 \over 2} (\detp \Delta)^{-d \over 2} Q_n P_n \bar{P}_n.
\ee
Here, the integral $[dm]$ runs over the moduli space of a Riemann surface with $n$ punctures and  genus g. $\phi_{\alpha}$ are holomorphic quadratic differentials on the punctured Riemann surface, and $\mu_{\alpha}$ are Beltrami differentials that parameterize infinitesimal motion on the moduli space.  We denote this moduli space by ${\cal M}_{g,n}$ and when we need to refer to the Riemann surface itself we use $M_{g,n}$. Apart from this measure, we also have the standard ghost-determinant $(\detp P_1^{\dagger} P_1)^{1 \over 2}$ and a power of the determinant of the scalar Laplacian, $(\detp\Delta)^{-{d \over 2}}$, which arises when we do the path integral over the worldsheet fields. Here $Q_n, P_n, \bar{P_n}$ are terms that come from the correlation functions of vertex operators, which are placed at the punctures,  and we give explicit expressions for these terms below when the vertex operators correspond to massless particles.
 
The main point that we would like to emphasize in this formula is that the integral over the positions of the vertex operators has been absorbed into the integral over the positions of the punctures of the Riemann surface. The equivalence of this prescription to the textbook Polyakov prescription is somewhat subtle because the punctured Riemann surface cannot be mapped to the unpunctured surface by a nonsingular Weyl transformation.

Nevertheless, it was shown in \cite{D'Hoker:1987pr,Sonoda:1987ra} that the formulation of string scattering on the moduli space of the unpunctured Riemann surface, ${\cal M}_{g,0}$ and the moduli space of the punctured Riemann surface ${\cal M}_{g,n}$ indeed gives rise to equivalent answers. We start by dividing the holomorphic quadratic differentials into two sets: $(\phi_{A}, \phi_i)$, where the differentials $\phi_{A}$ are holomorphic on the surface $M_{g,0}$ and $\phi_{i}$ which are meromorphic on $M_{g,0}$ with poles at the positions of the punctures. We can choose these two sets of differentials to have no overlap so that $(\phi_A, \phi_i) = 0$. Similarly, we can divide the Beltrami differentials into two sets: $\mu_i$ that move the punctures, and $\mu_{A}$ that change the other moduli. It is not difficult to see that we can also choose $\langle \mu_i, \phi_A \rangle = \langle \mu_A, \phi_i \rangle = 0$. 

This then leads to the expression 
\be
M(k_1 \ldots k_n) = {\cal N} g_s^{n-2 + 2 g} \int  [d m'] \prod d^2 z_i {\det \langle \mu_{A}, \phi_{B} \rangle  \over \big( \det \langle \phi_{A}, \phi_{B} \rangle \big)^{1 \over 2}}  {\det \langle \mu_{i}, \phi_{j} \rangle  \over \big( \det \langle \phi_{i}, \phi_{j} \rangle\big)^{1 \over 2}}  (\detp P_1^{\dagger} P_1)^{1 \over 2} (\detp\Delta)^{-d \over 2} Q_n P_n \bar{P}_n,
\ee
where we have divided the integral over the moduli, $[dm] = [dm'] \prod d^2 z_i$, into an integral over the moduli of the surface $M_{g,0}$ and an integral over the positions of the punctures. At this point the operator $P_1$, defined on a worldsheet with metric $g_{a b}$, as 
\be
\label{pdef}
(P_1 \delta V)_{a b} = {1 \over 2} \left(\nabla_{a} \delta V_{b} + \nabla_{b} \delta V_{a}  - g_{a b} \nabla_{c} \delta V^c \right),
\ee
still acts only on those vector fields that {\em vanish} at the positions of the punctures.

The main result of \cite{D'Hoker:1987pr}, which was further clarified in \cite{Sonoda:1987ra}, was that when the functional determinants are appropriately regulated, we have
\be
{\det \langle \mu_{i}, \phi_{j} \rangle  \over \det \langle \phi_{i}, \phi_{j} \rangle^{1 \over 2}}  (\detp P_1^{\dagger} P_1)^{1 \over 2} = \left(\detp \tilde{P}_1^{\dagger} \tilde{P}_1 \right)^{1 \over 2},
\ee
where $\tilde{P}_1$ is the same operator as \eqref{pdef} but with a domain that includes vector fields that do not vanish at the positions of the punctures. Therefore, on the right hand side above, we have the usual determinant that would have resulted from integrating out the ghost-fields on the surface $M_{g,0}$. This reduces the expression \eqref{polyakovpunctures} to the more familiar expression, which only involves quantities on the unpunctured Riemann surface:
\be
M(k_1 \ldots k_n) = {\cal N} g_s^{n-2 + 2 g} \int [d m'] \prod d^2 z_i {\det \langle \mu_{A}, \phi_{B} \rangle  \over \det \langle \phi_{A}, \phi_{B} \rangle^{1 \over 2}} (\detp \tilde{P}_1^{\dagger} \tilde{P}_1)^{1 \over 2} (\detp \Delta)^{-d \over 2} Q_n P_n \bar{P}_n.
\ee

However, the advantage of the expression \eqref{polyakovpunctures} is that it allows us to make contact with various results in the mathematical literature.  On the punctured surface, we can choose the so-called ``hyperbolic metric'' on the worldsheet so that, everywhere on the worldsheet, we have uniform scalar curvature: $R = -1$.  Note that it is possible to make this choice even for the tree-level amplitude because, for the $n$-punctured sphere, the Gauss Bonnet theorem reads 
\be
\int \sqrt{g} R d^2 z  = 2 \pi \chi(M_{g,n}) = 2 \pi (2 - 2 g - n).
\ee
Therefore, even for $g = 0$, there is no obstruction to choosing $R = -1$. This just  implies that the area of the worldsheet is $\int \sqrt{g} d^2 z = 2 \pi(n + 2 g - 2)$. 

We note that near a puncture at $z = 0$, the hyperbolic metric behaves like 
\be
\label{nearpuncture}
ds^2 \underset{z \rightarrow 0}{\longrightarrow} {d z d \bar{z} \over |z|^2 \log(|z|)^2},
\ee

With this choice of metric on the worldsheet, the measure on moduli space turns into the Weil-Petersson measure \cite{hubbard2006teichmuller}
\be
\dmwp = [d m] {\det \langle \mu_{\alpha}, \phi_{\beta} \rangle  \over \det \langle \phi_{\alpha}, \phi_{\beta} \rangle^{1 \over 2}},
\ee
where the inner-product between the quadratic differentials is taken with respect to the hyperbolic metric.

We now specialize to tree-level scattering so that we set $g = 0$. At tree-level, our expression for the string scattering amplitude now becomes
\be
\label{finalbosonic}
\mtr(k_1 \ldots k_n) = {\cal N} g_s^{n-2} \int \dmwp (\detp P_1^{\dagger} P_1)^{1 \over 2} (\detp \Delta)^{-d \over 2} Q_n P_n \bar{P}_n.
 \ee
where $\dmwp$ is the Weil-Petersson measure on the moduli space of the n-punctured sphere.

The final ingredient that we need is the correlation function of vertex operators. For massless states in the closed bosonic string theory, at tree-level,  these correlators are easy to write down explicitly. Moreover, they are Weyl invariant by themselves and so take on the same form when the metric on the worldsheet is hyperbolic, as they do when the worldsheet is flat.  For the closed bosonic string, we recall that the vertex operators for massless states are
\be
{\cal V}(z, \bar{z}) = (\epsilon \cdot \partial X) (\bar{\epsilon} \cdot \bar{\partial} X) e^{i k \cdot X},
\ee
where we have also specified the polarization vectors $\epsilon, \bar{\epsilon}$, and the anti-holomorphic polarization vector is just the complex conjugate of the holomorphic polarization vector. Physically, this means that we are considering the scattering of {\em linear combinations} of the graviton and the dilaton. 

The relevant correlation function can then be written
\be
\langle {\cal V}(z_1, \bar{z}_1) \ldots {\cal V}(z_n, \bar{z}_n) \rangle = Q_n P_n \bar{P}_n,
\ee
where
\be
\label{bosonicexprpq}
\begin{split}
&Q_{n}=\exp\left(-\frac{1}{2}\sum_{i\ne j}k_{i} \cdot k_{j}G_{ij}\right),  \\
&P_{n}=\mathcal{L}\left\{\exp\left(\sum_{i\ne j}\frac{1}{2}\epsilon_{i} \cdot \epsilon_{j}\partial_{i}\partial_{j}G_{ij}+k_{i} \cdot \epsilon_{j}\partial_{j}G_{ij}\right)\right\}.
\end{split}
\ee
$\bar{P}_{n}$ is the anti-holomorphic counterpart of $P_{n}$. $G_{ij}$ is the worldsheet Green's function: 
\be
\label{worldsheetgreen}
G_{i j} = \ln |z_i - z_j|^2.
\ee
The symbol $\mathcal{L}$ above is shorthand for a rule that instructs us to expand the exponential in $P_n$ and keep only the part that is linear in all the polarization tensors.

\paragraph{\bf Infrared divergences\\}
The moduli-space integral receives divergent contributions from the boundaries of moduli space where some closed geodesic on the Riemann surface pinches off and its length goes to zero. In the case of tree-level amplitudes, this corresponds to the situation where two punctures collide. 

These divergences can be regulated through a suitable $i \epsilon$ prescription as explained in \cite{Witten:2013pra}. Equivalently, as explained in \cite{Sen:2014dqa} one may divide the moduli-space integral into two regions: (1) the region that covers those Riemann surfaces that can be obtained by plumbing lower-dimensional surfaces together using the ``plumbing fixture'' and (2) the rest of the moduli space which can be understood as coming from a fundamental higher-point vertex.
The divergences mentioned above come from the first region of moduli space. Here, one can get rid of them by using field theoretic techniques to rewrite the integral as a sum over contributions coming from the propagation of intermediate particles. The full amplitude can then be obtained by including the contribution from the second part of moduli space where the integral is finite. 

However, both of these prescriptions will complicate our estimate of the size of the integral. So, here, we will follow a simple-minded procedure by regulating these divergences by placing a cutoff, 
$\ell_{0}$, on the length of the smallest simple closed geodesic on $M_{0, n}$. This cuts off moduli space near its boundaries, and we will work
in this cutoff moduli space.

We caution the reader that it is possible that this procedure is not justified. For example, it may happen that the contributions from the edges of moduli space cancel
off the contributions from the bulk of moduli space that we will focus on below. These cancellations are possible even though, in our analysis, we have chosen polarization vectors for the external particles that will make the integrand on the bulk of moduli space positive. But if we use the $i \epsilon$ prescription of \cite{Witten:2013pra}, this instructs us to adopt a contour in a complexified version of moduli space near the edges, and then the integrand is no longer positive. For these reasons, it would be nice to repeat the arguments below without this cutoff.

Keeping these caveats in mind, we now examine each of the terms that appear in \eqref{finalbosonic}.

\subsection{Volumes of Weil Petersson moduli spaces}
We will argue below that the main contribution to the growth of amplitude comes simply from the volume of moduli space
\be
V_{g,n} = \int \dmwp.
\ee
Volumes of the Weil-Petersson moduli spaces of punctured Riemann surfaces have been studied in the mathematics literature. For the $n$-punctured sphere, this volume was first calculated in \cite{zograf1993moduli}. Then, numerical techniques were used to  advance conjectures for the growth of these volumes at large genus \cite{zograf2008large}.  In \cite{mirzakhani2007simple},  an analytic recursion relation was developed to calculate the Weil-Petersson volumes for any value of $n,g$.  The asymptotic growth of this volume was then studied in \cite{mirzakhani2013growth}.

These papers found that,  for any fixed $n$, when $g$ becomes large 
\be
\label{wpexact}
V_{g,n} = (4 \pi^2)^{2 g + n - 3} (2 g - 3 + n)! {1 \over \sqrt{g \pi}} \left(1 + \Or[{1 \over g}] \right),
\ee
We do not need the subleading terms and are only interested in the leading asymptotic behaviour, which is given by 
\begin{equation}
\lim_{g+n\to\infty} \frac{\mathrm{log}(V_{g,n})}{(2g+n)\mathrm{log}(2g+n)}=1, \label{wp_volume}
\end{equation}
and holds when {\em either} $n$ or $g$ become large.

Note that putting $n=0$ in \eqref{wp_volume} we have,
\be
\label{largegvol}
\lim_{g\to\infty} V_{g,0} \propto  (2g)!
\ee
Here, as in the rest of this paper, when we use the symbol $\propto$ we mean that we have captured the leading growth of the physical quantity. For example, in \eqref{largegvol} we have dropped factors that may even grow exponentially with $g$ since these factors are subleading compared to the factorial growth.

This large-g growth in the volume of moduli space was independently obtained in the physics literature by using matrix model techniques  in \cite{Shenker:1990uf}. By combining these results with the analysis of \cite{Gross:1988ib}, it is possible to show that this growth implies that the vacuum amplitude in the closed bosonic string theory also grows as $(2g)!$ for large $g$. It is well known that this growth implies that nonperturbative effects in string theory appear with a strength of $\Or[e^{-1/g_{s}}]$. At weak coupling this is larger than the size of  nonperturbative effects in ordinary quantum field theory, which is expected to be  $\Or[e^{-1/g_{s}^{2}}]$. These stringy nonperturbative effects are related to the existence of D-branes 
in string theory.

In our case, we are interested in the growth of the volume at large $n$ with $g = 0$, which is given by
\be
\lim_{n \to \infty} {\log(V_{0,n}) \over n \log(n)} = 1 + \Or[{1 \over \log(n)}]
\ee
or in simpler notation
\be
\label{wpvol}
\lim_{n \to \infty} V_{0,n} \propto n!,
\ee
where we have kept only the leading part of the growth and dropped other factors, including those that are merely exponential in $n$.

We will now argue that the full scattering amplitude is dominated by the factorial growth \eqref{wpvol} since the other terms in the string scattering amplitude have sub-factorial growth. 

\subsection{Bounds on functional determinants}
The functional determinants that appear in \eqref{finalbosonic} can be related to special values of the Selberg zeta function and its derivatives. The Selberg zeta function is defined as 
\be
\label{selberg_zeta}
Z(s)=\prod_{\Theta}\prod_{k=1}^{\infty}\left(1-e^{-(s+k)\ell_{\Theta}}\right),  \quad \mathrm{Re(s)}>1. 
\ee 
Here the product labelled by $\Theta$ is over all simple  closed geodesics on the Riemann surface. The length of $\Theta$, $\ell_{\Theta}$, is measured with respect to the hyperbolic metric on the Riemann surface. In terms of $Z(s)$ we have \cite{D'Hoker:1988ta}
\be
 \label{det}
\begin{split}
&\mathrm{det}'\Delta =e^{-c_{0}\chi}\left(\frac{dZ}{ds}\right)_{s=1},  \\
&\mathrm{det}'(P_{1}^{\dagger}P_{1})=e^{-c_{1}\chi}Z(2),
\end{split}
\ee
where $\chi$ is the Euler characteristic of the Riemann surface under consideration. The constants $c_n$ are $\Or[1]$ numbers given by 
\be
\label{cgammadef}
c_\alpha = \sum_{0 \leq m \leq \alpha-{1 \over 2}} \Big[(2 \alpha - 2 m - 1) \log(2 \alpha - m) \Big] - (\alpha + {1 \over 2})^2 + 2 (\alpha - [\alpha]) (\alpha + {1 \over2}) \log(2 \pi) + 2 \zeta'(-1).
\ee
In particular, we have $c_0 \approx -0.58$ and $c_1 \approx -1.89$.

In most of the moduli space, we expect the Selberg zeta function to be well behaved. However, it is important to bound this term near the boundaries of moduli space and ensure that it cannot affect the factorial growth of the amplitude.

The behaviour of the Selberg zeta functions near the boundaries of moduli space, where the Riemann surface degenerates was studied in \cite{Gross:1988ib, D'Hoker:1988ta, wolpert1987asymptotics}. The basic method of estimating the behaviour of the Selberg zeta function near the boundaries of the moduli space is as follows. Let  $\Theta$ be a simple closed geodesic which gets pinched when two given punctures approach each other on the worldsheet and consider the collar region around $\Theta$, which is defined as 
\be
C(\Theta, r) = \{z: \rho(z,\Theta) < r\},
\ee
where $\rho(z,\Theta)$ is  the hyperbolic distance of a point $z$ from $\Theta$, and $r$ is referred to as the radius of the collar.

It turns out that as the length of this geodesic, $\ell \to 0$, the geometry of the Riemann surface excluding a collar of radius $r = \log(\ell^{-1})$ remains uniformly bounded.  The geodesics, which only remain in this part of the Riemann surface and do not intersect $\Theta$ are not affected by the degeneration of $\Theta$ and their contribution to the Selberg zeta function is simply a constant in the limit where $\ell \to 0$.  However the lengths of the geodesics which happen to intersect $\Theta$ tend to infinity since these now have to cross the collar region. As a result their contribution to $Z(s)$ is simply $1$ in the limit where $\ell \to 0$. 

This leaves behind the degenerating geodesic itself,  and its inverse, which has the same length. Their contribution to the infinite product can be calculated explicitly. This analysis allows one to estimate the behaviour of the Selberg zeta function and its derivatives as the moduli space degenerates \cite{D'Hoker:1988ta} and one finds that
\be
Z(s) \underset{\ell \rightarrow 0}{\longrightarrow} \ell^{-2s+1} e^{-{\pi^{2} \over 3 \ell}},
\ee

In our case, we recall that we have bounded the length of the smallest simple geodesic on the Riemann surface by a n-independent constant $\ell_0$. This  means that near the boundaries of this cutoff moduli space
\be
Z(2) \propto \ell_{0}^{-3} e^{-{\pi^{2} \over 3\ell_{0}}},
\ee
up to an $\ell_0$ independent constant. From the formulas above, we can also derive the behaviour of $Z'(1)$ near the boundary of the cutoff moduli space, which is given by
\be
Z'(1)\propto \ell_{0}^{-1} e^{-\pi^{2}/3\ell_{0}}.
\ee
Our interest in these formulas is restricted to the fact that, in the cutoff moduli space the Selberg zeta functions that appear in \eqref{det} are bounded away from $0$ and $\infty$ by finite constants. 

Now note that the only other terms that appear in \eqref{det} are $e^{-c_0 \chi}$ and $e^{-c_1 \chi}$. While these vary exponentially with $n$, since $\chi = 2 - n$, this behaviour is subleading compared to the factorial growth of the volume of moduli space. In fact since $c_0 < 0$ and $c_1 < 0$ both determinants in \eqref{det} {\em decay} with $n$.  

As far as the ghost determinant is concerned, the main result that we are interested in is that it is bounded below and does not {\em decay} as rapidly as a factorial.  However, we see that it also does not grow as a factorial and that
\be
\frac{\log\{\mathrm{det}'(P_1^{\dagger}P_1)\} }{n\log(n)}= 0+\mathcal{O}\left(\frac{1}{\log(n)}\right),
\ee
in the cutoff moduli space.
On the other hand,  to bound the amplitude from below we only need the result that the determinant of the scalar Laplacian is bounded {\em above} so that its inverse does not decay as rapidly as a factorial. However, we have the stronger result that
\be
\frac{\log\{\mathrm{det}'(\Delta)\} }{n\log(n)}= 0+\mathcal{O}\left(\frac{1}{\log(n)}\right),
\ee
in the cutoff moduli space.

Therefore the functional determinants in \eqref{det} do not alter the factorial growth shown in \eqref{wpvol}.

 \subsection{Bounds on Green's functions \label{boundgreen}}
We now argue that the term $Q_n$ also does not affect the factorial growth of the amplitude. 

Once again, we start by ensuring that its contribution
from the boundaries of moduli space is bounded. The region on the Riemann surface near the degeneration locus where two punctures on the worldsheet collide is conformally equivalent to a twice punctured disc. In \cite{hempel1988hyperbolic}, it was shown that on a twice punctured disc $\mathcal{D}\backslash \{-a,a\}$, the length, $\ell$ of the smallest closed geodesic separating the punctures $\{-a,a\}$ from the boundary of the disc $\partial\mathcal{D}$ is given by 
 \be
\ell = \frac{2\pi^{2}}{\log (1/a)},
\ee
as the punctures coalesce, i.e., as $a\to 0$. (See also \cite{beardon2012uniformisation}.)

Using this result, and the formula for the worldsheet Green's functions \eqref{worldsheetgreen} we see that on the cutoff moduli space, $G_{ij}$ can then be bounded as 
\be
|G_{ij}\ell_0| \leq C,
\ee
near the region where the punctures $i,j$ come close and where $C$ is a constant which is independent of the number of punctures, $n$.

It is also of interest to determine the contribution of the Green's function between two generic punctures on the punctured Riemann surface. We can set up holomorphic  coordinates on the punctured Riemann surface by setting the hyperbolic metric to be $ds^2 = e^{\rho} d z d \bar{z}$. The function  $\rho$ must have the appropriate logarithmic singularities at the positions of the punctures where the metric behaves like \eqref{nearpuncture}. Furthermore, we can fix the isometries of this metric by demanding that three of the punctures be at $z=0,z=1$ and $z=i$. 

Then, noting that the area of the Riemann surface is $2 \pi (n - 2)$, at a generic point in moduli space, we expect that we will have $|\log|z_i - z_j|| = \Or[\log(n)]$ for two generic punctures.

Next, note the energy per-particle scales as $E \propto {1 \over n^{\gamma}}$. Therefore the factor of $k_i \cdot k_j$ in the exponent of $Q_n$ has a magnitude of order ${1 \over n^{2 \gamma}}$.  Second although there are $\Or[n^2]$ terms in the exponent of $Q_n$,  the factor of $k_{i} \cdot k_{j}$  does not have a definite sign. More precisely, correlations between these terms come only from the momentum conserving delta function, but this single constraint tends to become unimportant in the large $n$ limit. The Green's function $G_{i j}$ also has varying signs on the moduli space, although it is bounded and has typical magnitude $\log(n)$ as explained above. 

Hence we expect that the sum of $\Or[n^2]$ terms, each of size $\Or[\log(n) n^{-2 \gamma}]$  will only contribute an $\Or[\log(n) n^{1 - 2 \gamma}]$ factor to the magnitude of the exponent, at least at most points in the moduli space. 

These arguments suggest that 
\be
\label{qnbound}
\frac{\log |\log Q_{n}|}{(1-2\gamma)\log(n)}  \leq  1+\mathcal{O}\left(\frac{\log(\log(n))}{\log(n)}\right),
\ee
on a significant fraction of the moduli space. 
This can be rewritten as
\be
\label{qngrowth}
|\log(Q_n)| \leq q \log(n) n^{1 - 2 \gamma},
\ee
where $q$ is some $\Or[1]$ factor on a significant part of moduli space. 

From \eqref{qngrowth} we see that, for any value of $\gamma > 0$, the possible suppression due to $Q_n$ is subleading compared to the factorial growth of the volume of the moduli space. 
Note that, in this subsection, we have been somewhat heuristic. We have also not bounded $Q_n$ everywhere in the cutoff moduli space, as we were able to do for the term involving the functional determinants.  In particular, we have not ruled out the possibility that $Q_n$ might become very 
large on some parts of moduli space, which might enhance the factorial growth of the amplitude. 

Nevertheless, insofar as the issue of proving a {\em lower bound} on the growth of the moduli-space integral in the cutoff moduli space is concerned, we believe that these arguments are correct.  Our conclusions above  are also verified very nicely by the numerical calculations of section \ref{numericalgrowth}. In particular, we direct the reader to figure \ref{figsaddle} which shows that \eqref{qngrowth} provides an excellent fit to our numerical results when $\log(Q_n)$ is evaluated at the saddle points of the moduli-space integral.

This analysis suggests that the factor of $Q_n$ does not modify the factorial growth of the amplitude that comes from the volume of moduli-space.

\subsection{Prefactors}
We are left with the prefactors $P_n$ and $\bar{P}_n$. These prefactors contain, within themselves, both a factorial number of terms, and also a product of $n$ Green's functions. Thus, in principle, this prefactor could either grow factorially or suppress the overall factorial growth of the volume.

A heuristic, and indirect,  argument that this prefactor does not alter the factorial growth of the amplitude is as follows. By unitarity, we know that massless amplitudes appear in the residues of the poles of tachyon amplitudes. If we can show that tachyon amplitudes grow factorially, then it is likely that these residues --- and, hence, the massless amplitudes --- also grow factorially and that the factors of $P_n$ do not suppress this growth.

If we had been considering tachyon amplitudes, then the analysis of the volume of moduli space and functional determinants would have been just as in the previous sections. The analysis of the factor $Q_n$ would also have been similar to subsection \ref{boundgreen} except that the typical components of the tachyon momenta would scale with ${1 \over \sqrt{\alpha'}}$. This corresponds to $\gamma = 0$ in the notation above. 

The analysis of this case is somewhat delicate since, on a large fraction of moduli space, we see from \eqref{qngrowth} that the suppression due to $Q_n$ may itself be as strong as ${1 \over n!}$. However, we would still expect that, at least on an exponentially small fraction of moduli space the inequality $|\log(Q_n)| < n \log(n)$ would hold. Since the volume of this fraction already grows factorially with $n$, this suggests that the final answer for tachyon amplitudes also grows factorially with $n$. 

By the argument above, this suggests that the factors of $P_n$ and $\bar{P}_n$ do not suppress the factorial growth of massless amplitudes.

However, this argument is only suggestive and not entirely satisfying. Therefore, here, we will borrow a result from section \ref{numericalgrowth}. The numerical analysis of section \ref{numericalgrowth} shows that $P_n$ does not decay as a factorial. The numerical analysis also suggests that $P_n$ does not grow factorially but we are unable to entirely rule out this latter possibility 
due to subtleties in the numerics  explained in section \ref{secnumericalerrors}. 

Nevertheless, this result tells us that  the factorial growth shown in \eqref{wpvol} provides at least a {\em lower bound} on the rate of growth of string amplitudes. 
 \subsection{Growth of tree amplitudes}
Combining the results above, we have
\be
{\log(\mtr(k_1 \ldots k_n)) \over (n-2) \log(g_s) + n \log(n)} \rightarrow 1,
\ee
in the limit where $g_s \rightarrow 0$ and $n \rightarrow \infty$. Said differently, we see that
\be
\label{factorialgrowth}
\mtr(k_1, \ldots k_n) \propto g_s^{n-2} n!,
\ee
where the $\propto$ sign indicates that we have dropped other terms that do not grow as fast as $n!$. This is the result that we wanted to prove.

\subsection{Superstring amplitudes}
We now describe how this analysis can be extended to the superstring. The analysis is largely parallel to the bosonic string, so we will be brief and not repeat all of our steps. We will also confine ourselves to tree-level amplitudes to avoid subtleties with the superstring moduli space at higher genus.

We start with the formulation of the superstring scattering amplitude as 
\be
\label{susypunctures}
\mtr(k_1 \ldots k_n) = {\cal N} g_s^{n-2} \int \dmwp  (\detp P_{1 \over 2}^{\dagger} P_{1 \over 2})^{-1 \over 2}  (\detp P_1^{\dagger} P_1)^{1 \over 2} (\detp \slashed{D})^{d \over 2} (\detp \Delta)^{-d \over 2} Q_n P_n \bar{P}_n,
\ee
which is the natural supersymmetric generalization \cite{DHoker:1986hhu} of \eqref{polyakovpunctures}. The determinant of the worldsheet Dirac operator arises from integrating out the worldsheet fermions, and in addition we also obtain a determinant by integrating out the superghosts. In the expression above, we have performed the integral over the odd moduli, which leaves behind the expression $Q_n P_n \bar{P}_n$. The only integral that remains is over the positions of the punctures. We have abused notation to use the same symbols $P_n, \bar{P}_n$ for the worldsheet correlators as in the bosonic string but in the case of the superstring the values of these correlators are different. This should not cause any confusion and which expression for $P_n$ needs to be used should be clear from the context.

We now describe this worldsheet correlator in more detail. We define
\be
\label{auxiliaryv}
V(\theta, \chi, z) =  \exp  \left(i \theta \chi \epsilon \cdot \partial X  + \theta k \cdot \psi - \chi \epsilon \cdot \psi \right),
\ee
which depends on the momentum $k^{\mu}$, a polarization vector $\epsilon^{\mu}$, and two auxiliary Grassmann variables $\theta, \chi$ \cite{Green:1987sp}.  
Then, we recall that the vertex operators for NS-NS sector operators in the type II superstring with polarization tensor $\epsilon_{\mu \nu} = \epsilon_{\mu} \epsilon_{\nu}$\footnote{For this choice of polarization tensors, which corresponds to the scattering of linear combinations of the graviton and the dilaton, tree-level amplitudes in the type I theory are the same as tree-level amplitudes in the type II theories.} are given in the $(-1,-1)$-picture and the $(0,0)$ picture by
\be
\begin{split}
&\mathcal{V}^{(-1,-1)}(z, \bar{z})= e^{-\phi - \bar{\phi}} e^{i k \cdot X} \int  d \bar{\chi} d \chi V(\theta = 0, \chi, z) \widetilde{V}(\bar{\theta} = 0, \bar{\chi},\bar{z}), \\
&\mathcal{V}^{(0,0)}(z, \bar{z})  =  e^{i k \cdot X}  \int    d \bar{\chi} d \bar{\theta}  d \chi d \theta  V(\theta, \chi, z) \widetilde{V}(\bar{\theta}, \bar{\chi}, \bar{z}), \\
\end{split}
\ee
where the operator $e^{-\phi - \bar{\phi}}$ arises from the bosonized superconformal ghost insertions and $\widetilde{V}$ is defined in analogy to \eqref{auxiliaryv} with all the left-moving fields
replaced with right-moving ones.

The $n$-point worldsheet correlation function that we need is then obtained by inserting 2 $(-1,-1)$ picture operators and $(n-2)$-(0,0) picture operators. This is given by
\be
\begin{split}
Q_n P_n \bar{P}_n &=\left\langle \mathcal{V}^{(-1,-1)}(z_{1},\bar{z}_{1})\mathcal{V}^{(-1,-1)}(z_{2},\bar{z}_{2})\mathcal{V}^{(0,0)}(z_{3},\bar{z}_{3})\cdots\mathcal{V}^{(0,0)}(z_{n},\bar{z}_{n})\right\rangle \non\\
&= \frac{1}{|z_{12}|^{2}} \int \prod_{i=3}^n d \theta_i d \bar{\theta}_i \prod_{j=1}^n d \chi_j  d \bar{\chi}_j \exp{I} \exp{\bar{I}} 
\end{split}
\ee
where
\be
I = \sum_{i \neq j} {1 \over 2} k_i \cdot k_j \ln(z_i - z_j)  - {\theta_i \theta_j k_i \cdot k_j \over z_i - z_j} + {\theta_i - \theta_j \over z_i - z_j} (\chi_i \epsilon_i \cdot k_j + \chi_j \epsilon_j \cdot k_i) - {\chi_i \chi_j \epsilon_i \cdot \epsilon_j \over z_i - z_j} - {\theta_i \theta_j \chi_i \chi_j \epsilon_i \cdot \epsilon_j \over (z_i - z_j)^2},
\ee
and it is understood that $\theta_1 = \theta_2 = 0$ in the expression for $I$.

Now the integral over the Grassmann variables may be done as follows. We separate the last term in $I$, which is quartic in the Grassmann variables, and then pull it down using a power series expansion of the exponent. Each power of this term soaks up some of the Grassmann integrals. The integral over the rest of the Grassmann variables is Gaussian and so can be performed in terms of a Pfaffian.
The result, after this  manipulation, can be seen to be $Q_n P_n \bar{P}_n$ where
\be
\begin{split}
\label{susyexprpq}
Q_n = &\prod_{i \neq j} |z_i - z_j|^{k_i \cdot k_j}, \\
P_{n} = &{1 \over z_{12}}\Bigg[\mathrm{Pf}(M_{12})+\sum_{\{k\}, \Pi_k} \left(\prod_{p=1}^{k}\frac{(\epsilon_{\pi_{2 p - 1}} \cdot \epsilon_{\pi_{2 p}})}{(z_{\pi_{2 p - 1}}-z_{\pi_{2 p}})^{2}}\right)\mathrm{Pf}\left(M_{12\pi_{1}(\pi_{1}+n)...\pi_{2 k} (\pi_{2 k} + n) }\right) \Bigg].
\end{split}
\ee
Here, the matrix $M$ is defined as
\be
\label{defm}
M = \begin{pmatrix}
A & -C^T \\ C & B
\end{pmatrix},
\ee
and the matrices $A,B,C$ are defined through
 \be
A_{ij}= \Bigg\{ \begin{array}{cc}
\frac{k_{i} \cdot k_{j}}{z_{i}-z_{j}} & i\neq j\\
0 & i=j\\
\end{array};
\quad
B_{ij}= \Bigg\{ \begin{array}{cc}
\frac{\epsilon_{i} \cdot \epsilon_{j}}{z_{i}-z_{j}} & i\neq j\\
0 & i=j\\ \end{array} ;
\quad 
C_{ij}= \Bigg\{ \begin{array}{cc}
\frac{\epsilon_{i} \cdot k_{j}}{z_{i}-z_{j}} & i\neq j\\
-\sum_p {\epsilon_i \cdot k_p \over z_i - z_p}   & i=j\\
\end{array},
\ee
and the notation $M_{i_1  \ldots i_n}$ means that the rows $i_1, \ldots i_n$ and the columns $i_1 \ldots i_n$ are removed before taking the Pfaffian. The sum over $\{k\}, \pi_k$ is over all  all distinct choice of $k$ pairs  from the range $(3 \ldots n)$ where $k$ runs from $1 \ldots (n-2)/2$. The pairs themselves are specified by $\{(\pi_1, \pi_2) \ldots (\pi_{2 k - 1}, \pi_{2 k}) \}$. 
For future reference we note that the Pfaffian in the last term in the sum above, with $k = (n-2)/2$, can be simplified using
\be
\label{lastterm}
\mathrm{Pf}\left(M_{1,2,3,3+n, \ldots n, 2 n} \right) = {\epsilon_1 \cdot \epsilon_2 \over z_{1 2}},
\ee
since only the rows and columns $(n+1)$ and $(n+2)$ are left after the deletion above.

The rest of the analysis for the type II superstring amplitude is entirely parallel to the analysis for the bosonic string that we displayed above. In particular, the functional determinants that appear
in the path-integral above can again be related to special values of the Selberg zeta function \cite{sarnak1987determinants,d1986determinants} through
\be
\begin{split}
&\detp P_{1 \over 2}^{\dagger} P_{1 \over 2} = \exp\left({- c_{1 \over 2} \chi} \right) Z({3 \over 2}), \\
&\detp \slashed{D} \slashed{D} = \exp\left(-c_{-{1 \over 2}} \chi \right)  {Z^{(2 N)}({1 \over 2}) \over (2 N)!}.
\end{split}
\ee
Here $N$ is the number of zero-modes of the Dirac operator and the constants $c_{\pm {1 \over 2}}$ are also given by \eqref{cgammadef}. It is not difficult to see, by a simple extension of the analysis above, that these functional determinants are also bounded away from factorial growth, and therefore do not affect the leading factorial behaviour of the amplitude.

Since $Q_n$ for the superstring has the same form as $Q_n$ for the bosonic string, our bounds on this term that we analyzed for the bosonic string also apply here.  To analyze the prefactors $P_n$ and $\bar{P}_n$,  the heuristic relation to tachyon amplitudes that was given for the bosonic string can also be utilized here. This is because the processes we are considering, in principle, also make sense in the type 0 string theories. The massless scattering amplitudes we are considering here can, therefore, be obtained by factorizing tachyon amplitudes in the type 0 theories. By using the bounds on functional determinants and the bound on $Q_n$ above, we can argue that tachyon amplitudes in the type 0 theories grow factorially. This suggests that at the poles of these amplitudes,  the residues, which include the massless amplitudes, also grow factorially.  

However, this argument is not watertight and, with just these arguments,  we cannot entirely rule out the possibility that the expansion of the Pfaffians leads to an additional term that also grows like $n!$ or alternately become very small. With the help of numerical analysis, in the next section, we will be able to rule out the possibility that $P_n \bar{P}_n$ decays as rapidly as ${1 \over n!}$. The numerical analysis also suggests that these terms {\em do not} grow as rapidly as $n!$ but this conclusion is less robust for reasons that we detail below.

Assuming this property of $P_n$, we find that \eqref{factorialgrowth} holds for the superstring amplitude as well.

\section{Numerical estimates of the growth of string amplitudes \label{numericalgrowth}}
In this section, we will turn to a numerical analysis of string scattering amplitudes. In our analysis above, we provided some evidence that string amplitudes grow at least as fast as $n!$ at large $n$. However, we were unable to deal precisely with the effect of the factor of $P_n \bar{P}_n$  that appears in \eqref{finalbosonic} and \eqref{susypunctures}.   By numerically estimating the growth of string amplitudes in this section, we will verify the factorial growth in an entirely independent manner and also check that the factors of $P_n$ and $\bar{P}_n$ do not seem to change this behaviour. Our conclusion that the factors of $P_n \bar{P}_n$ do not suppress the factorial growth is robust. However, our result that they do not enhance this growth is subject to some caveats as we describe in  section \ref{secnumericalerrors}.

In this section, it will be convenient to go back to the choice of a flat worldsheet metric. Therefore, our expressions for the bosonic and type II superstring amplitudes can both be written as
\be
\label{numericalexpr}
\mtr(k_1 \ldots k_n) = 4 \pi g_s^{n-2} \int \prod_{i=4}^{n} d^2 z_i Q_n P_n \bar{P}_n |{\cal G}|,
 \ee
where $P_n$ and $Q_n$ are given in \eqref{bosonicexprpq} for the bosonic string and \eqref{susyexprpq} for the type II superstring and ${\cal G}$ is the ghost contribution given by
\be
\label{ghostdet}
{\cal G} = z_{12 }^2 z_{2 3}^2 z_{1 3}^2.
\ee
Note that although the superstring also receives a contribution from the superghost insertions, we have already included them in \eqref{susyexprpq}.
 We have now also fixed the overall normalization of the scattering amplitude which can be determined by unitarity \cite{Polchinski:1998rq}. 

We note that this integral runs over $(n-3)$-complex dimensions. Moreover, as we mentioned above, it suffers from divergences when $z_i \rightarrow z_j$. As we explained above, these divergences can be regulated systematically. One procedure was outlined in  \cite{Witten:2013pra} --- which suggested a suitable $i \epsilon$ prescription ---  and another procedure was described in \cite{Sen:2014dqa} --- which described how the region of moduli space that led to the divergences could be isolated and dealt with using field theoretic techniques. However, in practice, both of these prescriptions are non-trivial to implement on a computer. Therefore, we will not directly attempt to perform the integral in \eqref{numericalexpr}.

Instead, we are able to proceed further as follows. Upon considering the string integral, we find that, even though individual energies are small, $\log(Q_n)$ becomes large at large $n$. The exponent in $Q_n$ has $\Or[n^2]$ terms and while we expect  these terms to cancel among each other, we still expect that the saddle is of order $n^{1 - 2 \gamma}$ as shown in \eqref{qngrowth}.  Therefore, even in the situation where the individual energies are small, we can approximate the amplitude by localizing the moduli-space integral onto the points where $\log(Q_n)$ is maximized. This procedure is not only numerically efficient, it also has the advantage that it sidesteps the issue of  divergences in the moduli-space integral.

Extremizing the exponent in $Q_n$ leads us to focus on the values of $z_i$ that satisfy 
\be
\label{scateq}
E_i = \sum_{j \neq i} {k_i \cdot k_j \over z_i - z_j } = 0, \forall i.
\ee
These equations were first discovered in \cite{Gross:1987ar,Gross:1987kza} in the study of high-energy string scattering. However they have recently turned out to be useful in the study of scattering in ordinary quantum field theories \cite{Cachazo:2013hca,Cachazo:2013iea}.

Note that solutions to the scattering equations are invariant under simultaneous $SL(2,C)$ transformations of the variables $z_i \rightarrow {a z_i + b \over c z_i + d}$ with $a d - b c = 1$. This invariance also appears in the string amplitude and can be gauge-fixed by setting $z_1, z_2, z_3$ to definite values. Modulo this gauge invariance, it is easy to see that there are actually $(n-3)!$ solutions of the scattering equations. This was proved in \cite{Cachazo:2013gna}. The proof is not difficult, and proceeds by induction. 

Assume that the number of distinct solutions to the scattering equations with $n-1$ particles is $(n-4)!$. Now consider the scattering equations with $n$ particles. We can use the $SL(2,C)$ freedom to set $z_1 = \infty$. Then  $z_1$ and $k_1$ drop out of the equations \eqref{scateq}.   We now deform the first and the last momenta through $k_n \rightarrow \alpha k_n$, while simultaneously deforming $k_1 \rightarrow k_1 + (1 - \alpha) k_n$. The deformation of $k_1$ has no effect since we have set $z_1 = \infty$. As we take $\alpha$ from $1$ to $0$, we see that the $\alpha k_n$ drops out of the scattering equations, $(E_1 \ldots E_{n-1})$  and this set of equations becomes an independent set of scattering equations for $n-1$ particles. By assumption, this set has $(n-4)!$ solutions. On the other hand, $E_n$ is independent of $\alpha$. This gives us a polynomial equation of order $(n-3)$ for $z_n$ that should have $(n-3)$ roots. Therefore for each of the $(n-4)!$ solutions to the equations $(E_1, \ldots E_{n-1})$  we now have $(n-3)$ solutions for $z_n$ leading to a total of $(n-3)!$ solutions for the full system $(E_1, \ldots E_n)$.  Now, as we deform $\alpha$ away from 0, and back towards 1, we can assume that these $(n-3)!$ solutions move continuously in the complex plane leading to $(n-3)!$ solutions for the original undeformed scattering equations.

It is rather remarkable that this is exactly the same number as the estimate of the volume of moduli space in \eqref{wpexact} at $g = 0$ including even the subleading terms in $n$. Therefore, there is {\em one solution of the scattering equations per unit volume of moduli space}. We do not understand the reason for this phenomenon.

Localizing on the solutions to the scattering equations, and after performing the integral about the Gaussian fluctuations about these saddle-points, our prescription for numerical evaluation of the amplitude becomes
\be
\label{saddlesum}
\mtr(k_1 \ldots k_{n}) = (4 \pi g_s)^{n-2}  \sum_{\{z_i\}} |{\cal G}| |{\cal J}|^{-1} Q_n P_n \bar{P}_n.
\ee
The sum goes over {\em all} inequivalent solutions of the scattering equations. Here $|{\cal J}|^{-1}$ is the Jacobian factor that results from integrating fluctuations about the saddle points and is given by ${\cal J} = \det(\partial_j E_{i})$. More explicitly, this matrix is
\be
\label{eijdef}
\partial_j E_{i} = \left\{ \begin{array}{ll}
{k_i \cdot k_j \over (z_i - z_j)^2},  &\quad i \neq j \\
-\sum_{q \neq i} {k_i \cdot k_q \over (z_i - z_q)^2}, &\quad i = j,
\end{array} \right.
\ee
and $i,j \in (4 \ldots n)$. The overall factor multiplying the amplitude arises because the integral over Gaussian fluctuations yields $(4 \pi)^{n-3} |{\cal J}|^{-1}$, and this factor combines with the $4 \pi g_s^{n-2}$ in \eqref{numericalexpr} to give $(4 \pi g_s)^{n-2}$.

We see immediately that we can also write our estimate as
\be
\label{numericalestimate}
\mtr(k_1 \ldots k_{n})  =  (n-3)! (4 \pi g_s)^{n-2} |{\cal G}| \langle |{\cal J}|^{-1} Q_n P_n \bar{P}_n  \rangle,
\ee
where by $\langle |{\cal J}|^{-1} Q_n P_n \bar{P}_n \rangle$, we mean the mean of this quantity across the set of solutions of the scattering equations. This mean can be estimated statistically by taking a {\em sample} of the set of saddle-points. Thus, by sampling over a large set of solutions to the scattering equations, we obtain an estimate for the full amplitude without having to find all $(n-3)!$ solutions to the scattering equations.

\subsection{Brief description of algorithm}
We now briefly describe the algorithms that we use here to solve the scattering equations and evaluate the integrand. We provide some additional details in Appendix \ref{appnum}. 

We choose a set of random external kinematics using a uniform measure in phase space. This can be done using standard algorithms \cite{kleiss1986new}  used to generate events in phenomenological calculations, as explained further in Appendix \ref{appnum}.

Given a set of external momenta, the first task in evaluating \eqref{numericalestimate} is to obtain solutions of the scattering equations. Our algorithm simply starts at a random point in $C^{(n-3)}$ (with some cutoffs that disallow very large initial values of $z$) and then uses a variant of the multi-dimensional Newton's method to seek the nearest solution. Multi-dimensional root-finding algorithms are not guaranteed to converge, so if the algorithm does not converge, it simply picks another starting point and flows to a nearby root. We caution the reader that although we pick our initial starting point using as uniform a distribution as possible, this does not mean that we are sampling the roots uniformly. This is because we do not have any prior estimates for the sizes and shapes of the basins of attraction of different roots. 

After having obtained a root, our next task is to evaluate the summand in \eqref{numericalestimate}. The terms ${\cal J}^{-1}$ and $Q_n$ are straightforward to evaluate. $Q_n$ can be obtained by summing over the ${n(n - 1) \over 2}$ terms in its exponent. The Jacobian factor, ${\cal J}^{-1}$ is a determinant of a $(n-3) \times (n-3)$ matrix, which can also be evaluated efficiently using LU decomposition \cite{press1982numerical}. However, note that  $P_n$ contains a large number of terms. Several of these terms are numerically expensive to evaluate, as we describe below. Therefore, in order to evaluate this term efficiently, we truncate the prefactor for the superstring as
\be
\label{truncatesusy}
P_n \bar{P}_n \approx |{\mathrm{Pf}(M_{12}) \over z_{12}}|^2 + \sum_{\pi} |\frac{\epsilon_{1}.\epsilon_{2}}{z_{12}^2}\prod_{i=1}^{(n-2)/2}\frac{\epsilon_{\pi_{2 i - 1}}.\epsilon_{\pi_{2 i }}}{(z_{\pi_{2 i - 1}}-z_{\pi_{2 i}})^{2}}|^2.
\ee
Here we note that the last term comes from \eqref{lastterm}. Here the sum over permutations runs over distinct pairings  $(3, \ldots n) \rightarrow \{(\pi_1, \pi_2) \ldots (\pi_{n-3}, \pi_{n-2})\}$.

For the bosonic string we truncate the prefactor as
\be
\label{truncatebos}
P_n \bar{P}_n \approx |\prod_{i} \sum_{j\neq i} {\epsilon_i \cdot k_j \over z_i - z_j}|^2  + \sum_{\pi} |\prod_{l=1}^{n \over 2} {\epsilon_{\pi_{2 l - 1}} \cdot \epsilon_{\pi_{2 l}} \over (z_{\pi_{2 l - 1}} - z_{\pi_{2 l }})^2}|^2.
\ee
Here the sum over permutations runs over all distinct pairings of all $n$ particles: $(1, \ldots n) \rightarrow \{(\pi_1, \pi_2) \ldots (\pi_{n-1}, \pi_{n})\} $. 

There are two truncations involved here. First, we see that the truncation reduces the $\Or[n^2]$ terms that appear in \eqref{susyexprpq} and \eqref{bosonicexprpq} to 2 terms each. We do not expect this to make any difference to the factorial growth, which is our main interest. Moreover, we expect that these two terms are the most important terms in the amplitude.  Naively, we might also expect that the term involving the polarization vectors is important at lower energies since it involves the fewest possible factors of the external momenta. On the other hand, we expect that the term involving the Pfaffian will be important at relatively higher energies for the superstring. Similarly, we expect that the term involving dot products of the momenta with the polarizations will be important at higher energies for the bosonic string. 

However, more significantly we have placed the permutation outside the absolute value sign in \eqref{truncatesusy} and \eqref{truncatebos} even though the original expressions involved the absolute-value squared of the sum. This is because we expect that the other terms in the sum will suffer cancellations, and therefore this part  --- which involves the sum of absolute-value squares in which all terms are positive  --- is likely to dominate. We cannot verify this assumption directly in our numerics since, in order to do so, we would have to check cancellation in the remaining terms to $\Or[{1 \over n!}]$ which is not feasible. It would be nice to understand this approximation better. 

Empirically we find for the bosonic string for $\gamma = 0$ (relatively higher energies) the first term in \eqref{truncatebos} starts dominating the second term after about $n \sim 60$. For $\gamma = {1 \over 24}$ (relatively lower energies)  the first term against starts dominating after about $n \sim 90$. From this analysis, it appears that, at large $n$, the string amplitude is well approximated by only the first term in \eqref{truncatebos}. However, we direct the reader to the paragraph on ``prefactor estimation'' in Appendix \ref{appnum} for a discussion of whether this is a genuine result or a numerical artifact.

For the superstring the second term in \eqref{truncatesusy} becomes unimportant even earlier --- after about $n \sim 30$ in the case where $\gamma = 0$ and after about $n \sim 70$ in the case where $\gamma = {1 \over 8}$.

Although the prefactor $P_n$ contains only $\Or[n^2]$ terms, the reason these terms are expensive to evaluate is as follows. Consider the term that involves the product of polarization vectors either in \eqref{truncatesusy} or \eqref{truncatebos}. We see that this involves a sum over all possible pairings of the polarization vectors. There are ${n! \over 2^{n \over 2} (n/2)!}$ distinct pairings, and so it cannot be evaluated directly. Therefore, to estimate this term, we sum over a set of random pairings and then multiply by the total number of possible pairings. This sampling process is expensive and it would be very numerically expensive to carry it out for all the other terms in \eqref{susyexprpq} and \eqref{bosonicexprpq}.

\subsection{Results}
We now describe the results of our numerical calculations. For the superstring, we evaluated 1000 solutions of the scattering equations for 500 different sets of random kinematics for each even value of $n \in [4,100]$. This constitutes $2.25 \times 10^7$ distinct solutions of the scattering equations.  For the bosonic string, we evaluated 500 solutions of the scattering equations for 500 different sets of random kinematics for each even value of $n \in [4,100]$.\footnote{Obviously, for $n=4,6,8$ there are only 1, 6, and 120 distinct solutions.} This constitutes $1.13 \times 10^7$ distinct solutions of the scattering equations. Both computations together took about 8000 hours of CPU time. 

\begin{figure}[!h]
\begin{center}
\includegraphics[height=0.3\textheight]{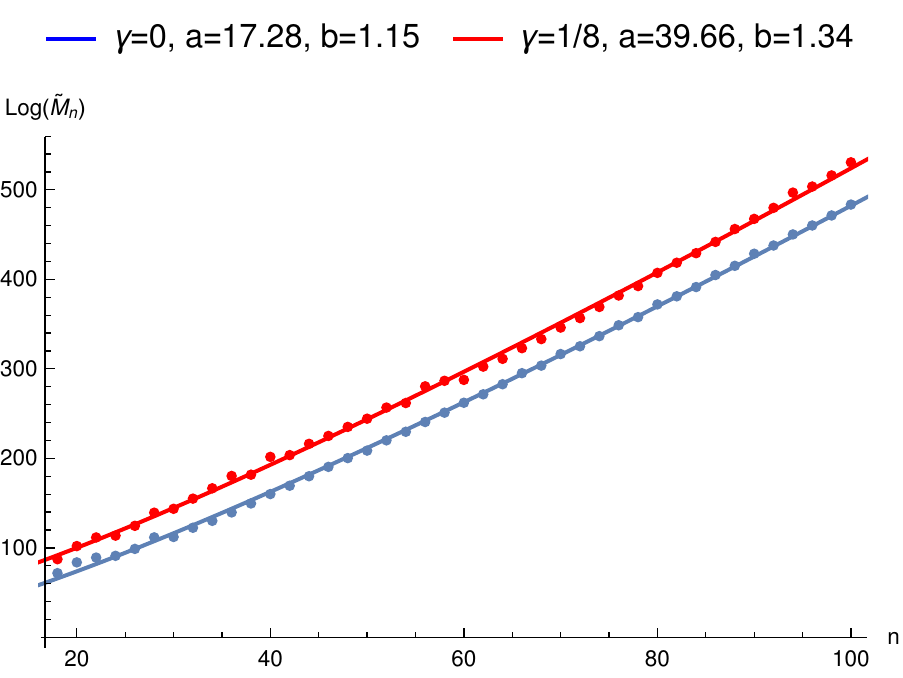}
\end{center}
\caption{\em Scattering amplitudes in the superstring for extreme values of $\gamma$. The solid lines show $a + b n + \log((n-3)!)$ \label{figsusyamp}}
\end{figure}

The results of our calculations for the type II superstring are displayed in figure \ref{figsusyamp}. In our calculation,  we are interested in values of $n$ that are as large as possible. Accordingly, in the figure we only show data for $n \in [18, 100]$ and discard the data for smaller values of $n$. This choice of starting point is simply based on empirical considerations since we find that the data does not obey the same nice trend for smaller values of $n$. On the y axis we plot 
\be
\label{tildemdef}
\log(\tilde{M}_n) = \log( \mtr(k_1 \ldots k_n) ) - (n-2) \log(4 \pi g_s) +   n \log(d-2),
\ee
where $d=10$ is the critical dimension for the superstring. The reason that the last factor $n \log(d-2)$ appears is to compensate for the fact that when we dot a set of  random polarization tensors with the amplitude, we expect to obtain a factor of ${1 \over (d-2)^n}$. Adding  $n \log(d-2)$ to the logarithm of the amplitude strips off this irrelevant kinematic factor and gives us a better sense of the magnitude of the amplitude.

We can fit the amplitude to the following expression
\be
\label{tildemfit}
\log(\tilde{M}_n) = a + b n + \log((n-3)!),
\ee
and the values of $a,b$ for two extreme possible choices of $\gamma$ are shown in figure \ref{figsusyamp}. These values are not of any direct relevance to our calculation. Our emphasis is simply on the leading $\log((n-3)!)$ term, and the graphs show that that it is an excellent fit to the amplitude.  In particular, what figure \ref{figsusyamp} shows is that the factor of $P_n$ does not suppress the amplitude or contribute another factorial term to the amplitude.

These results may be compared to the answer for the bosonic string, which we had previously discussed in \cite{Ghosh:2016fvm}. The graphs for the bosonic string are reproduced below in Figure \ref{figbosamp}.
\begin{figure}[!h]
\begin{center}
\includegraphics[height=0.3\textheight]{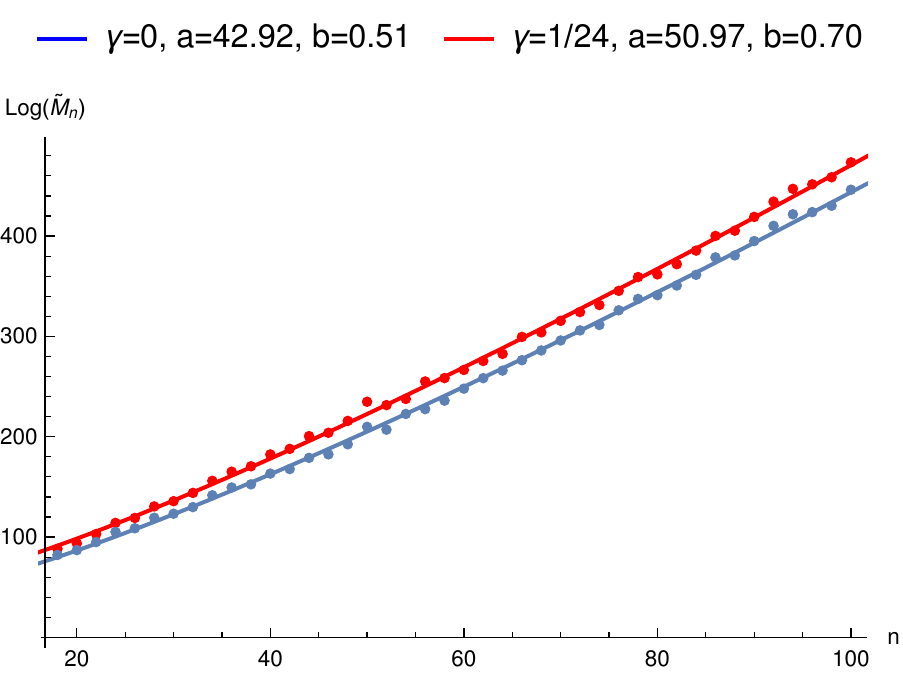}
\end{center}
\caption{\em Scattering amplitudes in the bosonic string for extreme values of $\gamma$. The solid lines show $a + b n + \log((n-3)!)$ \label{figbosamp}}
\end{figure}
Once again we plot $\log(\tilde{M}_n)$ as defined in \eqref{tildemdef} except that now we use $d = 26$, which is the critical dimension for the bosonic string. The best fit values for the fit \eqref{tildemfit}, are also displayed in figure \ref{figbosamp}. Note that the values of $a,b$ that appear here are slightly different from \cite{Ghosh:2016fvm} because, in order to retain consistency with the superstring case, we have discarded data for $n < 18$ here whereas we we had retained data for $n=12,14,16$ in \cite{Ghosh:2016fvm}. Once again we see that the factorial growth, which can be understood as coming from the volume of moduli space or from the number of solutions to the scattering equations, provides an excellent approximation to the amplitude.

Note the various values of $a,b$ are not comparable between the superstring and the bosonic string since the bosonic string amplitudes are evaluated in $d = 26$ whereas the superstring amplitudes are evaluated in $d=10$. The different dimensionality  leads to several subtle effects. For example, it alters the average value of kinematic invariants $k_i \cdot k_j$ even if we choose $|k_i|$ to scale in the same manner. 

\begin{figure}[!h]
\begin{subfigure}{0.5\textwidth}
\includegraphics[width=\textwidth]{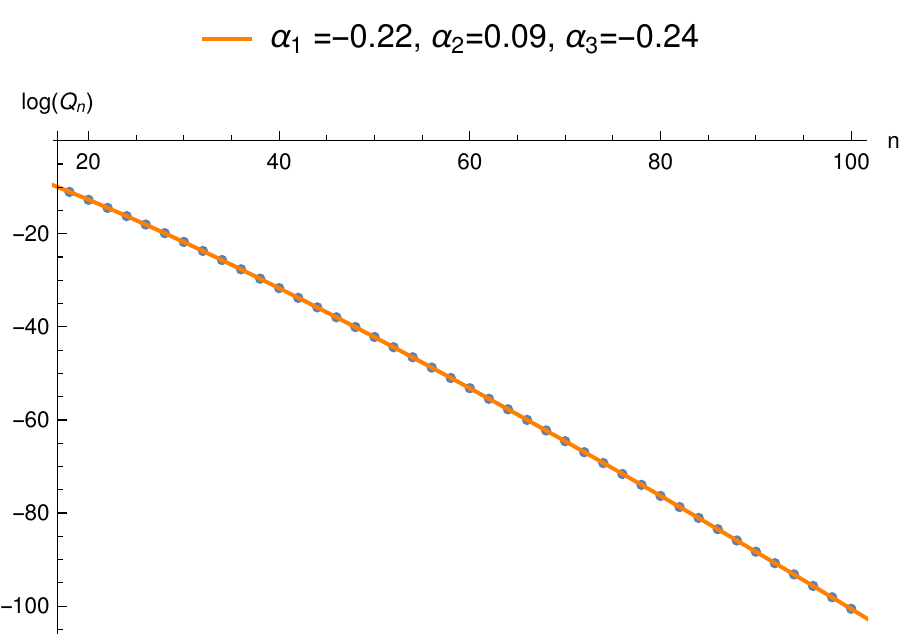}
\end{subfigure}
\begin{subfigure}{0.5\textwidth}
\includegraphics[width=\textwidth]{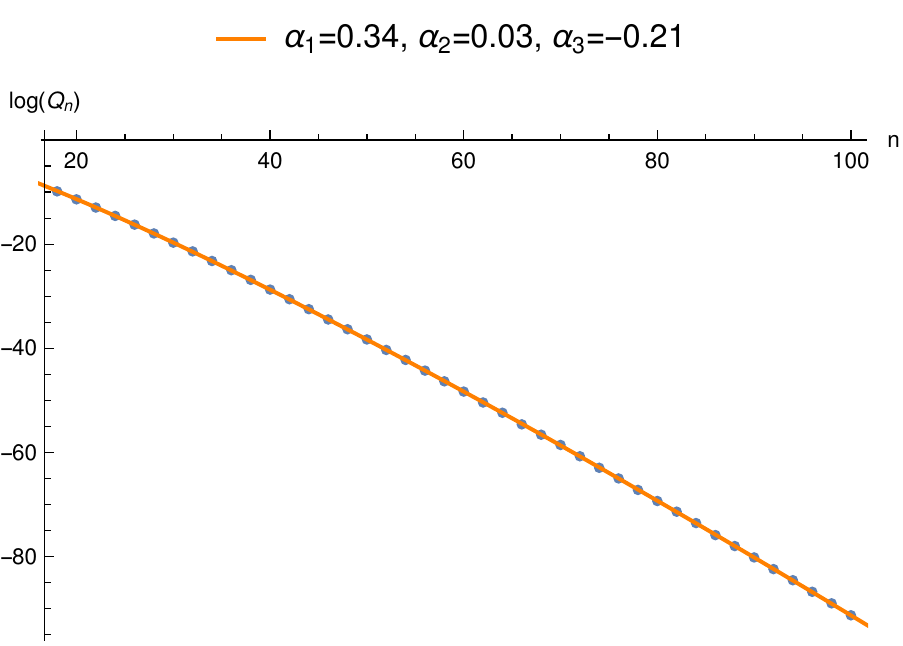}
\end{subfigure}
\caption{\em Plot of $\langle \log Q_n \rangle$ vs n for the bosonic string (left) and the superstring (right). The best fit values $\alpha_1, \alpha_2, \alpha_3$ defined in \eqref{qnfit} are also shown.  \label{figsaddle}}
\end{figure}

We can also use our numerical results to check our expectations about the growth of $Q_n$ outlined in section \ref{boundgreen}. Note that in section \ref{boundgreen}, we were concerned with a generic
point in moduli space whereas, by construction, our numerical results pertain to those points in moduli space where $\log(Q_n)$ is extremized. Nevertheless, in figure \ref{figsaddle} we show $\langle \log(Q_n) \rangle$, averaged over our numerical samples for $\gamma = 0$. Choosing a non-zero value for $\gamma$ would just change $\log(Q_n) \rightarrow {1 \over n^{2 \gamma}} \log(Q_n)$. For $\gamma = 0$, we find that we can fit
\be
\label{qnfit}
\langle \log(Q_n) \rangle = \alpha_1 + \alpha_2 n + \alpha_3 n \log (n).
\ee
We see from figure \ref{figsaddle} that, with the appropriate choice of parameters,  this is a perfect fit both for the bosonic string and for the superstring. This {\em precisely} justifies our analytic expectations from section \ref{boundgreen} 

It is a fact that, for every solution of the scattering amplitudes $\log(Q_n) < 0$. Physically, this is a sign of the good high-energy properties of string theory since it indicates that as we scale up the momentum, $Q_n \rightarrow 0$. On the other hand from the point of view of the scattering equations, viewed as a set of polynomial equations, it is somewhat miraculous that the roots always lead 
to a negative value for $\log(Q_n)$. This property is certainly not true everywhere on the moduli space. It would be nice to understand this property directly from the equations.

We now turn to a discussion of the possible errors in this result.

\subsection{Errors \label{secnumericalerrors}}
\begin{figure}[!h]
\begin{subfigure}{0.3\textwidth}
\includegraphics[width=\textwidth]{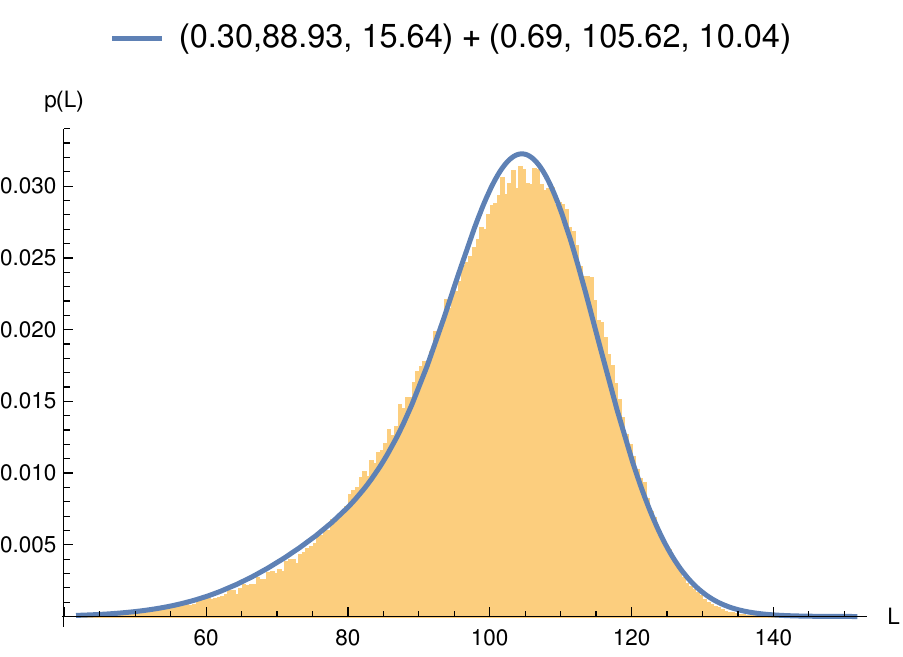}
\end{subfigure}
\begin{subfigure}{0.3\textwidth}
\includegraphics[width=\textwidth]{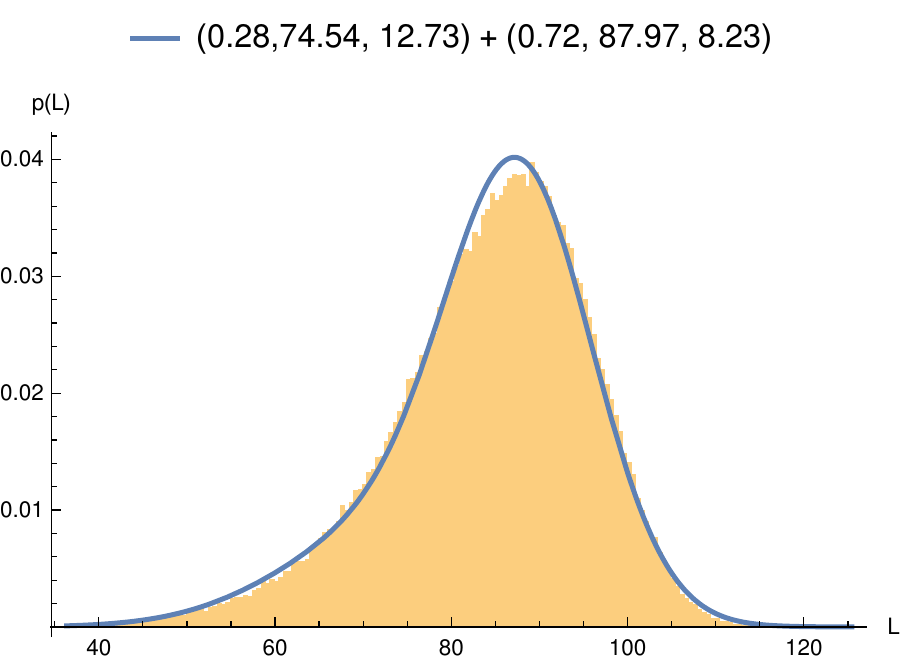}
\end{subfigure}
\begin{subfigure}{0.3\textwidth}
\includegraphics[width=\textwidth]{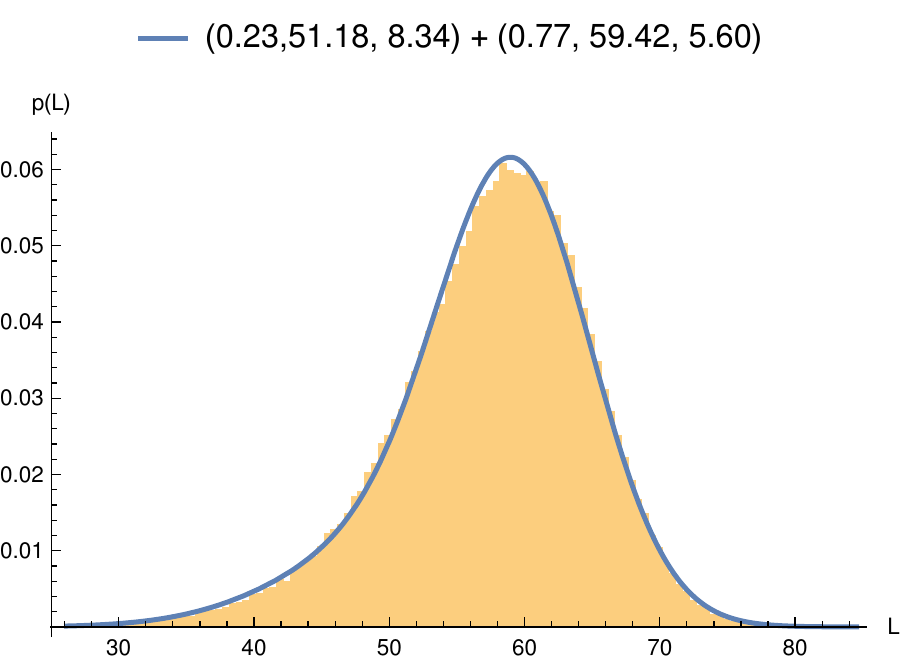}
\end{subfigure}
\caption{\em Histograms for the distribution of amplitudes for $n$= 100,80,50 respectively for the superstring. The legend shows the best fit sum of normal distributions as explained in the text. \label{fighistsusy}}
\end{figure}

\begin{figure}[!h]
\begin{subfigure}{0.3\textwidth}
\includegraphics[width=\textwidth]{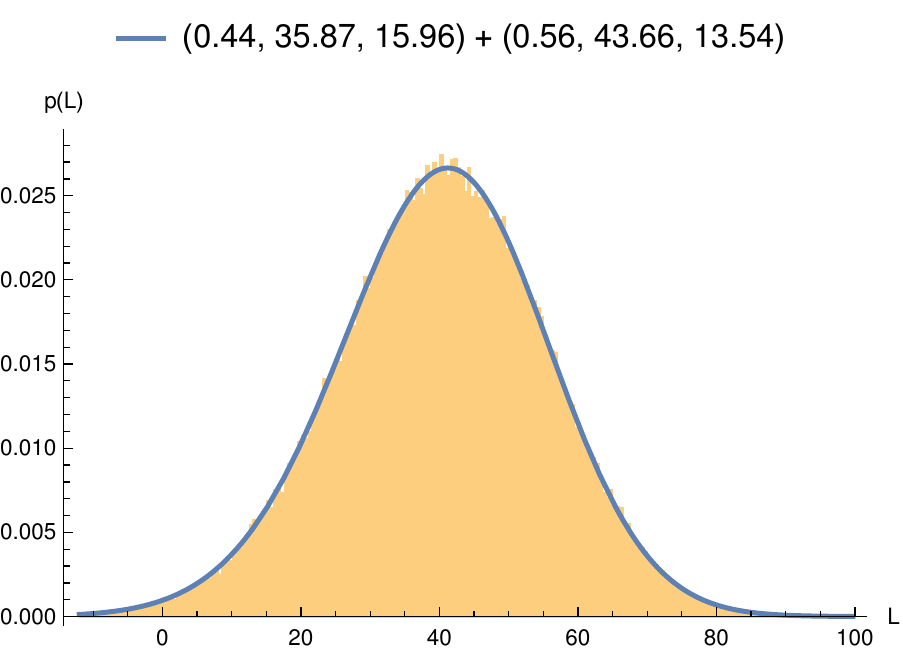}
\end{subfigure}
\begin{subfigure}{0.3\textwidth}
\includegraphics[width=\textwidth]{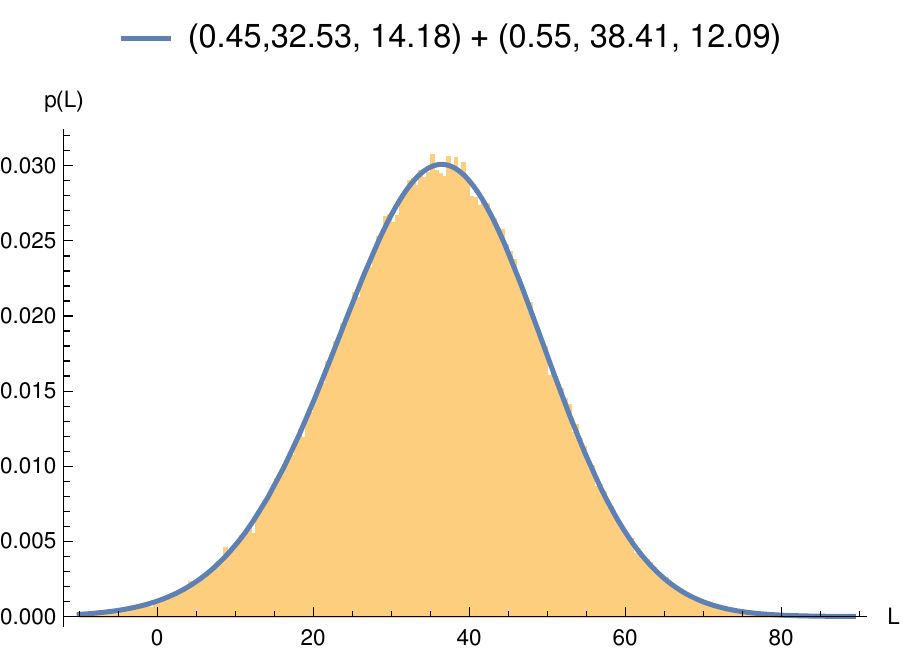}
\end{subfigure}
\begin{subfigure}{0.3\textwidth}
\includegraphics[width=\textwidth]{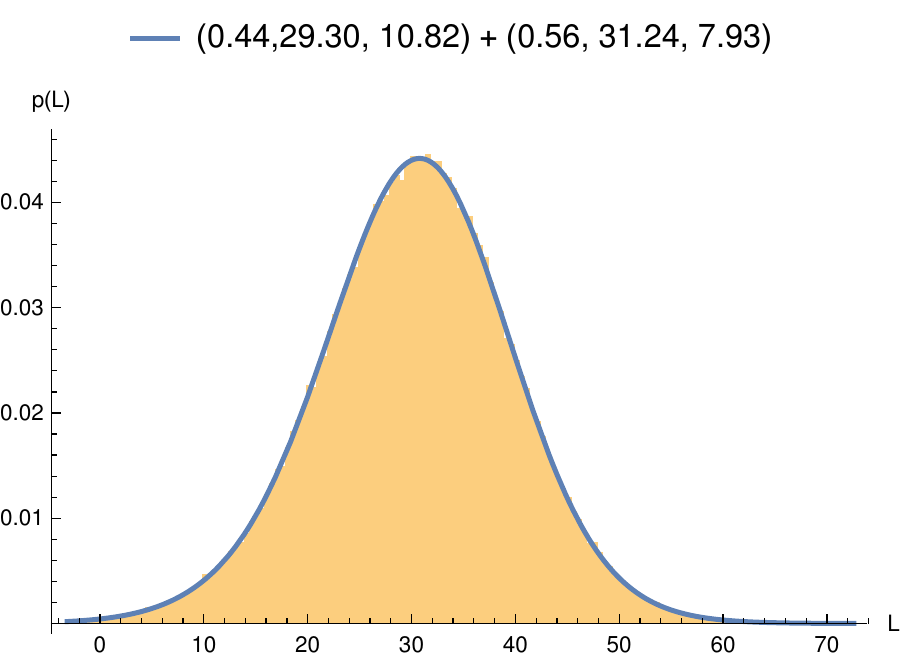}
\end{subfigure}
\caption{\em Histograms for the distribution of amplitudes for $n$= 100,80,50 respectively for the bosonic string. The legend shows the best fit sum of normal distributions as explained in the text. \label{fighistbos}}
\end{figure}

The main possible source of error in our answers arises as follows. We find, empirically, that our numerical samples obey a log-normal distribution. This is shown in Figure \ref{fighistsusy}, where we have plotted a histogram of the values obtained for the superstring for  
\[
L = \log\left(|{\cal G}| |{\cal J}|^{-1} Q_n P_n \bar{P}_n\right) + n \log(d-2).
\]
These values cover all the solutions to the scattering equations that we generated for $n = 50,80, 100$ respectively. As mentioned above, each histogram covers 500,000 distinct solutions of the scattering equations, since we generated $1000$ solutions for $500$ distinct sets of external momenta. It is convenient to combine all these sets into a single set for the discussion here.  Histograms for the same quantity and for the same values of $n$  are shown for the bosonic string in figure \ref{fighistbos} although, in this case, each histogram covers 250,000 samples for each $n$.

In each case, we find that the distribution of the amplitude is well-approximated by a sum of log-normal distributions. More precisely, the values of $L$ obtained in our numerical results are distributed according to the probability distribution
\be
\label{sumlognormal}
p(L) d L = {\alpha_1  \over \sqrt{2 \pi} \sigma_1} e^{-(L - \mu_1)^2 \over 2 \sigma_1^2} + {\alpha_2  \over \sqrt{2 \pi} \sigma_2} e^{-(L - \mu_2)^2 \over 2 \sigma_2^2} .
\ee
In figures \ref{fighistsusy} and \ref{fighistbos} the parameters that appear in the sum of distributions indicated above are displayed using the notation $(\alpha_1, \mu_1, \sigma_1) + (\alpha_2, \mu_2, \sigma_2)$. 

A log-normal distribution (or a sum of such distributions) creates difficulties for numerical sampling.  For example, if the distribution in \eqref{sumlognormal} had been an exact distribution for L then we would have had
\be
\label{meanm}
  (4 \pi g_s)^{2-n}  \langle \tilde{M}_n \rangle  = \int e^{L} p(L) d L = \alpha_1 e^{\mu_1 + {\sigma_1^2 \over 2}} + \alpha_2 e^{\mu_2 + {\sigma_2^2 \over 2}} 
\ee 
However, the probability distribution itself takes on an extremely small value at the point where this mean is attained. In fact the probability that a random variable distributed according to \eqref{sumlognormal} will be {\em above} some given value $\lambda$ is given by
\be
\label{labelmean}
{\cal P}(\lambda) = \int_{\lambda}^{\infty} p(L) d L = 1 - \frac{{\alpha_1}}{2} \text{erfc}\left(\frac{{\mu_1}-\lambda }{\sqrt{2} {\sigma_1}}\right)  - \frac{{\alpha_2}}{2} \text{erfc}\left(\frac{{\mu_2}-\lambda }{\sqrt{2} {\sigma_2}}\right).
\ee
The numerical values of this probability for $\lambda$ given by \eqref{meanm}, with the parameters for the distribution as given in figures \ref{fighistsusy} and \ref{fighistbos}, are given in the Table below. 
\begin{center}
\begin{tabular}{|c|c|c|}
\hline
n & ${\cal P}\big(  (4 \pi g_s)^{2-n} \langle\tilde{M}_n \rangle \big)$ (superstring) & ${\cal P}\big( (4 \pi g_s)^{2-n}\langle \tilde{M}_n \rangle \big)$ (bosonic string) \\ \hline
100 & $1.44 \times 10^{-15}$ & $4.44 \times 10^{-16}$ \\
80 & $5.26 \times 10^{-11}$ & $4.57 \times 10^{-13}$ \\
50 & $1.03 \times 10^{-5}$ & $2.09 \times 10^{-8}$
\\\hline
\end{tabular}
\end{center}
What this table indicates is that to obtain an accurate value for the mean of the amplitude, using random sampling we would need to obtain  about $10^{15}$ samples for the superstring and $10^{16}$ samples for bosonic string.

Clearly, this is not feasible using direct sampling. Therefore, it is clear that a better algorithm is required to identify the points where the amplitude is the largest and sample those accurately rather than simply using random sampling as we have done here. We are not aware of any such algorithm at the moment, and developing such an algorithm remains an important challenge to complete this program of numerically estimating string amplitudes.

\begin{figure}[!h]
\begin{subfigure}{0.5\textwidth}
\includegraphics[width=\textwidth]{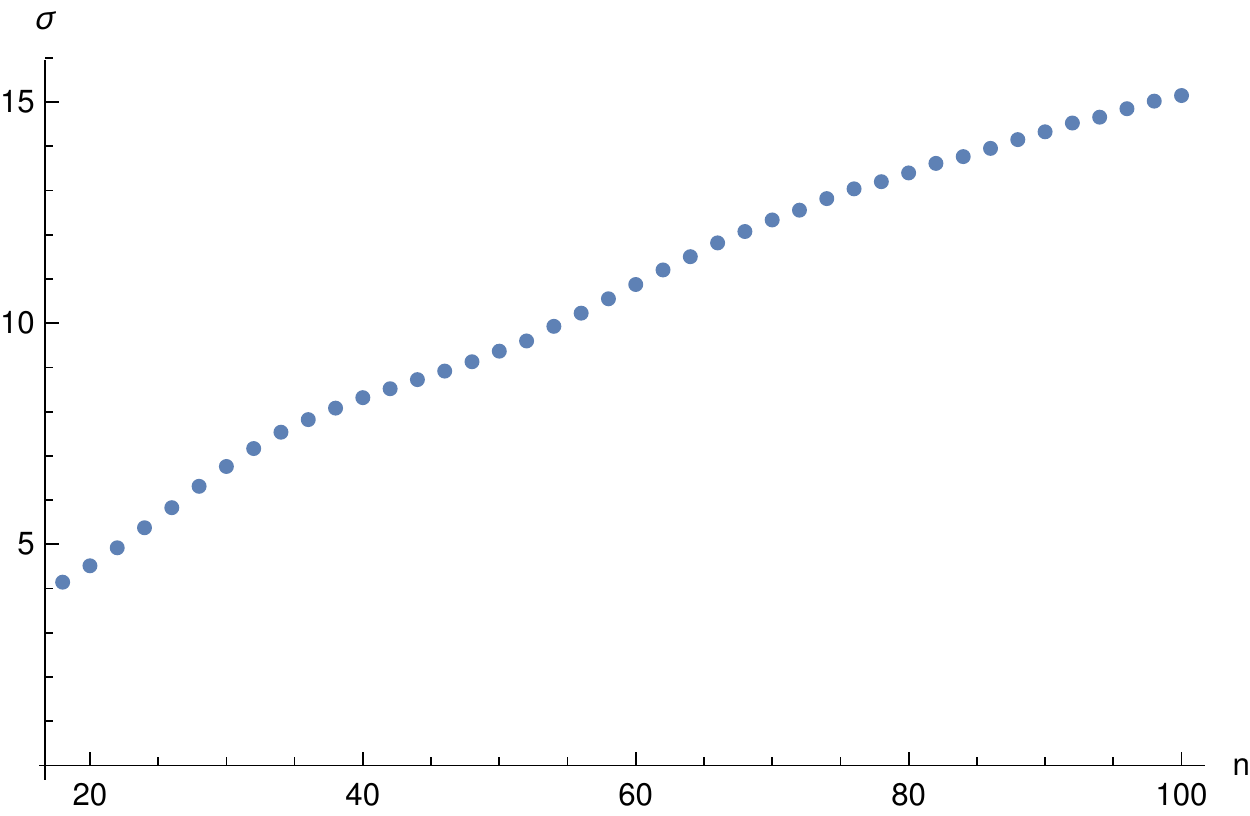}
\end{subfigure}
\begin{subfigure}{0.5\textwidth}
\includegraphics[width=\textwidth]{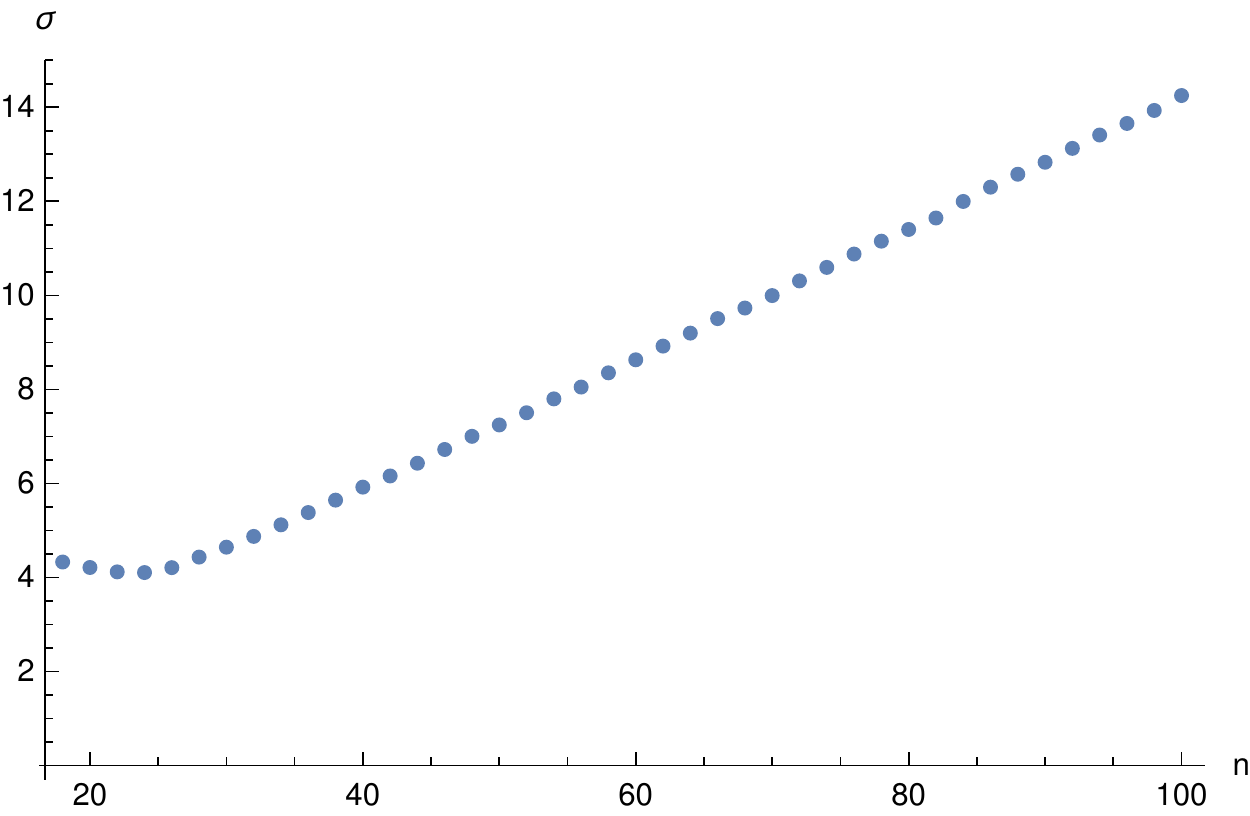}
\end{subfigure}
\caption{\em Standard deviation of the distribution of sampled values of $L$ for the bosonic string (left) and the superstring (right). \label{figstddev} }
\end{figure}

An order of magnitude estimate of the number of samples that would be 
required to accurately sample a log-normal distribution with mean $\mu$ and standard deviation $\sigma$ is given by $e^{{\sigma^2 \over 2}}$. In figure \ref{figstddev}, we provide a plot of the standard deviation of our sampled values of $L$ for the bosonic string and for the superstring respectively. These graphs are provided for the case $\gamma = 0$.

 The fact that this standard deviation does not grow very rapidly, indicates that this sampling error may not affect the leading factorial growth that we have found.

We emphasize that random numerical sampling tends to significantly {\em underestimate} the mean for log-normal distributions and not overestimate them. Therefore, in any case, our numerical results establish that the amplitude grows {\em at least} factorially. However, it is clearly important to bound these sampling errors more precisely and check whether the growth is exactly factorial or whether it is larger.

\subsection{Comments on pure gravity \label{puregravcomments}}
We conclude this section with some comments on scattering amplitudes in pure gravity. It appears to us that, surprisingly, numerically estimating amplitudes in a theory of pure gravity may be {\em harder} than estimating amplitudes in string theory. This is because although scattering amplitudes in gravity can also be localized to solutions of the scattering equations, pure-gravity amplitudes are given by holomorphic functions of the roots. The holomorphicity of the amplitude creates difficulties for numerical sampling, as opposed to the bosonic string and the type II superstring,  where the amplitude can be expressed as a sum of real positive terms at least for specific choices of the external polarizations.

 Note that the pure-gravity amplitude is given by a formula that is very similar to the formula for the superstring \eqref{saddlesum} In particular, we have 
\be
\label{puregravity}
\mtr(k_1 \ldots k_{n}) = (4 \pi g_s)^{n-2}  \sum_{\{z_i\}} {\cal G} {\cal J}^{-1} (P_n^{\text{grav}})^2.
\ee
On the right hand side, the terms match almost precisely with the superstring. We have ${\cal J} = \det(\partial_j E_i)$ which is shown explicitly in \eqref{eijdef}. The ghost term is also as defined in \eqref{ghostdet}. Moreover
\be
P_n^{\text{grav}} = {1 \over z_{12}} \text{Pf}\left(M_{12} \right),
\ee
where the matrix $M$ is defined in \eqref{defm}. We remind the reader that $M_{12}$ means that we remove the first two rows and columns. We note this is precisely the first term that appears in  $P_n$ for the superstring in \eqref{susyexprpq}. 

Of course, $P_n^{\text{grav}} \neq P_n$ and also the Koba-Nielsen factor $Q_n$ that appears in \eqref{saddlesum} does not appear in gravity. But an even more important difference between the superstring answer and the gravitational answer is that the superstring answer involves absolute values whereas the expression \eqref{puregravity} does not have absolute values, either on ${\cal G}$ or on ${\cal J}$. Moreover, we simply have the square of the term $P_n^{\text{grav}}$ in the pure-gravity answer and not its absolute-value squared as we have in the superstring. This is the reason that the pure-gravity answer must be obtained as a limit of the ``ambitwistor string'' rather than the usual superstring \cite{Mason:2013sva}.
 
The difficulty with parsing the holomorphic answer numerically is that  $(n-3)!$ contributions that come from the distinct solutions to the scattering equations
now involve several numerical cancellations since they are not simply the sum
over a set of positive terms. This makes the sum difficult to evaluate numerically.

For this reason, we have not carried out our numerical calculations for pure gravity. Nevertheless, we do expect that even in pure gravity the answer for the amplitude will scale like $n!$. In fact, this leading factorial growth may also be seen directly from the BCFW recursion relations \cite{Britto:2004ap,Britto:2005fq}.

The BCFW recursion relations state that that the $n$-point tree-level amplitude can be decomposed as a sum over products of lower point tree-level amplitudes. Schematically, we have 
\[
\mtr_n = \sum_{p=1}^{n-3} {\mtr_{p+2} \mtr_{n-p}  \over P_L^2}
\]
This recursion relation arises as follows. We mark two particular legs
for performing the BCFW extension. Of the remaining $n-2$ legs, we must
select $p$ legs on one side, and $n-p-2$ legs on the other. On one side, we get a $p+2$-point scattering amplitude after adding one of the original
marked legs, and an intermediate leg. On the other side, we similarly get
a $n-p$ point amplitude. This recursion relation is valid for $n \geq 4$.

To estimate the growth of the amplitude, we just want to calculate the number of terms in the BCFW recursion relations. Denoting this number by $N_n$, we see that it satisfies the recursion relations
\be
N_n = \sum_{p=1}^{n-3} \binom{n-2}{p} N_{p+2} N_{n-p}.
\ee
We define $N_3 = 1$, and we have $N_n = 0$ for $n \leq 3$.

Now, we consider the following generating function
\be
W(g) = \sum_{n=4}^{\infty} {N_n g^{n-2} \over (n-2)!}.
\ee
With the conventions above, but recalling that the recursion relation is valid only for $n \geq 4$, we then find that
\be
W(g)  = \sum_{n=4}^{\infty} \sum_{p=1}^{n-3} {N_{p+2} g^p \over p!}  {N_{n-p} g^{n-p-2} \over (n-p-2)!} .
\ee
We write $\tilde{p} = p+2$, and $\tilde{n} = n-p$, so that the sum above can be written as
\be
W(g)  = \sum_{\tilde{p}=3}^{\infty} \sum_{\tilde{n}=3}^{\infty} {N_{\tilde{p}} g^{\tilde{p}-2} \over (\tilde{p}-2)!}{N_{\tilde{n}} g^{\tilde{n}-2} \over (\tilde{n}-2)!} = (W(g) + g)^2.
\ee
We can now solve for 
\be
W(g)= {1 - 2g - \sqrt{1 - 4 g} \over 2}.
\ee
This provides an analytic formula for $N_n$:
\be
N_n = (-1)^{n-1} {1 \over 2} 4^{n-2} \binom{{1 \over 2}}{n-2} (n-2)!
\ee
Thus we see that $N_n$ also grows factorially with $n$.

Of course, even in the BCFW analysis there are cancellations and therefore this estimate cannot be used for direct numerical calculations either. However, this estimate of the number of terms in the BCFW relations reinforces the indications that pure gravity does not behave differently from string theory in any qualitative manner as far as the leading factorial growth of the amplitude is concerned.

However, note that in string theory,  it is important to have $\gamma > 0$ to ensure that the rapid falloff of Koba-Nielsen factor at high energies does not overwhelm the factorial growth. In pure gravity, there would be no such constraint. 

\section{Applications to the Information Paradox \label{secinfo}}
We now describe how the analysis above can be applied to the information paradox. First, we review the two forms of the paradox that we will discuss here --- the cloning paradox and the strong subadditivity paradox. We explain how an important --- and often unstated --- assumption that goes into formulating these paradoxes is that bulk locality holds even for very high point correlation functions, including $\Or[S]$-point correlators where the typical separation between points is the inverse of the Hawking temperature. We then describe how the loss of locality suggested by the breakdown of perturbation theory discussed above is precisely sufficient to resolve these paradoxes. 

At the end of this section, we return to the toy model of section \ref{sectoycomp} and describe how it can be used to set up some toy-models of the information paradox.  In this toy-model setting, the resolution to these paradoxes is also very clear. This toy-model analysis supports the idea that these paradoxes can be resolved by  recognizing that very complicated correlation functions in gravity may display the effects of a loss of bulk locality.

\subsection{The cloning and strong subadditivity paradoxes}

We now review the cloning and strong subadditivity paradoxes that, as we will see, are closely related. We will consider a Schwarzschild black hole in $d$ dimensions with a metric given by
\be
\label{schwarzschildd}
ds^2 = -(1 - {\mu \over r^{d - 3}}) d t^2 + (1 - {\mu \over r^{d-3}})^{-1} d r^2 + r^2 d \Omega_{d-2}^2 ,
\ee
where $\mu = {16 \pi G M  \over (d - 2) \Omega_{d-2}}$ and $\Omega_{d-2} = {2 \pi^{d - 1 \over 2} \over \Gamma\left({d - 1 \over 2}\right)}$ is the area of a unit $(d-2)$-sphere. So, the horizon radius is given by $r_h = \left({16 \pi G M \over (d - 2) \Omega_{d-2}}\right)^{1 \over d- 3}$. 

Strictly speaking, the metric \eqref{schwarzschildd} represents an eternal black hole, but it is correct at late times for black holes formed from collapse. Second the metric \eqref{schwarzschildd} does not
account for the effects of the back-reaction of Hawking radiation on the geometry. Nevertheless, it gives an excellent approximation to the geometry provided we consider time-scales that are short with respect to the evaporation time of the black hole. Whenever we refer to $M, r_h$ below, without qualification, we are referring to the mass and radius of the black hole upon formation.

\paragraph{\bf The Cloning Paradox \\}

The cloning paradox proceeds as follows. Consider a black-hole formed from collapse that gradually starts to evaporate. Now, a general argument due to Page \cite{Page:1993df} tells us that as energy is transferred from the black hole to the Hawking radiation, the von Neumann entropy of the Hawking radiation first increases and then starts to decrease. 

We can make this precise by defining a length scale $\delta \gg \lpl$ but $\delta \ll r_h$, and using this length scale to define a region outside the black hole, and well separated from the horizon:
\be
\text{region~A}: r_h + \delta < r < \infty.
\ee
For convenience below we will use $r_a = r_h + \delta$. Now to frame the result of Page in this notation we assume that on any spacelike slice drawn through the black hole spacetime, as in other local quantum field theories, the full set of operators in the theory factorizes as
\be
\label{hafact}
\alset = \bar{\alset}_A\otimes \alset_A, 
\ee
which emphasizes the tensor product decomposition of the full set of observables, $\alset$, into a factor corresponding to the set of operators localized in $A$, $\alset_A$  and a complementary factor,  $\bar{\alset}_A$. The assumption \eqref{hafact} is the same as assuming that the Hilbert space of the theory factorizes into a factor corresponding to the degrees of freedom on $A$ and a complementary factor.

With this assumption, we can define a density matrix for local regions. The result of Page concerns the von Neumann entropy of the region $A$
\be
S_{A} = -\tr(\rho_{A} \log(\rho_A)).
\ee
We are interested in the excess entropy over the entropy of the vacuum, so we define
\be
\label{saarea}
S_{A} = S_{A}^{\text{hawk}} + {\text{Area}(r_a) \over 4 G}, 
\ee
where the second term on the right hand side is given by $\text{Area}(r_a) = \Omega_{d-2} (r_a)^{d-2}$ and we are {\em assuming} that this is the vacuum von Neumann entropy of region $A$.

\begin{figure}[!h]
\begin{center}
\includegraphics[width=0.5\textwidth]{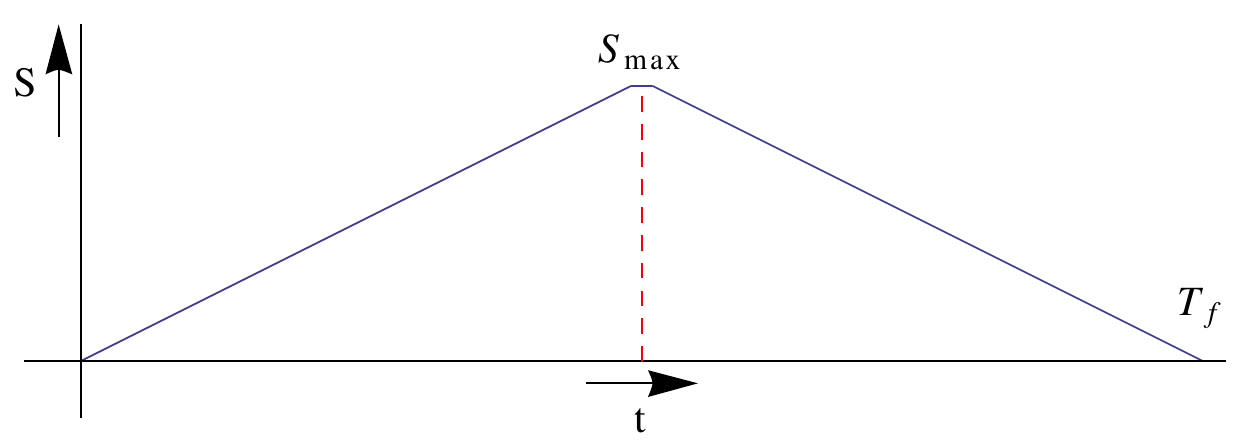}
\caption{\em Variation of the von Neumann entropy of the Hawking radiation with time. The ``Page time'' is shown by the dashed red line. \label{figpage}}
\end{center}
\end{figure}

Page then argued that, if we consider a black hole formed from the collapse of a pure state that then proceeds to evaporate completely then $S_{A}^{\text{hawk}}$ should start from zero, increase monotonically, turn around after a time called the ``Page time'' and then decrease
monotonically down to zero. This is shown schematically in figure \ref{figpage}.

The fact that the entropy starts to decrease after the Page time implies that the Hawking radiation outside starts to become pure. This can also be understood as the transfer of information from the interior of the black hole to the exterior. 

\begin{figure}[!h]
\begin{center}
\includegraphics[height=0.3\textheight]{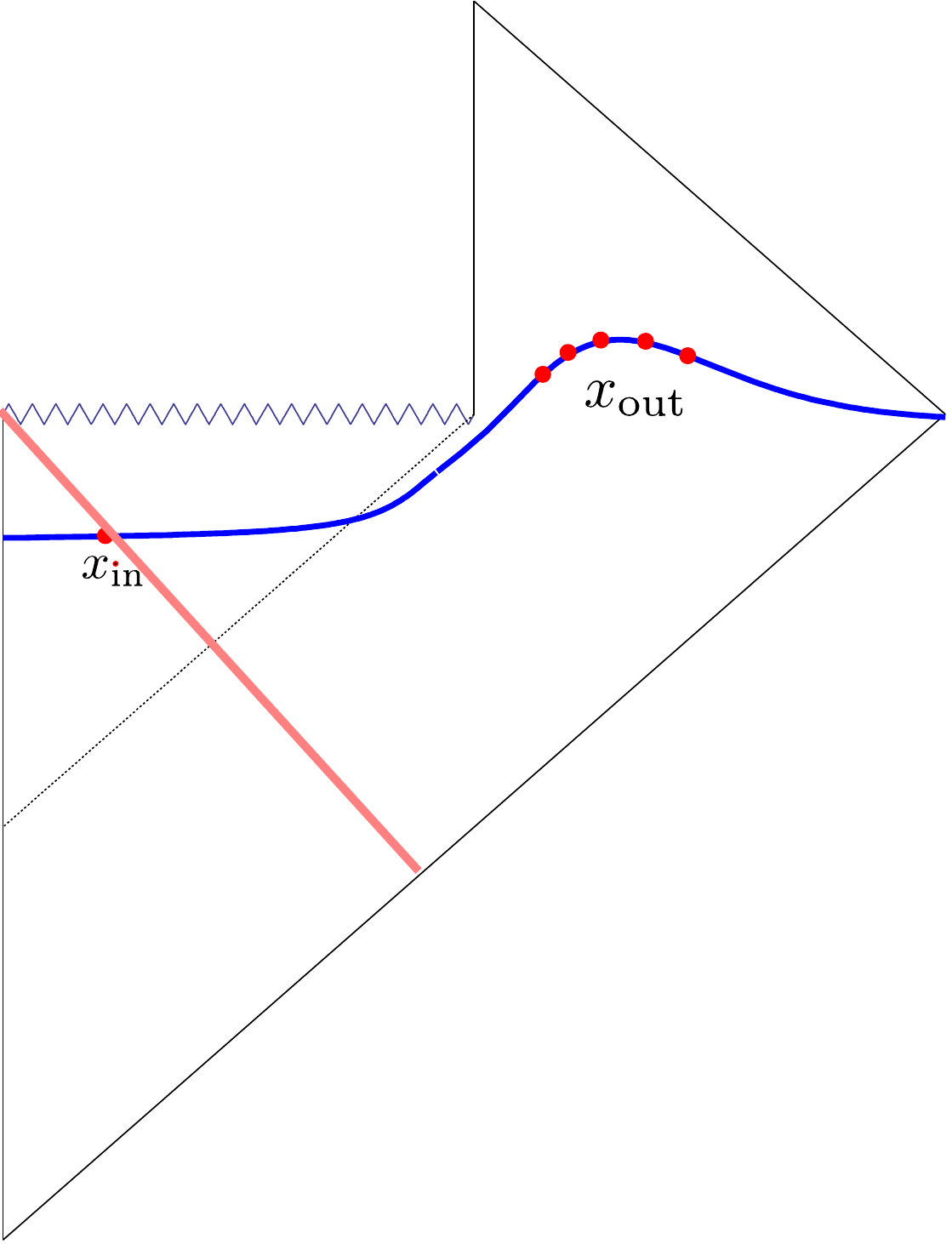}
\caption{\em A nice slice that captures the infalling matter and the late Hawking radiation. The cloning paradox is that a measurement at $x_{\text{in}}$ seems to contain the same information as a measurement at the points $x_{\text{out}}$.\label{fignicecloning}}
\end{center}
\end{figure}

To obtain the cloning paradox, we consider a black hole formed from collapse. Then, in the extended Penrose diagram,  it is possible to draw a nice slice that cuts through the infalling matter {\em and} captures a significant fraction of the late Hawking radiation. This slice is shown in figure \ref{fignicecloning}.

Now, by the argument above the section of the slice outside the black hole in the region $A$ contains a significant fraction of the information that was present in the infalling matter. However, we see that the infalling matter is itself present on another section of the slice. This seems to suggest that information has been duplicated or ``cloned'', in violation of the linearity of quantum mechanics. This is the ``cloning paradox'' \cite{Susskind:1993mu}. A closely related paradox was outlined by Hayden and Preskill \cite{Hayden:2007cs}. They considered a situation where after the black hole has been formed one throws in some {\em additional information}. They then argued that this additional information should return to the exterior in a ``scrambling time.'' This also leads to a cloning paradox since we can again draw a slice that captures the information thrown in, and the exterior Hawking radiation that appears to carry the same information. 

\paragraph{\bf The Strong Subadditivity Paradox \\}
A closely related paradox is the strong-subadditivity paradox. To set up this paradox, we consider an old black hole, {\em after} the Page time, but well before the end of evaporation.  At this point the black hole has shrunk so that its horizon radius is $r_h^{\text{old}}$ but we still have $r_h^{\text{old}} \gg \lpl$. We now define three regions $A,B,C$ as follows. 
\be
\label{abcdef}
\text{region~C}: r^{\text{old}}_h - \delta  < r < r^{\text{old}}_h; \quad
\text{region~B}: r^{\text{old}}_h < r < r^{\text{old}}_h + \delta; \quad
\text{region~A}: r^{\text{old}}_h + \delta < r < \infty
\ee
Note that in this discussion the region $A$ is now defined with respect to $r_h^{\text{old}}$ and not $r_h$. 
We also use $r^{\text{old}}_a \equiv r^{\text{old}}_h + \delta$ and use $r^{\text{old}}_c \equiv r^{\text{old}}_h - \delta$.

\begin{figure}
\begin{center}
\includegraphics[height=0.3\textheight]{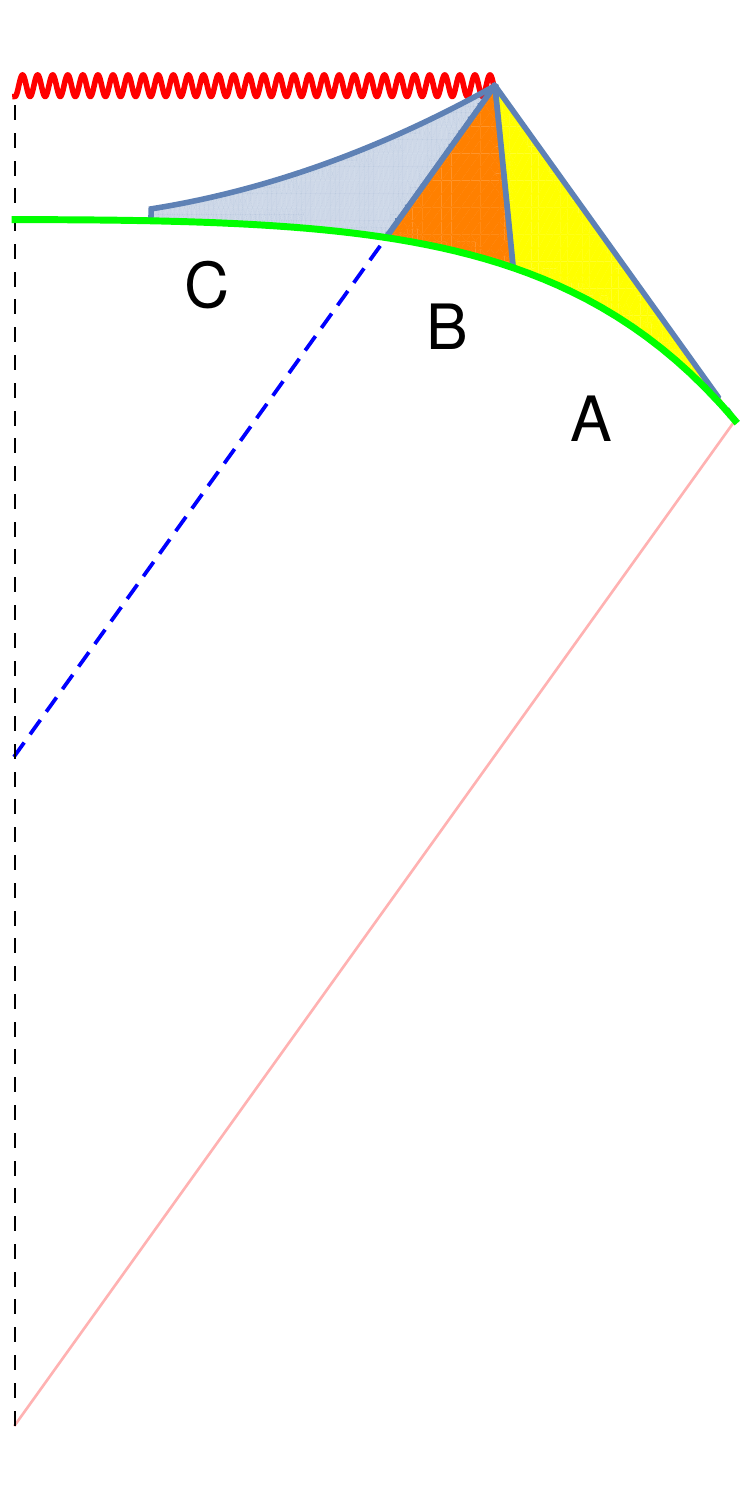}
\caption{\em A representation of the late-time parts of regions $A$ (yellow), $B$ (orange) and $C$ (blue) and their intersection with a spacelike slice (green).  \label{figabc}}
\end{center}
\end{figure}

 These regions are shown in Figure \ref{figabc}, where we also show a spacelike slice that cuts through all of them. Only the late-time parts of these regions --- the parts above the spacelike slice --- are shaded, since we are only interested in these regions at late times.

Now, to frame the strong subadditivity paradox we assume, analogously to \eqref{hafact} that  the set of operators in  the theory factorizes as
\be
\label{hfactorization}
\alset = \bar{\alset} \otimes \alset_C \otimes \alset_B \otimes \alset_A,
\ee
where $\bar{\alset}$ is another residual factor and the main point of \eqref{hfactorization} is to emphasize that the set of local operators corresponding to  the regions $A,B,C$ appear as tensor factors in the algebra of the full theory.

Subject to the assumption \eqref{hfactorization}, we may then consider the entropy of the regions $A,B,C$. This entropy can be written as a sum of the entropy of the Hawking radiation in this region, and the vacuum entanglement entropy. So we have
\be
\label{scarea}
\begin{split}
&S_{A} = S_{A}^{\text{hawk}} + {\text{Area}(r^{\text{old}}_a) \over 4 G}, \quad S_{B} = S_{B}^{\text{hawk}} + {\text{Area}(r^{\text{old}}_a) \over 4 G} +  {\text{Area}(r^{\text{old}}_h) \over 4 G}; \\
&S_{C} = S_{C}^{\text{hawk}} + {\text{Area}(r^{\text{old}}_c) \over 4 G} +  {\text{Area}(r^{\text{old}}_h) \over 4 G},
\end{split}
\ee
We can also consider the entropies of pairs of regions, which gives us
\be
\label{sbcarea}
S_{BC} = S_{BC}^{\text{hawk}} + {\text{Area}(r^{\text{old}}_a) \over 4 G}  +  {\text{Area}(r^{\text{old}}_c) \over 4 G}; \quad S_{AB} = S_{AB}^{\text{hawk}} + {\text{Area}(r^{\text{old}}_h) \over 4 G}.  
\ee

Now the process of Hawking radiation involves pair-creation in the region $BC$. The Hawking particles that exit into region $B$ are entangled with their pairs in region $C$ that fall into the black hole. On the other hand, if one views the Hawking radiation only in $B$ or only in $C$, then this radiation appears almost thermal. This argument leads to the inequality  
\be
\label{hawkingpairs}
S_{BC}^{\text{hawk}} < S_{C}^{\text{hawk}}.
\ee
Now we note that the Hawking radiation in $B$ will emerge into region $A$. If we are in the phase of Hawking evaporation after the Page time, then the radiation in $B$ purifies the radiation in $A$. This leads to the inequality
\be
\label{pagepurify}
S_{AB}^{\text{hawk}} < S_{A}^{\text{hawk}}.
\ee

However, a general theorem concerning von Neumann entropies --- called the strong subadditivity of von Neumann entropy \cite{Lieb:1973cp,lieb1973prl}  --- states that
\be
\label{strongsubadditivity}
S_{AB} + S_{BC} \geq S_{A} + S_{C}.
\ee
When we add the area terms in \eqref{sbcarea} and \eqref{scarea} the strong subadditivity inequality becomes
\be
\begin{split}
S^{\text{hawk}}_{AB} + {\text{Area}(r^{\text{old}}_h) \over 4 G} +  S^{\text{hawk}}_{BC} +  {\text{Area}(r^{\text{old}}_a) \over 4 G}  + {\text{Area}(r^{\text{old}}_c) \over 4 G} \geq &S^{\text{hawk}}_{A} + {\text{Area}(r_a^{\text{old}}) \over 4 G} + S^{\text{hawk}}_{C} \\ &+  {\text{Area}(r^{\text{old}}_c) \over 4 G} +  {\text{Area}(r^{\text{old}}_h) \over 4 G}.
\end{split}
\ee
The area terms neatly cancel in this inequality leaving us just with
\be
S^{\text{hawk}}_{AB} + S^{\text{hawk}}_{BC} \geq S^{\text{hawk}}_{A} + S^{\text{hawk}}_{C}.
\ee
However, we see that the inequality above is in contradiction with \eqref{pagepurify} and \eqref{hawkingpairs}. This is the ``strong subadditivity paradox'' \cite{Mathur:2009hf,Almheiri:2012rt}.

We note that both the cloning and the strong subadditivity paradox involve assumptions of bulk locality. In the cloning paradox, we assume that operators defined on the nice slice, at the point where the nice slice intersects the infalling matter, commute with the operators that measure the same information in the Hawking radiation. In the strong subadditivity paradox, we assume that the Hilbert space factorizes as \eqref{hfactorization}. We now analyze the protocol for determining the von Neumann entropy and extracting information from Hawking radiation more closely. We will find that this process necessarily requires the measurement of very high-point correlators, which invalidates the assumption of locality made in framing both paradoxes.

\subsection{Protocols for extracting information \label{secprotocol}}
The key to resolving the paradoxes above lies in setting up a precise protocol to extract information from the emitted Hawking radiation. We can consider a family of  observers who stay fixed at a very large value of $r$ and measure the outgoing Hawking radiation as it crosses them at large values of $t$. Such observers may also have injected information into the system in the far past, as in the Hayden-Preskill scenario mentioned above. For simplicity, we will restrict ourselves to s-wave measurements here, since most of the energy emitted in Hawking radiation from a Schwarzschild black hole emerges in s-waves. We will also be interested in a limit where the mass of the black hole becomes very large compared to the Planck mass so that $M \gg M_{\text{pl}}$ and the entropy also becomes very large, $S \gg 1$.

In this subsection we would like to argue that this family of observers needs to measure $S$-point correlators of the emitted Hawking quanta to extract information that is relevant for the cloning and the strong subadditivity paradoxes. However, our arguments in this subsection will not be entirely precise and we will be forced to rely on some discretizations and plausible assumptions. Improving this analysis of protocols for extracting information from the Hawking radiation remains an important objective for future work.

The correlation functions measured by these observers are 
\be
C(u_1, \ldots u_s, v_{s+1} \ldots v_n) = \text{lim} \int \langle T\left\{\phi(t_1, \vect{x_1}) \ldots \phi(t_s, \vect{x}_{s}), \phi(t_{s+1}, \vect{x}_{s+1}), \ldots \phi(t_n, \vect{x}_{n}) \right\} \rangle  \prod d\Omega_p, 
\ee
where the limit is taken so that
\be
\begin{split}
&r_m = |\vec{x}_m| \rightarrow \infty, t_1 \ldots t_s \rightarrow \infty, t_{s+1} \ldots t_n \rightarrow -\infty \\ 
&u_m = t_m - r_m; ~ v_m = t_{m} + r_{m},
\end{split}
\ee
where we keep $u_1 \ldots u_s$ finite and $v_{s+1} \ldots v_n$ finite. 
The integrals are over the sphere at infinity to extract s-wave information. We can easily generalize this to consider higher spherical harmonics.

We should also emphasize that we are only interested in the {\em connected} part of the position space correlator. This is because the part of a high-point correlator that factorizes into a product
of lower-point correlators does not contain any fresh information. In momentum space, it is easy to exclude the disconnected part of the scattering amplitude by simply focusing on external kinematics where momentum is not conserved by any subset of the particles. However, in position space, the restriction to the connected part of the correlator must be imposed by hand by subtracting off the disconnected parts. We will assume that this has been done in the analysis below.

Although, from a physical perspective,  these correlators are more natural, they contain {\em precisely} the same information as S-matrix elements. In fact a simple application of the LSZ formula tells us that they are Fourier transforms of $S$-matrix elements.
\be
\label{corrtosmat}
C(u_i, v_j) = {N^n e^{i \pi s d \over 2} \over r^{n {d - 2 \over 2}}}  \int M (k_1, \ldots k_n) (2 \pi)^d \delta^d(\sum k_p) e^{-i \sum E_i u_i + i \sum E_j v_j}  \prod d \hat{k}_p d E_p  (E_p)^{d/2 - 2},
\ee
where the on-shell $d$-momenta $k_i = (-E_i, E_i \hat{k}_i)$ for $i = 1 \ldots s$ and $k_j = (E_j, E_j \hat{k}_j)$ for $j = s+1 \ldots n$ and $p = 1 \ldots n$. Here, the normalization factor, $N$, and the phase factor $e{i \pi s d \over 2}$ are both irrelevant.  The leading suppression by a power of ${1 \over r}$ indicates the asymptotic falloff of the correlator, and will also not be relevant. This formula is proved in Appendix \ref{appcorrsmat}. In the case under consideration we are only interested in correlators measured at {\em future null infinity}, and so we will suppress the $v$ coordinates now.

The task of setting up a protocol to extract information is to determine which set of correlation functions we must measure at future null infinity to pin down the density matrix of the emitted radiation. We will proceed in a simple minded fashion. First, we choose the origin of coordinates so that the Hawking radiation starts to emerge at $u = 0$. Then, we allow the observer at infinity to measure correlators at $u = 0, {\alpha \over T}, {2 \alpha \over T} \ldots {n_{\text{max}} \alpha \over T}$, where $T$ is the temperature of the Hawking radiation and $\alpha$ is an order $1$ number that controls how finely we can make measurements. We can choose $\alpha \ll 1$, but we do not scale $\alpha$ with the entropy of the black hole so that ${\log(\alpha) \over \log(S)} \ll 1$.  We assume that the radiation continues till ${n_{\text{max}} \alpha \over T}$, and we calculate $n_{\text{max}}$ below.  Here, we are assuming that the measurement of correlators at finer separation is not relevant for extracting information from Hawking radiation which has a characteristic energy $T$. However, it would be nice to make this protocol more precise.

In terms of the horizon radius, the Hawking temperature is given by
\be
\label{tempd}
T =  {d - 3 \over 4 \pi r_h}.
\ee
The entropy of the black hole is given by
\be
\label{entropyd}
S = {\Omega_{d - 2} r_h^{d-2} \over 4 G}.
\ee
Note that the total time required for the black hole to evaporate entirely is given by 
\be
t_{\text{evap}} = {K S \over T},
\ee
where $K =  {1 \over \sigma \Omega_{d-2}} \left({4 \pi \over d - 3} \right)^{d-2} $, and $\sigma$ is the Stefan-Boltzmann constant. Therefore  the largest value of $u$ that is relevant in this discussion is ${K S \over T}$ and we have $n_{\text{max}} = {K S \over \alpha}$.

Now the task of extracting information from the outgoing Hawking radiation boils down to the task of determining its density matrix, $\rho_A$. However,  we note that each correlation function measured by our family of observers gives them only partial information about the density matrix $\rho_{A}$. When appropriately discretized, we expect this density matrix to have dimension $e^{S} \times e^{S}$, where $S$ is the entropy of the black hole. If we are given a set of ${\cal D}$ observables corresponding to the region $A$: ${\alset}_A = \{\al_1 \ldots \al_{\cal D}\}$, then these observables give us information about matrix elements of this density matrix
\be
\langle \al_i \rangle = \tr(\rho_{A} \al_i).
\ee
To extract the full density matrix, we require ${\cal D} = e^{2 S}$ distinct observables. 

In fact, as explained in \cite{Papadodimas:2012aq,Papadodimas:2013kwa}, if we have only a small number of observations, so that ${\cal D} \ll e^{2 S}$,  then our information about properties of the density matrix $\rho_{A}$ is very limited. For example, a thermal density matrix and a pure state density matrix may appear almost identical when the number of observations is small. Therefore, we often require accurate information about all components of the density matrix even if, in the end, we are only interested in a single number like its von Neumann entropy. 

Now, here, the set of observables comprises  correlation functions at null infinity measured at discrete points as outlined above. If we measure up to $p$-point correlators, then the number of distinct
observables is given by
\be
{\cal D} = \sum_{j=2}^p \begin{pmatrix}n_{\text{max}} \\  j \end{pmatrix}.
\ee
At large $S$, setting ${\cal D} = e^{2 S}$, we see that we must consider at least $p = \eta S$ point correlators, where $\eta \log(\eta) + (1 - \eta) \log(1 - \eta) = {-{2 \alpha \over K}}$. The precise value of $\eta$ depends on our choice of allowed spacing in our measurements, which is controlled by $\alpha$.  However, what is important for us is simply that $p = \Or[S]$. In words, this leads to the following conclusion. {\em To extract information from the outgoing Hawking radiation, we need to measure $\Or[S]$-point correlation functions of the outgoing Hawking quanta.} Furthermore by \eqref{corrtosmat} these correlators are sensitive to S-matrix elements that involve $S$ particles with typical energies $E_i = \Or[T]$.\footnote{Note that the exponent in the Fourier transform that appears in \eqref{corrtosmat} appears to become large when we take energies to be of order the Hawking temperature because the $u$ values can become large in units of ${1 \over T}$. However, this is just an overall large phase and does {\em not} imply that these energies are unimportant in the Fourier transform. We need energies of size ${1 \over T}$ to distinguish between a correlator with an insertion at ${i \alpha \over T}$: $C(u_1, \ldots {i \alpha \over T} \ldots u_s, v_j)$ and another correlator with an insertion at ${(i+1) \alpha \over T}$ and all other insertion points unchanged: $C(u_1, \ldots {(i+1) \alpha \over T} \ldots u_s, v_j)$.}

This conclusion is also relevant for the von Neumann entropies that appear in \eqref{pagepurify}. The von Neumann entropy, $S_A$ depends on the density matrix $\rho_A$. If we use an approximation to this density matrix, which corresponds to  using a limited set of observables, then we do not expect \eqref{pagepurify} to hold. This is because we expect that the entanglement between the Hawking radiation in $B$ and the old Hawking radiation in $A$ can only be measured by considering very complicated operators in $A$. If such complicated operators are excluded from the set of allowed observables in $A$, then we may not find that $S_{A B} < S_{A}$. Therefore, the arguments in this subsection tell us that in order for \eqref{pagepurify} to be hold, we must use the density matrix $\rho_A$ obtained by  measuring a set of observables, $\alset_A$, that contains $S$-point correlators of simple local field operators.

\subsection{Resolving the cloning and strong subadditivity paradoxes}
Once we have the conclusion above in hand, the resolution of both the cloning and the strong subadditivity paradoxes follows naturally. We have already explained that these paradoxes rely on an assumption
of bulk locality. However, the measurement protocol that we have described above relies on measuring an $\Or[S]$ point correlation function with separations of ${\alpha \over T}$. Using the conversion \eqref{corrtosmat}, this simply corresponds to an $S$-matrix element with $\Or[S]$ insertions with typical energy $T$. 

We now see from \eqref{tempd} and \eqref{entropyd} (where we recall that  $8 \pi G = \lpl^{d-2}$)  that as we scale the mass of the black hole, $M \gg M_{\text{pl}}$, so that $S \rightarrow \infty$, the energy and number of insertions precisely satisfy
\be
{\log(S) \over (2 - d) \log(T \lpl)} \rightarrow 1.
\ee
Comparing this with \eqref{thresholdlpl} we see that  string perturbation theory breaks down {\em precisely} for $\Or[S]$ measurements, with energy $\Or[T]$.  By the arguments of section \ref{secpertloc}, we see that this is also the limit where we might expect a loss of locality.

In particular, this implies that that the assumption that goes into the strong subadditivity paradox \eqref{hfactorization} is not justified for such measurements. It also tells us that, in this limit, the assumption of locality required for the cloning paradox fails: measurements of $\Or[S]$-point correlators of the Hawking radiation outside the black hole may not commute with measurements of the infalling matter inside the black hole, even though both of these measurements are made on the same spacelike slice.  

Thus we see that the breakdown of perturbation theory that we have described above appears precisely in time to invalidate the assumption of locality that went into the strong subadditivity and cloning paradoxes.

\subsection{Toy models of the Information Paradox \label{sectoyinfo}}
It is amusing that we can use our toy model of black hole complementarity to 
also construct toy models of the information paradox, which are resolved in precisely the manner described above. 
\begin{figure}[!h]
\begin{center}
\includegraphics[height=0.3\textheight]{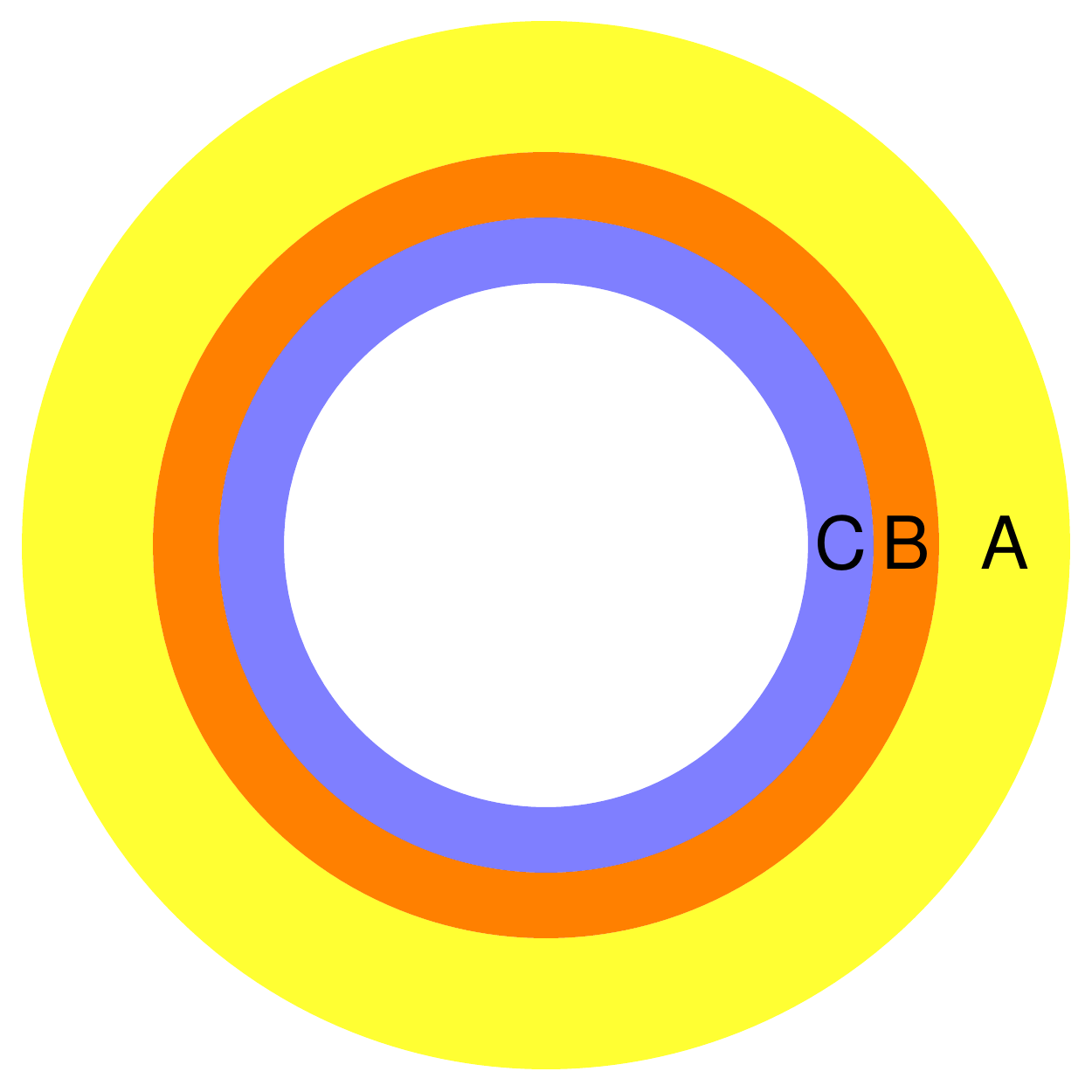}
\caption{\em A schematic description of the intersection of the regions A, B, C with a constant global time slice in empty global AdS \label{emptyadsabc}}
\end{center}
\end{figure}

Consider a constant-time slice of AdS that we can divide into three-regions as shown. More precisely we define
\be
\text{region~C}: r_c < r < r_0; \quad
\text{region~B}: r_0 < r < r_a; \quad
\text{region~A}: r_a < r < \infty.
\ee
The difference with the black hole case is that now the boundaries of all three regions are timelike. So in figure \ref{emptyadsabc} we show a constant global-time cross-section of empty AdS that displays these regions.

Now in section \ref{sectoycomp} we showed that local operators in $C$ could be written entirely in terms of a very complicated set of operators in $A$. This immediately leads to a version of the {\em cloning paradox}. Consider a qubit in region $C$ that can be manipulated by a local operator $\phi_C$. Since this local operator can also be written in terms of operators that are in A, we find that this qubit can also be manipulated by some operator $\phi_A$. We thus appear to have a cloning paradox, where the {\em same information} is present both in $C$ and in $A$.

In this setup, the resolution to this paradox is trivial. We can make a distinction between operators in $C$ and operators in $A$ only when locality holds approximately. Locality holds provided that we restrict ourselves to ``small algebras'', where we do not consider very complicated polynomials in the field operators. If we start considering such polynomials then, as shown in section \ref{sectoycomp}, a local operator in $C$ can be equated to a complicated polynomial of operators in $A$. Therefore, there is no cloning of information. Rather what we are seeing is a loss of locality when we consider very complicated operators in AdS.

We can also produce a version of a paradox related to strong-subadditivity. The strong-subadditivity of entropy is closely related to the ``monogamy'' of entanglement. We see, from the figure, that the region $B$ is entangled with the region $C$ because they are close together in AdS. However, as we explained above, all the information in $C$ is also present in $A$. Therefore, the region $B$ is also entangled with the region $A$. Thus, we seem to have a loss of the monogamy of entanglement --- with $B$ being entangled both with $C$ and with $A$.

Once again, the resolution to this paradox is very simple in this case. The regions $A,B,C$ are approximately independent local regions only when we consider low-order polynomials of simple operators localized in these regions. At a fine-grained level, the bulk Hilbert space does not contain tensor factors ${\cal H}_{A} \otimes {\cal H}_{B} \otimes {\cal H}_{C}$. In fact, we know from AdS/CFT that the entire Hilbert space is contained in ${\cal H}_{A}$. Therefore this seeming paradox is also resolved by recognizing that bulk locality is not exact.

\section{Discussion}
In the analysis above, we studied S-matrix elements in string theory and argued that they could not be computed perturbatively when the number of external particles became too large, even if each of these particles carried only a small amount of energy in string units.

This is an interesting limit because of the relation between the breakdown of perturbation theory for S-matrix elements and the loss of bulk locality. As we explained in section \ref{secpertloc}, these two thresholds are related because, in any theory of gravity, the Gauss law leads to small perturbative nonlocal commutators. When perturbation theory breaks down, we must consider the possibility that these commutators become significant  at leading order. 

As we reviewed in section \ref{sectoycomp}, these arguments --- which  relate the loss of locality to the combination of the Gauss law and the breakdown of perturbation theory --- are known to be correct at least in one entirely controlled setting:  empty AdS. In that setting, \cite{Banerjee:2016mhh} showed how an operator anywhere in AdS could be rewritten as a sufficiently complicated combination of operators that were all spacelike separated from the original operator.

We showed how these potential nonlocal effects have significant consequences for some versions of the information paradox. In particular, we considered the cloning paradox that appears because the information present in the infalling matter also appears to be present in the outgoing Hawking radiation. We argued that the usual framing of this paradox ignores the fact that to parse this information outside the black hole, we need to  measure $\Or[S]$-point correlators in the outgoing Hawking radiation, where $S$ is the entropy of the black hole. But the breakdown of gravitational perturbation theory, which we linked to a loss of locality, occurs precisely for $\Or[S]$-point correlators. We used this to argue that the complicated observable that is required to perform this measurement is not guaranteed to commute with the operator in the interior that acts on the infalling matter, even though these operators are spacelike separated. 

A similar loophole exists in the strong subadditivity paradox. The strong subadditivity paradox is framed in terms of the von Neumann entropies of different parts of the black hole spacetime. In a theory of gravity, we need to be careful about how to define the von Neumann entropy of a spacetime region. This entropy depends on how we truncate the ``algebra'' of operators localized in this region. We argued that, to obtain a paradox, it was necessary to define the von Neumann entropy in a fine-grained manner where we expand the algebra of operators to include polynomials that are very high order in the simple local operators from the region. But such high order polynomials may not commute on a spacelike slice. Therefore, with this choice of local algebra,  it is not valid to assume that the von Neumann entropies of geometrically disjoint regions will satisfy the same properties as von Neumann entropies of direct-product Hilbert spaces.

As additional evidence for the resolutions above, we also considered toy models of the strong subadditivity and cloning paradoxes that appear in empty AdS. In this setting, the resolution to these paradoxes is clear and indeed relies on the loss of locality that we have described above. 

Nevertheless, while we believe that these arguments are persuasive, we should indicate several points at which they can be made firmer. First, while Gauss-law commutators are clearly also present in flat space, and while our arguments for the breakdown of string perturbation theory are robust, we do not have the same controlled understanding of nonlocal effects in flat space or about black hole backgrounds. It would be nice to make this more precise.

Second, when we applied our analysis of the breakdown of perturbation theory to black hole evaporation, we were forced to use several ad-hoc discretizations to analyze our measurement protocols.  Also, it is possible to set up other protocols for extracting the information in Hawking radiation. For example, one protocol that seemingly avoids the need for measuring high-point correlators is simply to measure a lower-point function to arbitrarily high accuracy. However, then we see that to extract information we need to measure this lower-point correlation function to exponential accuracy in the entropy of the black hole \cite{Papadodimas:2012aq,Papadodimas:2013kwa}.  This observation is sensitive to  {\em very high} order terms in perturbation theory in ${E \over M_{\text{pl}}}$, where $E$ is the energy of a typical Hawking quantum. Just as perturbation theory breaks down for a very large number of particles, it is well known that it also breaks down at very high loop order. Therefore, even in a measurement protocol that involves the measurement of low-point correlators to very high accuracy, we would expect nonperturbative and ---  by the arguments above ---  nonlocal effects to be important. It would be nice to frame this entire analysis of measurement protocols within a unified framework that makes it clear that extracting information from Hawking radiation necessarily involves some nonperturbative physics.

We would also like to caution the reader on a few points related to our discussion of the information paradox. Our discussion of the information paradox here does not address the question of reconstructing the interior of large black holes in AdS/CFT. This has been the subject of considerable recent discussion \cite{Almheiri:2013hfa,Marolf:2013dba} and involves several interesting and partially unresolved puzzles \cite{Raju:2016vsu,gravresearch2017}. Here we are only concerned with the information paradox for evaporating black holes. Second, we note that some alternative resolutions to these paradoxes have been proposed in \cite{Mathur:2008kg,Bena:2015dpt}, which do not rely on the nonlocal effects described here. In forthcoming work, we hope to provide a comparative analysis of these resolutions.

Finally, we should also mention another important interpretation of the breakdown of perturbation theory that we have described. In this paper, and especially in section \ref{secinfo} we considered this breakdown in a setting where a black hole already exists in spacetime, and where we are making complicated measurements in the outgoing Hawking radiation. 

However, it is also possible to consider high-point S-matrix elements in the absence of a black hole. Simple dimensional analysis then tells us that at the threshold of perturbative breakdown, we expect the {\em formation} of black holes.

This can  be seen as follows. If the momentum components of each particle are of order $E$ then each particle is delocalized over a length of at least $\Or[1/E]$ and so we expect that the smallest black hole that can be made by many such particles also has radius $\Or[1/E]$. The mass of such a black hole in d dimensions is $\Or[M_{\text{pl}}^{d-2}/ E^{d-3}]$ which requires $\Or[(M_{\text{pl}}/ E)^{d-2}]$ particles of energy $E$. But this is precisely the value of $n$ at which our breakdown occurs. So we see that the breakdown of perturbation theory that we have described is also related to black-hole formation. This analysis of the threshold of black-hole formation, and  its relation to the loss of locality was also discussed in \cite{Giddings:2006vu}.

Separate from the issue of the breakdown of perturbation theory, the analysis of string scattering at large $n$ is of intrinsic interest.  It would be very interesting, for example, to examine the limit where $n$ becomes large but we are still below the threshold of perturbative breakdown and ask whether string perturbation theory simplifies in this limit.  

Finally, it would be nice to have a better understanding of the nonperturbative effects that become important beyond the perturbative threshold.  The recognition that string perturbation theory breaks down at large genus spurred the discovery of D-branes, and we would like to understand if the large-n breakdown is related to similar interesting nonperturbative phenomena.

\section*{Acknowledgments}
We are grateful to Nima Arkani-Hamed, Clay Cordova, Steve Giddings, Rajesh Gopakumar, David Gross, Subhojoy Gupta, Gautam Mandal, Samir Mathur, Shiraz Minwalla,  Yasunori Nomura,   Ashoke Sen, Aninda Sinha, Sandip Trivedi, Edward Witten and especially R. Loganayagam and Kyriakos Papadodimas for useful discussions. We would also like to acknowledge discussions with the participants of the the ICTS discussion meeting, String Theory: Past and Present,  ICTS/STTP/2017/01, the ICTS program, Fundamental Problems in Quantum Physics, ICTS/Prog-FPQP/2016/11, and the Bangalore Area Strings Meeting, ICTS/Prog-BASM/2016/07. We would like to acknowledge the hospitality of the Strings 2017 conference (Tel Aviv), the National Strings Meeting (Pune),  the program on Black Holes and Emergent Spacetime at Nordita (Stockholm), and the Mandelstam theoretical physics school and workshop (Durban). Our numerical computations were performed on the high-performance computing cluster at ICTS: {\em Mowgli.} We would also like to thank the Astrophysical Relativity group at ICTS for giving us access to their computational facilities.

\appendix

\section*{Appendix}
\section{Additional details about the numerical algorithm \label{appnum}}
\paragraph{Root Finding \\}
Our root-finding algorithm is a generalization of the Newton algorithm.  We use the GNU Scientific Library's ``hybridj'' function \cite{galassi2009gnu}. 
We are interested in solving the equations $E_i = 0$ specified in \eqref{scateq}. The basic multi-dimensional Newton's algorithm starts with an initial guess $z_i$. Each step updates this to $z'_i = z_i - J^{-1}_{i j} z_j$ where $J_{i j} = \partial_i E_j$.   This algorithm was improved by Powell \cite{powell1970hybrid} who pointed out that the steps should be restricted to $|z_i' - z_i| \leq \delta$ where $\delta$ is a dynamically determined trust region. If the step is outside the trust region, then Powell's hybrid algorithm uses a combination of the Newton step and gradient flow by taking a step along
\[
z'_i - z_i = - \alpha J^{-1}_{i j} E_j - \beta \nabla_i \sum_j |E_j|^2.
\]
Here $\alpha, \beta$ are fixed by demanding that the step minimize the norm of the function $\sum_j |E_j|^2$, with the constraint that the step also stay inside the trust region.  After the step is completed, the algorithm then determines if $|E_i(z_i')| \leq |E_i(z_i)|$. If this is the case, the step is accepted and $\delta$ is increased. Otherwise the step is rejected and $\delta$ is decreased. 

In our case, we choose a starting point by choosing a random value for each $z_i$ satisfying $-50 \leq \text{Re}(z_i) \leq 50$ and $-50 \leq \text{Im}(z_i) \leq 50$. We then use the algorithm detailed above to attempt to find a root.

Even the simple one-dimensional Newton method is not guaranteed to converge. The convergence properties of multi-dimensional problems are, in general, even worse. However, we were pleasantly surprised to find
that the the hybrid algorithm converges rather well to a root of the scattering equations, starting with a random initial guess as detailed above.  This indicates that most points in our starting region  are within the basin of attraction of some root.

Some versions of the hybrid algorithm scale the trust region with a factor that is dynamically determined using the Jacobian but we found, empirically, that the unscaled hybrid method works better than the scaled hybrid method. The vanilla Newton's method does not converge at all.

Note that the algorithm only solves $(n-3)$ equations. This may sometimes lead to ``false roots'' where  some $z_i$ drifts off to $\infty$ thereby causing the corresponding $S_i \rightarrow 0$ within numerical errors. These false roots can be detected by checking if the roots that have been found also satisfy the other $3$ equations. 

A judicious gauge choice is required to ensure that the number of false solutions is kept within limit. We find that the algorithm converges well if we choose $z_1 = 10^6, z_2 = 0.0, z_3 = 1.0$. We also find the best performance by then solving the $(n-3)$ equations $(E_2 \ldots E_{n-1}) = 0$. 

\paragraph{Random Number Selection \\}
To ensure that we sample the space of roots as uniformly as possible, the choice of random number generator is also important. In fact hidden correlations in pseudo-random number generators used for Monte Carlo sampling can sometimes lead to significant errors in the final result \cite{ferrenberg1992monte}. We use the random number generator ``gsl\_rng\_ranlxs2'' provided with the GNU Scientific Library. This uses a variant of the RANLUX algorithm of Luscher \cite{Luscher:1993dy}.

\paragraph{Momentum Selection \\}

It is also important to correctly sample scattering amplitudes in momentum space. We are interested in generating external kinematics that are uniformly distributed in phase space according to the measure \eqref{phasespacemeasure}. For this, we can use the RAMBO algorithm \cite{kleiss1986new} which is used to generate events in particle physics. This algorithm works well for massless particles where it can generate external kinematics precisely according to the distribution \eqref{phasespacemeasure}. The algorithm proceeds as follows. 
\begin{enumerate}
\item
First we pick a set of $n/2$ momenta corresponding to ingoing particles using the following procedure for each momentum. We choose the energy $q_0$ from the distribution 
\be
p(q_0) = {1 \over \Gamma(d-2)} (q_0)^{d-3} e^{-q_0}.
\ee
The other components of the momentum are picked by picking a vector on a sphere $\vec{n} \in \Omega_{d-1}$ and then choosing the momentum to be
\be
(q_0, q_0 \vec{n})
\ee
We will call this set of momenta $q_i$.
\item
Let the center of mass-momentum of the momenta so obtained by $\vec{P}$. Now, we simultaneously boost all the momenta formed so that the new center of mass-momentum becomes $0$.
\item
We now rescale the momenta so that the total energy becomes ${n E \over 2}$. The transformations corresponding to this boost and rescaling are as follows.
\be
\label{boostrescale}
k_{i}^{0}=x(g q_{i}^{0}+\vec{b} \cdot \vec{q}_{i}), \quad \vec{k}_{i}=x(\vec{q}_{i}+\vec{b}q_{i}^{0}+a(\vec{b} \cdot \vec{q}_{i})\vec{b}),
\ee
where 
\be
\begin{split}
&\vec{b}=-\frac{\vec{Q}}{M}, \quad x=\frac{n E}{2 M}, \quad g= \frac{Q^{0}}{M}, \\
& a=\frac{1}{1+g}, \quad Q^{\mu}=\sum_{i=1}^{n \over 2}q_{i}^{\mu}, \quad M=\sqrt{Q^{2}}.
\end{split}
\ee
\item
The momenta $k_1 \ldots k_{n \over 2}$ obtained above can be taken to be the physical incoming momenta. 
We now pick another set of $n/2$ momenta corresponding to outgoing particles and use the same procedure to ensure that their center of mass momentum is $0$ and total center of mass energy is ${n E \over 2}$.
\end{enumerate}
It was shown in \cite{kleiss1986new} that this procedure leads to momenta that are uniformly distributed in phase space.

\paragraph{Polarization tensor selection \\}
As explained in the text, we choose polarization tensors $\epsilon_{\mu \nu} = \epsilon_{\mu} \epsilon_{\nu}$ which are simply outer-products of real polarization vectors. For a momentum given by $(k_0, k_0 \vec{n})$, we first perform an orthogonal transformation in the spatial directions so that the components of this momentum  become $\vec{n}^1 = 1$ and $\vec{n}^i = 0$ for $i = 2 \ldots d-1$. We then choose a random unit vector with components $\epsilon^1 = 0$ and satisfying $\sum_{i = 2}^{d-1} \epsilon^{i} \epsilon^{i} = 1$. We then perform the inverse orthogonal transformation so that the momentum is rotated back to its original value. By acting with the same transformation on the polarization vector, we generate a transverse polarization vector that has no timelike component. We then simply take the outer product of this polarization vector with itself to obtain a polarization tensor for the spin-2 particles.

\paragraph{Prefactor estimation \\}
The primary difficulty in estimating the prefactor $P_n$ originates in evaluating the term that involves the Wick contractions. We find that the most stable solution is obtained by simply selecting a random sample of pairings and by multiplying the mean of this sample with the total number of pairings. Note that trying to identify the Wick contraction that gives the largest contribution is equivalent to pairing the $n$-points so that the product of the distances between pairs is the minimum. This is similar to a traveling salesman problem, and therefore cannot be solved for large $n$ in any computationally efficient manner. 

In our computations, we sum over a sample of 1000 samples of different possible pairings, each time we need to evaluate the term $P_n$. 

The log normal distribution of the samples of amplitudes that we discussed in section \ref{secnumericalerrors} is also visible in these samples. Therefore it is possible that a better algorithm that efficiently locates nearest neighbours might increase the significance of this term in the numerical results.

\paragraph{Parallelization \\}
Our calculations can be entirely parallelized, since every solution of the scattering equations is independent of every other solution. In principle, it is possible that two random choices of initial points lead to the same solution, and therefore while generating random parallel samples of solutions, it is important to check that we are not generating any duplicate solutions. In practice we have never found any duplicate solutions for $n > 10$.  Therefore,  we were able to perform our calculations efficiently by using a number of CPUs running in parallel in a large computational cluster. In advance of our final computations, we also generated some preliminary data-sets on a desktop workstation for which we found that GNU parallel \cite{Tange2011a} was a useful tool.

\section{Asymptotic correlators and the S-matrix \label{appcorrsmat}}
In this appendix, we review how correlators of fields at null infinity can be rewritten as Fourier transforms of S-matrix elements using the LSZ formula. 

Consider a scalar field, $\phi(t, \vect{x})$. As above, in this appendix we will use  $\vect{x}$ --- to indicate spatial vectors. Now, as in section \ref{secinfo} we wish to consider the following correlator
\be
C(u_1, \ldots u_s, v_{s+1} \ldots v_n) = \text{lim} \int \langle T\left\{\phi(t_1, \vect{x}_{1}) \ldots \phi(t_s, \vect{x}_{s}) \phi(t_{s+1}, \vect{x_{s+1}}), \ldots \phi(t_n, \vect{x}_{n}) \right\} \rangle \, \prod d\Omega_p, 
\ee
with
\be
\begin{split}
&r_m = |\vec{x}_m| \rightarrow \infty, t_1 \ldots t_s \rightarrow \infty, t_{s+1} \ldots t_n \rightarrow -\infty, \\
&u_m = t_m - r_m; ~ v_m = t_{m} + r_{m},
\end{split}
\ee
where we keep $u_i$ finite for $i = 1 \ldots s$ and $v_i$ finite for $i = s+1 \ldots n$ and $p = 1 \ldots n$.

This is a s-wave correlator with some points taken the past null infinity, and some others taken to future null infinity. The spherical integrals are taken over the $(d-2)$-sphere at infinity. We will only consider massless scalar particles here. As we have mentioned, it is not difficult to generalize this to higher spherical harmonics.
Now, the time-ordered correlator can be written as the Fourier transform
of the momentum space time-ordered Green's function so the correlator of interest becomes 
\be
C(u_i, v_j) = \lim \int G(\omega_1, \vec{k}_1 \ldots \omega_n, \vec{k}_n) e^{i \sum_q \omega_q t + \vec{k}_q \cdot \vec{x}_{q}} \prod d \omega_q \, d^{d-1} \vec{k}_q \, d\Omega_p.
\ee
Here, to lighten the notation we have suppressed the distinct range of indices on $u,v$ although it is understood that $i = 1 \ldots s$ and $j = s +1 \ldots n$. We have also split up the momentum
of each insertion into an energy and a spatial component as $(\omega_i, \vect{k}_i)$. Note that these momenta do {\em not} have to be on-shell. 

We can do the position space spherical integrals as follows. For each coordinate $\vect{x}_{m}$, we can choose coordinates so that $\vec{k}_m \cdot \vect{x}_{m} = k_m r_m \cos \theta_m$. Here, through a slight abuse of notation we have used $k_m \equiv |\vec{k}_m|$ although at other times we also use $k_m$ to denote the full on-shell $d$-momenta. We hope that this will not lead to confusion. We also have $d \Omega_m = {2  \pi^{d-2 \over 2}  \over \Gamma({d  - 2\over 2})} (\sin \theta_m)^{(d-3)} d \theta_m$. We can then do the remaining angular integrals through
\be
\int_0^{\pi} e^{i k_m r_m \cos \theta_m } (\sin \theta_m)^{d-3} d \theta_m = \sqrt{\pi } \Gamma \left(\frac{d-2}{2}\right) \, _0\tilde{F}_1\left(;\frac{d-1}{2};-\frac{k_m^2 r_m^2}{4}\right),
 \ee
where $\tilde{F}$ is the regularized hypergeometric function. 
Moreover, we are only interested in the limit that $r_m \rightarrow \infty$. In this limit, expanding the hypergeometric function and keeping only the leading terms we find that
\be
\begin{split}
\int_0^{\pi} e^{i k r  \cos(\theta)} \sin(\theta)^{d-3} d \theta 
\underset{r \rightarrow \infty}{\longrightarrow} i 2^{\frac{d}{2}-2}  (k r)^{-d/2 + 1} \left(e^{i k r - i \pi {d\over 4}}   - e^{-i k r +  i \pi {d \over 4}} \right) \Gamma\left(\frac{d-2}{2}\right).
\end{split}
\ee
For convenience define the constant
\be 
N = -{1 \over 2}  (2 \pi)^{{d \over 2}}  e^{- i \pi {d \over 4}},
\ee
which comes from combining the various constants that appear above and we have added an extra factor of $\pi i$ using some foresight.

We find that our correlator reduces to 
\be
\begin{split}
C(u_i, v_j)  &= \lim  {1 \over r^{n({d \over 2} - 1)}} \int G(\omega_1, \vec{k}_1 \ldots \omega_n, \vec{k}_n) \prod_p \Bigg[  \, {d^{d-1} \vect{k}_p \, d \omega_p  \over k_p^{{d \over 2} - 1}}  \\ &\times \left({N \over \pi i} e^{i {\omega_p + k_p \over 2} (t_p + r_p) + i {\omega_p - k_p \over 2} (t_p - r_p)} - {N^* \over  \pi i} e^{i {\omega_p + k_p \over 2} (t_p - r_p) +i {\omega_p - k_p \over 2} (t_p + r_p)} \right) \Bigg].
\end{split}
\ee
We change variables to $k^+_p = {\omega_p + k_p \over 2}; \quad k^{-}_p = {\omega_p - k_p \over 2}$. We also rewrite the momentum space measure as
\be
{d \omega d^{d-1} \vect{k}} = 2 d k^+ d k^{-} \left(k^+ - k^{-} \right)^{d - 2} d \hat{k}.
\ee
We note that when we do this, we need to be careful about the limits of integration over $k^+$ and $k^{-}$ since $k$  must be positive and so $k^+ \geq k^{-}$.
This leads to
\be
\begin{split}
C(u_i, v_j) = \lim   {1 \over r^{n({d \over 2} - 1)}} \int &G(\omega_1, \vec{k}_1 \ldots \omega_n, \vec{k}_{n}) \prod_p \left({2 N \over  \pi i} e^{i k_p^+ v_p + i k_p^{-} u_p} - {2 N^* \over \pi i} e^{i k_p^{+} u_p + i k_p^{-} v_p} \right)\\ 
& \times (k_p^+ - k_p^-)^{d-2 \over 2}  d k_p^+ \, d k_p^- \, d \hat{k}_p. 
\end{split}
\ee
Now, we need the following formula
\be
\lim_{k \rightarrow \infty} \int_a^{\infty} {f(x) \over x - i \epsilon} e^{i k x} d x = (2 \pi i ) f(0) \theta(-a),
\ee
where the order of limits is that, after doing the integral, $\epsilon$ is taken to zero {\em before} $k$ is taken to infinity. 
To prove this, simply consider the contour that goes up in an arc from $x = \infty$, and comes down vertically on $a$ through the complex $x$ plane. This contour encloses the pole at $x = i \epsilon$, if $a < 0$, and not otherwise. Other poles in the function $f(x)$ for $\text{Im}(x) > 0$ do not contribute in the limit where $k \rightarrow \infty$. Also, the part of the contour that is not on the real line does not contribute in the limit where $k \rightarrow \infty$.  Repeating this, for other cases, which will be useful below, we get the following additional identities:
\be
\begin{split}
&\lim_{k \rightarrow \infty} \int_a^{\infty} {f(x) \over x + i \epsilon} e^{i k x} d x = 0, \quad \lim_{k \rightarrow -\infty} \int_a^{\infty} {f(x) \over x - i \epsilon} e^{i k x} d x = 0;  \\ &\lim_{k \rightarrow -\infty} \int_a^{\infty} {f(x) \over x + i \epsilon} e^{i k x} d x = -2 \pi i f(0) \theta(-a). 
\end{split}
\ee

Now, the key point is that the Green's function has a pole when the momenta go on shell, with a residue that is the S-matrix element. In a scheme, where the wave-function renormalization factors are unity (this is simply a question of the normalization of
the two point function), we have that
\be
G(\omega_1, \vec{k}_{1} \ldots \omega_n, \vec{k}_{n}) = {\widehat{M}(k_1 \ldots k_n) \over \prod  (4 k_p^+ k_p^- + i \epsilon)} + F(\omega_p, \vec{k}_{p}),
\ee
where $F$ is some function that does not have a pole in the limit where  $k_p^+ k_p^- = 0$ and the momenta that enter the scattering amplitude are now on-shell. We use $\widehat{M}$ to denote the amplitude with the momentum-conserving delta function included
\be
\widehat{M}(k_1, \ldots k_n) = M(k_1, \ldots k_n) (2 \pi)^d \delta(\sum_{p=1}^n k_p)
\ee
Now, as we take the points to either future or past null infinity, for each point either $u_p \rightarrow -\infty$ or $v_p \rightarrow \infty$. In each of these limits, the momentum space integrals simplify as follows.
\be
\lim_{u_p \rightarrow -\infty} \int_{-\infty}^{\infty}  d k_p^{-} \int_{k_p^-}^{\infty}  d k_p^+ {\widehat{M}(k_1 \ldots k_n) (k_p^{+} - k_p^{-})^{d - 2 \over 2} \over (4 k_p^+ k_p^- + i \epsilon)} e^{i k_p^+ u_p + i k_p^- v_p}  = 0.
\ee
This is because, to pick up a contribution, we need the pole in $k_p^+$ to be in the lower half plane, and this requires $k_p^- > 0$. However, then by the identity above, the integral vanishes. Similarly, we find
\be
\begin{split}
&\lim_{u_p \rightarrow -\infty} \int_{-\infty}^{\infty}  d k_p^{+} \int_{-\infty}^{k_p^+}  d k_p^- {\widehat{M}(k_1, \ldots k_n)(k_p^+ - k_p^-)^{d - 2 \over 2} \over  (4 k_p^+ k_p^- + i \epsilon)} e^{i k_p^- u_p + i k_p^+ v_p} \\  &= \int_{0}^{\infty} d k_p^+ {\pi i \over 2} (k_p^+)^{{d \over 2} - 2} \widehat{M}(k_1, \ldots k_n) e^{i k_p^+ v_p},
\end{split}
\ee
and,
\be
\begin{split}
&\lim_{v_p \rightarrow \infty} \int_{-\infty}^{\infty} d k_p^{-} \int_{k_p^{-}}^{\infty} d k_p^+ {\widehat{M}(k_1, \ldots k_n) (k_p^{+} - k_p^{-})^{d - 2 \over 2} \over  (4 k_p^+ k_p^- + i \epsilon)} e^{i k_p^- u_p + i k_p^+ v_p}  \\ &= -e^{i \pi d \over 2} \int_{-\infty}^{0} d k_p^- {\pi i \over 2} (k_p^+)^{{d \over 2} - 2} \widehat{M}(k_1 \ldots k_n) e^{i k_p^- u_p},
\end{split}
\ee
while
\be
\lim_{v_p \rightarrow \infty} \int_{-\infty}^{\infty} d k_p^{+} \int_{-\infty}^{k_p^+} d k_p^- {\widehat{M}(k_1, \ldots k_n)\over (4 k_p^+ k_p^- + i \epsilon)} e^{i k_p^- v_p + i k_p^+ u_p}  = 0.
\ee
Putting all of this together, we find the final formula
\be
\begin{split}
C(u_i, v_i) = {N^n e^{i \pi s d \over 2} \over r^{n {d - 2 \over 2}}}  \int \prod_{i=1}^n &d \hat{k}_p d E_p  (E_p)^{d/2 - 2} \widehat{M}(k_1, \ldots k_n) e^{-i \sum_{p=1}^s E_p u_p + i \sum_{p=s+1}^n  E_p v_p},
\end{split}
\ee
where now the integral over the energies $E_i$ goes over $(0, \infty)$ and
\be
k_i \equiv (-E_i, E_i \hat{k}_i) \quad \text{for}~i = 1 \ldots s
\ee
and
\be
k_i \equiv (E_i, E_i \hat{k}_i) \quad \text{for}~i = s+1 \ldots n.
\ee
This is the formula that is used in section \ref{secinfo}. 

\section{Volume of phase space \label{phasespacevol}}
Now we compute the volume of phase space using the formulas in Appendix A of \cite{kleiss1986new}. The only difference with their analysis is that we work in $d$ dimensions whereas they were exclusively concerned with $d = 4$. We consider $m = {n \over 2}$ particles with total $d$-momentum $P = (Q, 0, \ldots)$. Recall that in the text we had $Q = {n E \over 2}$ and $m = {n \over 2}$ but we will work with general parameters here. 

Then, on grounds of symmetry it must be true that the volume of phase space is  $V_m = v_m Q^{(d-2)m - d}$. We determine $v_m$ by recursion. We have
\be
V_m = \int \prod_{j=1}^m {d^{d} p_j  \over (2 \pi)^{d}} (2 \pi) \delta(p_j^2) (2 \pi)^d \delta^d( \sum p_j - P).
\ee
We insert $1$ in the integral through
\be
1 = \int (2 \pi)^d \delta^d (\sum_{j=1}^{m-1} p_j - q) {d^d q \over (2 \pi)^d} (2 \pi) \delta(q^2 - w^2) {d w^2 \over (2 \pi)}.
\ee
Using Lorentz invariance, and in the presence of the delta functions above, we can write the integral over $m-1$ particles as
\be
 \int \prod_{j=1}^{m-1} {d^{d} p_j  \over (2 \pi)^{d}} (2 \pi) \delta(p_j^2) (2 \pi)^d \delta^d( \sum p_j - q) =  v_{m-1} w^{{(d-2) (m-1) - d}}.
\ee
Inserting this we find that 
\be
V_m = v_{m-1} w^{{(d-2) (m-1) - d}} \int {d^{d} p_m  \over (2 \pi)^{d}} (2 \pi) \delta(p_m^2) (2 \pi)^d \delta^d( q + p_m - P) (2 \pi) \delta(q^2 - w^2) {d w^2 \over (2 \pi)} {d^d q \over (2 \pi)^d}.
\ee

The last integral over $p_m$ and $q$  can be done as follows. First, we simply do the $q$-integral using the delta functions. The measure over $p_m$ can be written as 
\be
{d^{d-1} p_m \over (2 \pi)^{d-1} 2 |p_m|} = {\Omega_{d-2} \over 2 (2 \pi)^{d-1}} |p_m|^{d-3} d |p_m|.
\ee
The remaining integral becomes
\be
{\Omega_{d-2} \over 2 (2 \pi)^{d-1}}  \int d |p_m| |p_m|^{d-3}   v_{m-1}  w^{{(d-2) (m-1) - d}}  \delta(|p_m| + \sqrt{|p_m|^2 + w^2} - Q) {1 \over 2 \sqrt{|p_m|^2 + w^2}}.
\ee
The delta function is solved by
\be
|p_m| = {Q^2 - w^2 \over 2 Q},
\ee
at which point we also have (after including the Jacobian from the delta function) and doing the integral over $|p_m|$
\be
\begin{split}
V_m &= v_{m-1} {\Omega_{d-2} \over 4 (2 \pi)^{d-1}} \int_0^Q (2 w ) d w 2^{(d-3)} Q^{2-d} 2^{3-d} (Q^2 - w^2)^{(d-3)} w^{(d-2)(m-1) - d}  \\
&= v_{m-1} {\Omega_{d-2} \over 4 (2 \pi)^{d-1}} Q^{(d-2) m - d}\frac{2^{3-d} \Gamma (d-2) \Gamma \left(\frac{1}{2} (d-2)
   (m-2)\right)}{\Gamma \left(\frac{1}{2} (d-2) m\right)}.
\end{split}
\ee
This leads to the recursion relation
\be
v_m = v_{m-1} {\Omega_{d-2} \over (2 \pi)^{d-1}}\frac{2^{1-d} \Gamma (d-2) \Gamma \left(\frac{1}{2} (d-2)
   (m-2)\right)}{\Gamma \left(\frac{1}{2} (d-2) m\right)}.
\ee

We can also compute $V_2$ using 
\be
V_2 =  \int {d^{d} p_1  \over (2 \pi)^{d}} (2 \pi) \delta(p_1^2) (2 \pi)^d \delta^d( p_1 + p_2 - P) (2 \pi) \delta(p_2^2) {d^d p_2 \over (2 \pi)^d},
\ee
which just gives 
\be
V_2 = {\Omega_{d-2} \over (2 \pi)^{d-1}}  \int |p_1|^{(d-4)} {d |p_1| \over 4} (2 \pi ) \delta(2 |p_1| - w) = {\Omega_{d-2} \over 8 (2 \pi)^{d-2}} \left({w \over 2} \right)^{d-4},
\ee
or
\be
v_2 = {\Omega_{d-2} \over (2 \pi)^{d-2} 2^{d-1}}.
\ee

The recursion relation we have is of the form
\be
v_m = C {\Gamma(k (m-2)) \over \Gamma( k m)} v_{m-1},
\ee
and this is solved by
\be
v_m  = C^{m-2} {\Gamma(k) \Gamma(2 k)  \over \Gamma(k m) \Gamma(k(m-1))} v_2 .
\ee 
Specializing to the constants we have we find that 
\be
\begin{split}
v_m &= \left({2^{1-d} \Gamma (d-2) \Omega_{d-2} \over (2 \pi)^{d-1}} \right)^{m-2}  \frac{\Gamma\left( \frac{1}{2} (d-2) \right)  \Gamma (d-2)}{\Gamma \left(\frac{1}{2} (d-2) m\right) \Gamma \left(\frac{1}{2} (d-2) (m-1)\right)} v_2 \\
&=  \left({2^{1-d} \Gamma (d-2) \Omega_{d-2} \over (2 \pi)^{d-1}} \right)^{m-1}  2 \pi \frac{\Gamma\left( \frac{1}{2} (d-2) \right) }{\Gamma \left(\frac{1}{2} (d-2) m\right) \Gamma \left(\frac{1}{2} (d-2) (m-1)\right)}.
\end{split}
\ee
If we now substitute $\Omega_{d-2} = {2 \pi^{d-1 \over 2}  \over \Gamma({d - 1 \over 2})}$ and use the identity $\Gamma(x) \Gamma(x+{1 \over 2})  = 2^{(1 - 2 x)} \sqrt{\pi} \Gamma(2 x)$ and do some simplifications, we find that 
\be
v_m = (2 \pi) (4 \pi)^{-(m-1) d \over 2} {\Gamma({d - 2 \over 2})^{m} \over \Gamma \left(\frac{1}{2} (d-2) m\right) \Gamma \left(\frac{1}{2} (d-2) (m-1)\right)}.
\ee
The full phase space volume can then be written as
\be
V_m  =  \left({Q \over m} \right)^{(d-2)m - d} (2 \pi) (4 \pi)^{-(m-1) d \over 2} {\Gamma({d - 2 \over 2})^{m} m^{(d-2) m - d} \over \Gamma \left(\frac{1}{2} (d-2) m \right)  \Gamma \left(\frac{1}{2} (d-2) (m-1)\right)}.
\ee
Putting $m = {n \over 2}$ and $Q = {n E \over 2}$, we find the formula \eqref{phasespacefactor} used in the text.

\bibliographystyle{JHEP}
\bibliography{references}

\end{document}